\documentclass[twocolumn]{aastex63}
\usepackage[normalem]{ulem}
\usepackage{makecell}
\usepackage{amsmath}
\usepackage{soul}
\usepackage{float}
\usepackage[caption = false]{subfig}
\usepackage{graphicx}

\newcommand{\halpha}{{\rm H}\alpha}
\newcommand{\haew}{{\rm H}\alpha \,{\rm EW}}
\newcommand{\lhalbol}{L_{{\rm H}\alpha}/L_{\rm bol}}

\newcommand{\grp}{{G_\mathrm{RP}}}

\newcommand{\nagecal}{$892$ }
\newcommand{\nmgmem}{$871$ }
\newcommand{\nnewmgmem}{$7$ }
\newcommand{\nlitsearch}{$89,270$ }
\newcommand{\nlitsearchothers}{$15,540$ }
\newcommand{\nlitsearchgaia}{$86,918$ }
\newcommand{\percentingaia}{$97\%$}
\newcommand{\nhighqualparallax}{$27,201$ }
\newcommand{\nmdwarfs}{$25,720$ }
\newcommand{\nnotspt}{$6\%$ }
\newcommand{\nrepeated}{$726$ }
\newcommand{\nuniquesources}{$24,330$ }
\newcommand{\ncompatible}{$24,202$ }
\newcommand{\naccretors}{$45$ }
\newcommand{\doublestarsremoved}{$6$ }

\newcommand{\percenthighqualparallax}{$31\%$}
\newcommand{\nnotacc}{$24,166$ }
\newcommand{\nnotinmg}{$23,274$ }
\newcommand{\wdmpairs}{$61$ }
\newcommand{\wdmpairshighchancealignment}{$22$ }
\newcommand{\wdmpairslowchancealignment}{$39$ }
\newcommand{\wdmpairsage}{$21$ }
\newcommand{\nrepearedagecalibrators}{$155$ }
\newcommand{\nactivecalibrators}{$856$ }
\newcommand{\changedtext}[1]{#1}

\received{August 17 2020}
\accepted{March 30 2021}
\submitjournal{The Astronomical Journal}

\begin{document}

\title{Calibration of the $\halpha$ Age-Activity relation for M dwarfs}
\shorttitle{Age activity relation}
\shortauthors{Kiman et al.}

\author[0000-0003-2102-3159]{Rocio Kiman}
\affil{Department of Physics, Graduate Center, City University of New York, 365 5th Ave, New York, NY 10016, USA}
\affil{Department of Astrophysics, American Museum of Natural History, Central Park West at 79th St, New York, NY 10024, USA}
\affil{Hunter College, City University of New York, 695 Park Ave, New York, NY 10065, USA}
\email{rociokiman@gmail.com}

\author[0000-0001-6251-0573]{Jacqueline K. Faherty}
\affil{Department of Astrophysics, American Museum of Natural History, Central Park West at 79th St, New York, NY 10024, USA}

\author[0000-0002-1821-0650]{Kelle L. Cruz}
\affil{Department of Physics, Graduate Center, City University of New York, 365 5th Ave, New York, NY 10016, USA}
\affil{Department of Astrophysics, American Museum of Natural History, Central Park West at 79th St, New York, NY 10024, USA}
\affil{Hunter College, City University of New York, 695 Park Ave, New York, NY 10065, USA}
\affil{Center for Computational Astrophysics, Flatiron Institute, 162 5th Avenue, New York, NY 10010 USA}

\author[0000-0002-2592-9612]{Jonathan Gagn\'e}
\affiliation{Plan\'etarium Rio Tinto Alcan, Espace pour la Vie, 4801 av. Pierre-de Coubertin, Montr\'eal, Qu\'ebec, Canada}
\affiliation{Institute for Research on Exoplanets, Universit\'e de Montr\'eal, D\'epartement de Physique, C.P.~6128 Succ. Centre-ville, Montr\'eal, QC H3C~3J7, Canada}

\author[0000-0003-4540-5661]{Ruth Angus}
\affil{Department of Astrophysics, American Museum of Natural History, Central Park West at 79th St, New York, NY 10024, USA}
\affil{Center for Computational Astrophysics, Flatiron Institute, 162 5th Avenue, New York, NY 10010 USA}
\affil{Department of Astronomy, Columbia University, 116th St \& Broadway, New York, NY 10027, USA}

\author[0000-0002-7224-7702]{Sarah J. Schmidt}
\affil{Leibniz-Institute for Astrophysics Potsdam (AIP), An der Sternwarte 16, 14482, Potsdam, Germany}

\author[0000-0003-3654-1602]{Andrew W. Mann}%
\affiliation{Department of Physics and Astronomy, The University of North Carolina at Chapel Hill, Chapel Hill, NC 27599, USA} 

\author[0000-0001-8170-7072]{Daniella C. Bardalez Gagliuffi}
\affil{Department of Astrophysics, American Museum of Natural History, Central Park West at 79th St, New York, NY 10024, USA}

\author[0000-0002-3252-5886]{Emily Rice}
\affil{Department of Physics, Graduate Center, City University of New York, 365 5th Ave, New York, NY 10016, USA}
\affil{Department of Astrophysics, American Museum of Natural History, Central Park West at 79th St, New York, NY 10024, USA}
\affil{Macaulay Honors College, City University of New York, 35 W. 67th street, New York, NY 10024, USA}

\begin{abstract} 
In this work, we calibrate the relationship between $\halpha$ emission and M dwarf ages. We compile a sample of $\nagecal$ M~dwarfs with $\halpha$ equivalent width ($\haew$) measurements from the literature that are either co-moving with a white dwarf of known age ($\wdmpairsage$ stars) or in a known young association ($\nmgmem$ stars)\footnote{The sample is available to download from \href{https://doi.org/10.5281/zenodo.4659293}{Zenodo}.}. In this sample we identify $\nnewmgmem$ M dwarfs that are new candidate members of known associations. By dividing the stars into active and inactive categories according to their $\haew$ and spectral type (SpT), we find that the fraction of active dwarfs decreases with increasing age, and the form of the decline depends on SpT. Using the compiled sample of age-calibrators we find that $\haew$ and fractional $\halpha$ luminosity ($\lhalbol$) decrease with increasing age. \changedtext{$\haew$ for SpT$\leq {\rm M}7$ decreases gradually up until $\sim$\,$1$\,Gyr. For older ages, we found only two early M dwarfs which are both inactive and seem to continue the gradual decrease. We also found 14 mid-type out of which 11 are inactive and present a significant decrease of $\haew$, suggesting that the magnetic activity decreases rapidly after $\sim$\,$1$\,Gyr.} We fit $\lhalbol$ versus age with a broken power-law and find an index of $-0.11^{+0.02}_{-0.01}$ for ages $ \lesssim 776$\,Myr. The index becomes much steeper at older ages however a lack of field age-calibrators ($\gg 1$\,Gyr) leaves this part of the relation far less constrained. Finally, from repeated independent measurements for the same stars we find that $94\%$ of these has a level of $\haew$ variability $\leq 5$\,${\rm \AA}$ at young ages ($<1$\,Gyr).

\end{abstract}


\date{\today}

\section{Introduction}
\label{sec:intro}

M dwarfs are the coolest and most abundant stars in the Milky Way \citep{Gould1996,Bochanski2010}.
As the lifetime of M dwarfs is longer than the current age of the Universe \citep[e.g., ][]{Fagotto1994,Laughlin1997} those that we find throughout the Galaxy span a wide range of ages. Therefore, M~dwarfs are a rich stellar population for a statistical analysis of the Milky Way evolution, dynamics and composition \citep[e.g., ][]{Gizis2002,Faherty2009,Bochanski2007a,Bochanski2010,Jones2011}. 
In addition, M~dwarfs are attractive targets to study exoplanet populations because the occurrence of small rocky exoplanets is higher for M~dwarfs than any other spectral type.  Furthermore, it is easier to detect small planets around low-mass stars than around higher-mass stars due to the large reflex motion \citep[e.g., ][]{Mulders2015,Dressing2015,Shields2016}. 

M~dwarfs are intrinsically faint, especially towards later and cooler spectral types, and measuring their fundamental properties can therefore be challenging, even for the nearest ones \citep[e.g., ][]{Ribas2017}. 
However, several fundamental properties of M~dwarfs have been studied extensively.
For instance, using spectroscopy, photometry, and astrometry, effective temperature \citep[e.g., ][]{Ness2015,Mann2015,Birky2020}, radius \citep[e.g., ][]{Kesseli2018}, luminosity \citep[e.g., ][]{Reid2002}, metallicity \citep[e.g., ][]{Bochanski2013,Newton2014,Schmidt2016} and mass \citep[e.g., ][]{Boyajian2012,Mann2019} measurements have been studied.
Age, however, is one of the most difficult fundamental properties to evaluate, especially for M~dwarfs \citep{Soderblom2010}.

Current age-dating methods used for higher-mass stars cannot be applied to low-mass stars. 
Asteroseismology \citep[e.g., ][]{Chaplin2014} is a common age-dating method for giant stars but cannot be applied to M~dwarfs because their acoustic oscillations have extremely small amplitudes and short timescales \citep{Rodriguez2016}.
Isochrones from stellar evolution models are also not efficient to estimate M~dwarf ages due to the extremely slow and small changes in luminosity after $1$\,Gyr, making the isochrones very similar for older ages \citep{Chabrier1997}. 
Furthermore, there are not sufficient empirical calibrations to validate isochrones for low-mass stars, especially at young ages. As a consequence, model isochrones still suffer from significant systematic errors that are unexplored, and are inaccurate for precise age determinations \citep{Baraffe2015}.
Empirical methods such as gyrochronology \citep[e.g., ][]{Skumanich1972,Barnes2003,Barnes2007,Angus2015,VanSaders2016} are either based on the Sun, or calibrated on higher-mass stars, and do not yield precise age estimates for M~dwarfs \citep[e.g., ][]{Angus2019}.
As current available methods cannot be used to estimate M~dwarf ages, empirically-calibrated relations for age-related properties are needed.

Magnetic activity, age and rotation period are known to be correlated for solar type stars \citep[e.g., ][]{Skumanich1972,Barry1988,Soderblom1991,Mamajek2008}. 
These stars have a radiative core and a convective envelope and they do not rotate as a rigid body.
As solar type stars rotate, a magnetic dynamo is generated in between the two layers, which is responsible for their magnetic field \citep{Parker1955}.
Given that rotation and age are correlated for solar-type stars \citep[e.g., ][]{Skumanich1972,Barnes2003,Barnes2007,Angus2015,VanSaders2016}, their magnetic activity is correlated with age as well.
For the lowest-mass stars (spectral type $>{\rm M}3$) the correlation between magnetic activity, age and rotation is not well understood because these cool stars are fully convective \citep{Chabrier1997} and do not have an interface to produce a dynamo. 
However, previous studies of M dwarfs indicate that magnetic activity, rotation and age are correlated  \citep[e.g., ][]{Eggen1990,Fleming1995,Delfosse1998,Mohanty2003,West2004,Reiners2012,West2015,RIEDEL2017,Newton2017,Kiman2019,Angus2019}. 

A well-studied magnetic activity indicator in M dwarfs is the $\halpha$ emission line  \citep{Hawley1996,West2008a,West2008c} which is generated by collisional excitations when magnetic field lines heat the dense chromosphere \citep{Stauffer1986}.
Therefore, $\halpha$ equivalent width $(\haew)$ is an indirect measurement of the chromospheric magnetic activity of a star.
The fact that M~dwarf magnetic fields are driven by their rotation means that the rotation-age correlation should translate into a $\haew$-age correlation \citep{Newton2017}. 
As a consequence, $\haew$ could be used as an age indicator for M~dwarfs.
For low-mass stars, previous studies have confirmed that $\halpha$ emission is correlated with age by using kinematics as an age-indicator \citep{Gizis2002,West2008a,Kiman2019}. 
\citet{West2008c} found a functional form for the age-activity relation from a sample of M~dwarfs with $\haew$ by modeling the relation between kinematics and age. However, to date, there has not been a study which calibrates empirically the age-activity relation for M~dwarfs.

A first necessary step to calibrate the $\halpha$ age-activity relation is to collect M~dwarfs with known ages, calculated with methods independent of the magnetic activity. 
The second data release of \textit{Gaia} \citep{GaiaCollaboration2016,Collaboration2018,Arenou2018,Lindegren2018a} plays a key role in age-dating stars. 
The \textit{Gaia} DR2 catalog contains $\sim$\,$1.3$ billion sources with a five-parameter astrometric solution: positions, parallaxes ($\pi$), and proper motions ($\mu$) with unprecedented precision, down to a magnitude of $G = 21$. The parallax uncertainties ($\sigma_{\rm \pi}$) are between  $0.04-0.7$~milliarcsecond, and for proper motion ($\sigma_{\rm \mu}$) between $0.06-1.2 $\,${\rm mas/yr}$, depending on the magnitude of the star.
With these high quality measurements, \textit{Gaia} made it possible to identify new stars that belong to age-calibrated moving groups and new associations \citep[e.g., ][]{Gagne2018a,Faherty2018,Kounkel2019,Roser2020}, and new co-moving pairs of stars \citep[e.g., ][]{Oh2017a,El-Badry2018a}.
Both young associations and co-moving pairs can be sources for M~dwarf ages.

Moving groups and other coeval associations are ensembles of stars born from the same molecular cloud with common space velocities and a small spread of ages \citep{Bell2015}. 
As ages of the young associations are well calibrated down to a precision of a few Myr \citep{Soderblom2010,Bell2015}, the age of the association can be used to build a set of age-calibrated M~dwarfs, if the stars can be identified as members.

Binary stars are born from the same molecular cloud at the same time \citep{Bodenheimer2011}. Therefore, we can calibrate the age of an M~dwarf by constraining the age of a co-mover. For example, we can estimate a white dwarf's total age adding the white dwarf cooling age, and the progenitor star main sequence age \citep{Fouesneau2018}. The white dwarf cooling age and mass are strongly constrained by cooling tracks, from theoretical models \citep[e.g., ][]{Bergeron1995,Fontaine2001,Bergeron2019}. 
Using the mass of the white dwarf and the semi-empirical initial-to-final mass relations \citep{Cummings2018}, we can estimate the mass of the progenitor star.
Finally with the mass of the progenitor star, we can calculate a main-sequence age, combined with the length of the pre- and post- (but pre-white dwarf) main-sequence stages \citep[obtained with MESA models,][]{Dotter2016}.

The aim of the present study is to calibrate the $\halpha$ age-activity relation for M~dwarfs using a sample of age-calibrators. 
We describe how we compiled a sample of M~dwarfs with $\haew$ measurements in the literature in Section~\ref{sec:sample}. We show how we obtained the age-calibrators from that sample by identifying M~dwarfs co-moving with a white dwarf with known age or members of known young associations in Section~\ref{sec:sampleages}. The final table of age-calibrators in described in Table \ref{table:columnsagecal}.
In Section~\ref{sec:binariesandlhalbol} we describe the calculation of the fractional $\halpha$ luminosity from $\haew$ for the age-calibrators, which is a key parameter to calibrate the age-activity relation. In this section we also describe a search for known unresolved binaries which could bias our calibration of the age-activity relation. 
In Section~\ref{sec:activityfraction} we show how we divided the sample of age calibrators into active or inactive objects according to their $\haew$ measurement and photometry, and how we studied the relation between the active fraction and age for different spectral types. 
In Section~\ref{sec:relation} we discuss the relation between both $\haew$ and fractional $\halpha$ luminosity with age and fit it using a Markov chain Monte Carlo (MCMC) algorithm. 
In Section~\ref{sec:comparisonstudies} we compare our results for the age-activity relation with literature results for $\halpha$, X-ray and UV.
Finally, in Section \ref{sec:conclusions} we discuss our results and summarize our work and conclusions. All the code used in this work is available on Zenodo\footnote{\url{https://doi.org/10.5281/zenodo.4660208}} and GitHub\footnote{\url{https://github.com/rkiman/M-dwarfs-Age-Activity-Relation}}.

\section{Identifying M dwarfs in the literature with $\halpha$ measurements}
\label{sec:sample}
 
\subsection{Compiling the literature search sample}
\label{subsec:literaturesearch}

In order to empirically calibrate the age-activity relation for M~dwarfs, we began by collecting M$0$-M$9$ dwarfs from the literature with a reported $\haew$. 
In \citet{Kiman2019} they compiled one of the largest published samples of M~dwarfs with $\haew$, including measurements from \citet{West2011} and \citet{Schmidt2015}. 
This sample contains $74,216$ M~dwarfs, out of which $486$ were removed since they have a spectroscopically identified but unresolved white dwarf companion \citep{West2011,Schmidt2015} which could increase the magnetic activity of the star \citep{Skinner2017}. 
The remaining $73,730$ M~dwarfs from \citet{Kiman2019} make up the largest part of our literature search sample.

The M~dwarfs in \citet{Kiman2019} are assumed to be primarily field stars. We complemented this sample with studies of M~dwarfs in known star forming regions, clusters, moving groups, or co-moving with a white dwarf, as well as all the studies in the literature that have measured $\haew$ for cool dwarfs. All the studies we checked are listed or mentioned in Table~\ref{table:age_cal}.
In total, we identified \nlitsearch stars from the literature ($73,730$ from \citet{Kiman2019} and \nlitsearchothers from other studies)\footnote{The sample is available to download from \href{https://doi.org/10.5281/zenodo.4659293}{Zenodo}.}. 
From the total number of stars in our sample, we found that \nlitsearchgaia stars (\percentingaia) have photometric and astrometric information from \textit{Gaia} DR2 \citep{GaiaCollaboration2016,Collaboration2018}. 
We applied the quality cuts described in \citet{Kiman2019} to select the best astrometric and photometric data from \textit{Gaia} DR2 (the Sub Red sample).
These cuts not only select best quality photometry, parallaxes and proper motions, but also remove possible unresolved binaries. 
We refer to Section 2.3 of \citet{Kiman2019} for more details. 
By applying these quality cuts we were left with \nhighqualparallax M~dwarfs (\percenthighqualparallax) in our literature search sample. In Section \ref{sec:sampleages}, we describe the cross-match with \textit{Gaia} in more detail.
We use the spectral type classification from the literature when available and estimate the spectral type for the rest of the stars (\nnotspt of the sample) using their \textit{Gaia} red color $(G-G_{\rm RP})$ and the relation in \citet{Kiman2019}.
Although most of the studies we compiled were focused on M~dwarfs, some contained higher mass stars, mostly G and K dwarfs, which were removed with a cut in spectral type keeping \nmdwarfs M~dwarfs ($\geq {\rm M}0$).

From our literature search sample, \nrepeated M~dwarfs have between two and six measurements of $\haew$.
We identified duplicates through a position search within a $2\farcs0$ radius. 
These duplicated stars are indicated in the column $\texttt{star\_index}$ in Table~\ref{table:columnsagecal}. 
If two or more stars have the same number in this column, then they likely are the same star.
In total we found that our literature search sample has \nuniquesources unique M~dwarfs with good data from \textit{Gaia} DR2 and $\haew$ measurements. 

\begin{deluxetable*}{lcl}[ht!]
\tablewidth{290pt}
\tabletypesize{\scriptsize}
\tablecaption{Columns in the sample of age calibrators, available as a fits file. \label{table:columnsagecal}}
\tablehead{\colhead{Column name} & \colhead{Units} & \colhead{Description} 
}\startdata 
$\texttt{ra}$&deg&Original R.A. from the source of the $\halpha$ measurement\\
$\texttt{dec}$&deg&Original Decl. from the source of the $\halpha$ measurement\\
$\texttt{spt}$&&Spectral Type\\
$\texttt{gaia\_source\_id}$&...&Unique Gaia source identifier (unique within DR2)\\
$\texttt{ra\_gaia}$&deg&R.A. in Gaia DR2 epoch\\
$\texttt{dec\_gaia}$&deg&Decl. in Gaia DR2 epoch\\
$\texttt{pmra}$&mas ${\rm yr}^{-1}$&Proper motion in R.A. direction in Gaia DR2\\
$\texttt{pmra\_error}$&mas ${\rm yr}^{-1}$&Standard error of proper motion in R.A. direction in Gaia DR2\\
$\texttt{pmdec}$&mas ${\rm yr}^{-1}$&Proper motion in decl. direction in Gaia DR2\\
$\texttt{pmdec\_error}$&mas ${\rm yr}^{-1}$&Standard error of proper motion in decl. direction in Gaia DR2\\
$\texttt{parallax}$&mas&Parallax in Gaia DR2\\
$\texttt{parallax\_error}$&mas&Standard error of parallax in Gaia DR2\\
$\texttt{phot\_g\_mean\_flux}$&electron ${\rm s}^{-1}$&$G$ band mean flux\\
$\texttt{phot\_g\_mean\_flux\_error}$&electron s$^{-1}$&Error on $G$ band mean flux\\
$\texttt{phot\_g\_mean\_mag}$&mag&$G$ band band mean magnitude\\
$\texttt{phot\_rp\_mean\_flux}$&electron ${\rm s}^{-1}$&Integrated $G_{\rm RP}$ mean flux\\
$\texttt{phot\_rp\_mean\_flux\_error}$&electron s$^{-1}$&Error on the integrated $G_{\rm RP}$ mean flux\\
$\texttt{phot\_rp\_mean\_mag}$&mag& Integrated $G{\rm RP}$ mean magnitude\\
$\texttt{phot\_bp\_mean\_flux}$&electron ${\rm s}^{-1}$&Integrated $G_{\rm BP}$ mean flux.\\
$\texttt{phot\_bp\_mean\_flux\_error}$&electron s$^{-1}$&Error on the integrated $G_{\rm BP}$ mean flux\\
$\texttt{phot\_bp\_mean\_mag}$&mag&Integrated $G{\rm BP}$ mean magnitude\\
$\texttt{g\_corr}$&mag&$G$ magnitude corrected for extinction\\
$\texttt{rp\_corr}$&mag&$G_{\rm RP}$ magnitude corrected for extinction\\
$\texttt{ewha}$&${\rm \AA}$&Equivalent width $\halpha$ for the compatible literature search\\
$\texttt{ewha\_error}$&${\rm \AA}$&Equivalent width $\halpha$ error for the compatible literature search\\
$\texttt{ewha\_all}$&${\rm \AA}$&Equivalent width $\halpha$ for all the stars\\
$\texttt{ewha\_error\_all}$&${\rm \AA}$&Equivalent width $\halpha$ error for all the stars\\
$\texttt{lhalbol}$&...&Fractional $\halpha$ luminosity\\
$\texttt{lhalbol\_error}$&...&Fractional $\halpha$ luminosity error\\
$\texttt{age}$&yr&Age of the star\\
$\texttt{age\_error\_low}$&yr&Lower bound on the confidence interval of the estimated age\\
$\texttt{age\_error\_high}$&yr&Upper bound on the confidence interval of the estimated age\\
$\texttt{group\_num}$&...&Number identifying the young association. $0$ indicates white dwarf companion\\
$\texttt{group\_name}$&...&Young association the star belong to\\
$\texttt{star\_index}$&...&Number indicating repeated stars. Same stars have the same number\\
$\texttt{source\_num}$&...&Number indicating the source of the $\halpha$ measurement\\
$\texttt{source\_ref}$&...&Source of the $\halpha$ measurement\\
$\texttt{potential\_binary}$&...&$1$ if it is a potential binary, $0$ if not.\\
\enddata 
\end{deluxetable*}

\begin{deluxetable*}{lcccccc}[ht!]
\tablewidth{290pt}
\tabletypesize{\scriptsize}
\tablecaption{Age Calibrators summary. \label{table:age_cal}}
\tablehead{    \colhead{Reference \tablenotemark{a}}     & \colhead{Spectral}     & \multicolumn{2}{c}{$N$ of M dwarfs}     & \colhead{OC \tablenotemark{c}}     & \multicolumn{2}{c}{Ages from}     \\ & Resolution & Total & Compatible \tablenotemark{b}     &  &moving group & white dwarf 
}\startdata 
\citet{Kiman2019}&1800&73729&73729&0&46&21\\ 
LG11\tablenotemark{d}&1000&2504&134&2&48&-\\ 
\citet{Jeffers2018}&62000, 48000, 40000&2133&2&2&22&-\\ 
\citet{Douglas2014}&3300, 4000&1906&50&1&264&-\\ 
\citet{Lepine2013}&2000,4000&1577&1&2&4&-\\ 
\citet{Riaz2006}&1750&1098&4&2&65&-\\ 
\citet{Ansdell2015}&1000, 1200&794&35&2&31&-\\ 
\citet{Gaidos2014}&~1200&582&59&2&2&-\\ 
\citet{Fang2018}&1800&561&1&1&159&-\\ 
\citet{Newton2017}&3000&456&14&2&2&-\\ 
\citet{Terrien2015}&2000&351&13&2&2&-\\ 
\citet{Reid1995}&2000&343&4&2&1&-\\ 
\citet{Schneider2019}&32000&336&10&2&26&-\\ 
\citet{Bouy2009}&Multiple&227&1&2&135&-\\ 
\citet{Kraus2014}&35000&205&2&2&88&-\\ 
\citet{Shkolnik2009}&60000&184&7&2&9&-\\ 
\citet{Alonso-Floriano2015}&1500&179&8&2&3&-\\ 
\citet{Slesnick2008}&1250&145&69&2&23&-\\ 
\citet{Malo2014}&1750&120&1&2&45&-\\ 
\citet{Torres2006}&50000, 9000&114&2&2&18&-\\ 
\citet{Shkolnik2017}&35000, 58000&106&3&2&24&-\\ 
\citet{Elliott2016}&85000, 48000&83&2&2&3&-\\ 
\citet{Reiners2010}&31000&73&2&1&2&-\\ 
\citet{Slesnick2006}&1250&65&19&2&18&-\\ 
\citet{Jayawardhana2006}&60000&52&1&2&20&-\\ 
\citet{Riedel2014}&Multiple&50&3&2&5&-\\ 
\citet{Hawley1996}&2000&31&1&2&1&-\\ 
\citet{Song2003}&24000&25&1&2&3&-\\ 
\citet{Rodriguez2014}&3000, 7000&23&12&2&12&-\\ 
\enddata 
\tablenotetext{a}{Compatible catalogs without age calibrators: \citet{Gizis2002}, \citet{Mochnacki2002}, \citet{Reid2002}, \citet{West2011}, \citet{Song2004}, \citet{Lyo2004}.
 Catalogs with overlap but not compatibles: \citet{Mohanty2005}, \citet{Shkolnik2011}, \citet{Lawson2002}.
 Catalogs without overlap: \citet{Frasca2018}, \citet{Bayo2012}, \citet{Reid2007}, \citet{Cruz2002}, \citet{Feigelson2003}, \citet{Gizis2000}, \citet{Gizis1997}, \citet{Murphy2010}, \citet{Phan-Bao2006}, \citet{Bochanski2005}, \citet{Mohanty2003}, \citet{Reiners2008}, \citet{Lepine2009}, \citet{Reiners2007}, \citet{Lepine2003}, \citet{Martin1996}, \citet{Ivanov2015}, \citet{Stauffer1997}, \citet{Tinney1998}.
 Other catalogs checked: \citet{Lodieu2005}.}
\tablenotetext{b}{Compatible with                        \citet{Kiman2019}.}
\tablenotetext{c}{Order of compatibility. Order $1$                       is compatible with \citet{Kiman2019}. Order $2$ is                        compatible with at least one order $1$ catalog.}
\tablenotetext{d}{\citet{Lepine2013,Gaidos2014} with                        additional data observed in                        an identical manner.}
\end{deluxetable*}

\subsection{Identifying compatible $\halpha$ measurements from the literature}
\label{subsec:overlapping}

The $\haew$ is calculated by dividing the flux under the emission line of $\halpha$ by the flux of the continuum. 
Both sources of $\haew$ for \citet{Kiman2019} \citep{West2011,Schmidt2015} used the same definition for the emission line: $6557.61-6571.61{\rm \AA}$, and the surrounding continuum: $6530-6555{\rm \AA}$ and $6575-6600 {\rm \AA}$. Both \citet{West2011} and \citet{Schmidt2015} measured the $\haew$ from spectra with $R\sim 1800$ resolution. 
Our literature search sample, however, contains objects with measurements of $\haew$ from different spectral resolutions (shown in Table~\ref{table:age_cal}) that were calculated with slightly different definitions of the line and the continuum. 
Such diversity of approaches and results could cause inconsistencies in our analysis. 
To account for the differences between $\haew$ measurements, we followed a procedure similar to the one described in \citet{Newton2017}. 
We only used the $\haew$ from catalogs that were \textit{compatible} with the \citet{Kiman2019} sample, as that is the largest component of our literature search sample (see Section~\ref{subsec:literaturesearch}). 
We considered a given catalog \textit{compatible} if it had at least one star in common and if at least $90\%$ of the stars in common had a difference in $\haew$ smaller than $3{\rm \AA}$
($|\Delta \haew |<3{\rm \AA}$)
with the measurement in \citet{Kiman2019}. 
We chose $3{\rm \AA}$ as the limit because this is the typical $\halpha$ variability for ${\rm M}4-{\rm M}5$ identified by \citet{Lee2010} in a spectroscopic survey of $43$ M~dwarfs in the range ${\rm M}3.5-{\rm M}8.5$. 
This cut assumes that $90\%$ of the M~dwarfs have small variability. We note that this might bias against variable $>{\rm M}5$ dwarfs, because variability increases for later types \citep{Lee2010}.

We defined the criterion described above as \textit{first order compatibility} and it is indicated with a $1$ in the column ``OC" of Table~\ref{table:age_cal}. 
We could only find three catalogs from the literature search that had overlapping stars with \citet{Kiman2019}: \citet{Douglas2014}, \citet{Fang2018} and \citet{Reiners2010}.
Therefore we decided to iterate upon our method and search for further studies compatible with these three catalogs to increase the number of stars we consider \textit{compatible}.
We repeated the procedure described above to find what we defined as \textit{second order compatibility} catalogs. 
These second order of compatibility catalogs do not have stars in common with \citet{Kiman2019}, but they do have stars in common with at least one of the order $1$ catalogs.

In all we found \ncompatible unique M~dwarfs ($99.6\%$ of the sample) with $\haew$ measurements that were either in \citet[][order 0]{Kiman2019} or first or second order compatible, meaning that we can use them together to calibrate the age-activity relation.
We excluded the $128$ remaining M~dwarfs from our analysis.


\subsection{Removing potentially accreting M~dwarfs}
\label{subsubsec:accretors}

To characterize the age-activity relation we are interested in chromospheric $\halpha$ emission, however this spectral line could in some systems result from accretion. 
To distinguish between the two types of emission we used an empirical criterion which depends on the star's $\haew$ and spectral type, developed by \citet{White2003} based on a sample of low-mass T Tauri stars. 
They proposed that a T Tauri star is classical, meaning accreting, if $\haew \geq 10{\rm \AA}$ for ${\rm K}7-{\rm M}2.5$ stars, $\haew \geq 20{\rm \AA}$ for ${\rm M}3-{\rm M}5.5$, and $\haew \geq 40{\rm \AA}$ for ${\rm M}6-{\rm M}7.5$ stars.

Based on the \citet{White2003} criterion, we removed \naccretors stars which are $\halpha$ outliers from our analysis that are possibly accreting.
It should also be noted that by using only $\haew$ to discard possible accretors we are also removing $\haew$ outliers such as non-accreting stars whose measurement of $\halpha$ was taken during a flare.
In Table~\ref{table:accretors} we list the $\haew$ outliers, possibly accreting, identified in this study with their 2MASS name, spectral type, the $\haew$ and $\Delta \haew$ above the limit defined by \citet{White2003}.
Given that the criteria only goes to ${\rm M}7.5$, we did not remove any later spectral types according to their $\haew$.
Therefore we are likely to have some contamination from accreting later spectral types.

\begin{deluxetable}{lccc}[ht!]
\tablewidth{290pt}
\tabletypesize{\scriptsize}
\tablecaption{Short sample of $\haew$ outliers,    possibly accreting according to the     criterion from \citet{White2003}. The full sample can be found online.    \label{table:accretors}}
\tablehead{    \colhead{2MASS Name}     & \colhead{SpT}     & \colhead{$\haew$}     & \colhead{$\Delta \haew$\tablenotemark{a}} 
}\startdata 
J16075567-2443267 &M5.5 & $47.3\pm0.1$ & $27.3$ \\ 
J05353004+0959255 &M5.6 & $22.82\pm0.62$ & $2.82$ \\ 
J05334992+0950367 &M2.3 & $14.01\pm0.51$ & $4.01$ \\ 
J12350424-4136385 &M2.5 & $13.6\pm0.2$ & $3.6$ \\ 
J02591904-5122341 &M5.4 & $32.11\pm0.1$ & $12.11$ \\ 
J16104996-2212515 &M5.5 & $23\pm0.1$ & $3$ \\ 
J04480085+1439583 &M5.0 & $73.6\pm0.1$ & $53.6$ \\ 
J04262939+2624137 &M6.0 & $97.7\pm9.8$ & $57.7$ \\ 
J05340393+0952122 &M2.2 & $30.2\pm1.14$ & $20.2$ \\ 
J12071089-3230537 &M4.3 & $114.8\pm0.5$ & $94.8$ \\ 
\enddata 
\tablenotetext{a}{Delta above the $\haew$ limit.}
\end{deluxetable}

\section{Identifying M~dwarfs in young associations and co-moving with white dwarfs}
\label{sec:sampleages}

\subsection{Identifying young association members}
\label{subsec:movinggroups}

As mentioned in Section~\ref{subsec:literaturesearch}, we complemented \citet{Kiman2019} with M~dwarfs which were classified as members of young associations by previous studies and/or had $\haew$ measurements.
However, most of these studies were completed prior to the release of \textit{Gaia} DR2, which provided kinematics of an unprecedented quality.

In light of the new astrometric improvement of the \textit{Gaia} DR2 survey, we decided to re-assess the likelihood of membership for each source and/or identify new members.
To obtain proper motions and parallaxes for
our literature search sample, we used TOPCAT \citep{Taylor2005} and a $2$ arcsecond radius to match to \textit{Gaia} DR2 objects.
We found \percentingaia~(\nlitsearchgaia out of the original \nlitsearch sources) of our literature search sample in \textit{Gaia} DR2. 
As mentioned in Section \ref{subsec:literaturesearch}, we applied the quality cuts described in \citet{Kiman2019}, to obtain the best astrometric and photometric data, and to remove potential binaries, and we were left with $\nhighqualparallax$ M~dwarfs.
To look for mismatches we used the \nrepeated M~dwarfs in our sample with duplicated measurements of $\haew$ (see Section \ref{subsec:literaturesearch}) and compared their \textit{Gaia} source id.
All the duplicated stars had the same \textit{Gaia} source id, except for \doublestarsremoved stars which we think are potentially unresolved M~dwarf binaries because after visually inspecting these stars we found that in the image there was only one star with two \textit{Gaia} sources. 
These \doublestarsremoved stars were removed from our analysis.

To confirm or identify members of young associations from our literature search sample, we used the BANYAN~$\Sigma$ bayesian membership classification algorithm \citep{Gagne2018}\footnote{The IDL version is available at \url{https://github.com/jgagneastro/banyan_sigma_idl}. 
Also the Python version at \url{https://github.com/jgagneastro/banyan_sigma}, and the web portal is available at \url{http://www.exoplanetes.umontreal.ca/banyan/banyansigma.php.}}.
BANYAN~$\Sigma$ models young associations with a multivariate gaussian density. 
This gaussian has a total of six dimensions: three galactic positions (XYZ) and three space velocities (UVW). 
BANYAN~$\Sigma$ also models the galactic field within $300$\,pc by combining $10$ multivariate gaussians. 
To calculate the probability of a star belonging to a young association or the field, BANYAN~$\Sigma$ compares observables such as position and proper motion to the multivariate Gaussian model in a Bayesian classification likelihood, and marginalizes over radial velocities and distances when they are not available. 
The marginalization integrals are solved with an analytical solution, making the code more precise and efficient. 
In total, there are $27$ associations modeled in BANYAN~$\Sigma$ within $150$\,pc. A summary of the associations in BANYAN~$\Sigma$ is in Table~\ref{table:ya_gagne}.

\begin{deluxetable*}{lccccc}[ht!]
\tablewidth{290pt}
\tabletypesize{\scriptsize}
\tablecaption{Young Associations in Banyan $\Sigma$ \citep{Gagne2018} and summary of M~dwarfs in young associations used in this work. \label{table:ya_gagne}}
\tablehead{\colhead{Association} & \colhead{Short name} & \colhead{Age (Myr)} & \colhead{Age Ref.\tablenotemark{a}} & \colhead{Total members \tablenotemark{b}}& \colhead{New members} 
}\startdata 
Taurus                               & TAU& $1.5\pm 0.5$&$1$& $3$&- \\ 
$\rho$ Ophiuchi                & ROPH&$<2$&$2$& -&-\\
$\epsilon$ Chamaeleontis &EPSC& $3.7\pm 4.6$&$3$&$9$&$1$ \\ 
Corona Australis                &CRA&$4-5$&$4$& -&-\\
TW Hya                              &TWA& $10.0\pm 3.0$&$5$& $10$&- \\ 
Upper Scorpius                  &USCO& $10.0\pm 3.0$&$6$& $167$&$1$ \\ 
118 Tau                            &118TAU&$\sim 10$&$7$& -&-\\
Upper CrA                         &UCRA&$\sim 10$&$8$& $1$&- \\ 
$\eta$ Chamaeleontis       &ETAC& $11.0\pm 3.0$&$5$& $1$&- \\ 
Lower Centaurus Crux      &LCC& $15.0\pm 3.0$&$6$& $20$&$1$ \\ 
Upper Centaurus Lupus    &UCL& $16.0\pm 2.0$&$6$& $4$&$2$ \\ 
32 Orionis                        &THOR&$22^{+4}_{-3}$&$5$& -&-\\
$\beta$ Pictoris               &$\beta$PMG& $24.0\pm 3.0$&$5$& $54$&$1$ \\ 
Octans                             &OCT& $35.0\pm 5.0$&$3$& $1$&- \\ 
Argus                              &ARG& $40-50$&$9$ &$1$&-\\
Columba                          &COL& $42.0\pm 6.0$&$5$& $12$&- \\ 
Carina                              &CAR& $45.0\pm 11.0$&$5$&$6$&- \\ 
Tucana-Horologium association &THA& $45.0\pm 4.0$&$5$&$97$&- \\ 
Platais 8                          &PL8&$\sim 60$&$10$& -&-\\
Pleiades cluster               &PLE&$112\pm 5$&$11$&$106$&- \\ 
AB Doradus                     &ABDMG& $149.0\pm 51.0$&$5$& $36$&$1$ \\ 
Carina-Near                    &CARN& $200.0\pm 50.0$&$12$& $3$&- \\ 
core of the Ursa Major cluster &UMA& $414.0\pm 23.0$&$13$&  $1$&- \\ 
$\chi ^1$ For                  &XFOR&$\sim 500$&$14$& -&-\\
Coma Berenices               &CBER& $562.0\pm 98.0$&$15$& $9$&- \\ 
Praesepe cluster\tablenotemark{c} &PRA & $650.0\pm 50.0$&$16$&$251$&- \\ 
Hyades cluster                 & HYA& $750.0\pm 100.0$&$17$& $79$&- \\ 
\enddata 
\tablenotetext{a}{$(1)$\citet{Kenyon1995}, $(2)$\citet{Wilking2008}, $(3)$\citet{Murphy2015}, $(4)$\citet{Gennaro2012}, $(5)$ \citet{Bell2015},  $(6)$\citet{Pecaut2016}, $(7)$\citet{Mamajek2016}, $(8)$\citet{Gagne2018}, $(9)$\citet{Zuckerman2018}, $(10)$\citet{Platais1998}, $(11)$\citet{Dahm2015}, $(12)$\citet{Zuckerman2006}, $(13)$\citet{Jones2015}, $(14)$\citet{Pohnl2010}, $(15)$\citet{Silaj2014}, $(16)$\citet{Douglas2019,Gao2019}, $(17)$\citet{Brandt2015}}
\tablenotetext{b}{\changedtext{Total number of stars with ${\rm H}\alpha$ compatible (See section \ref{subsec:overlapping}).}}
\tablenotetext{c}{Praesepe is not included in the current public version of Banyan $\Sigma$ (v1.2, September 6 2018).}
\end{deluxetable*}

Using BANYAN~$\Sigma$ we calculated the probability that any given M~dwarf belongs to a known association according to their position, proper motion, parallax, and radial velocity when available from \textit{Gaia} DR2. We used this code to analyze the \nnotacc single, compatible, not accreting stars in our sample (See Section \ref{sec:sample}).

To remove as many false positives as possible without loosing true positives, we used a $90\%$ cutoff in the membership probability output by BANYAN~$\Sigma$ as suggested by \citet{Gagne2018}. 
In total we found \nmgmem M~dwarfs which yielded $90\%$ probability in a known young association. 
The remaining \nnotinmg stars were rejected as young association members by our cut and/or have a high probability ($>90\%$) to be field stars according to BANYAN. In addition, $21$ objects were removed because they were rejected as young association members by \citet{Gagne2018}. These objects were rejected mostly because their literature radial velocity was inconsistent with the one of the group.

\begin{figure}[ht!]
\begin{center}
\includegraphics[width=\columnwidth]{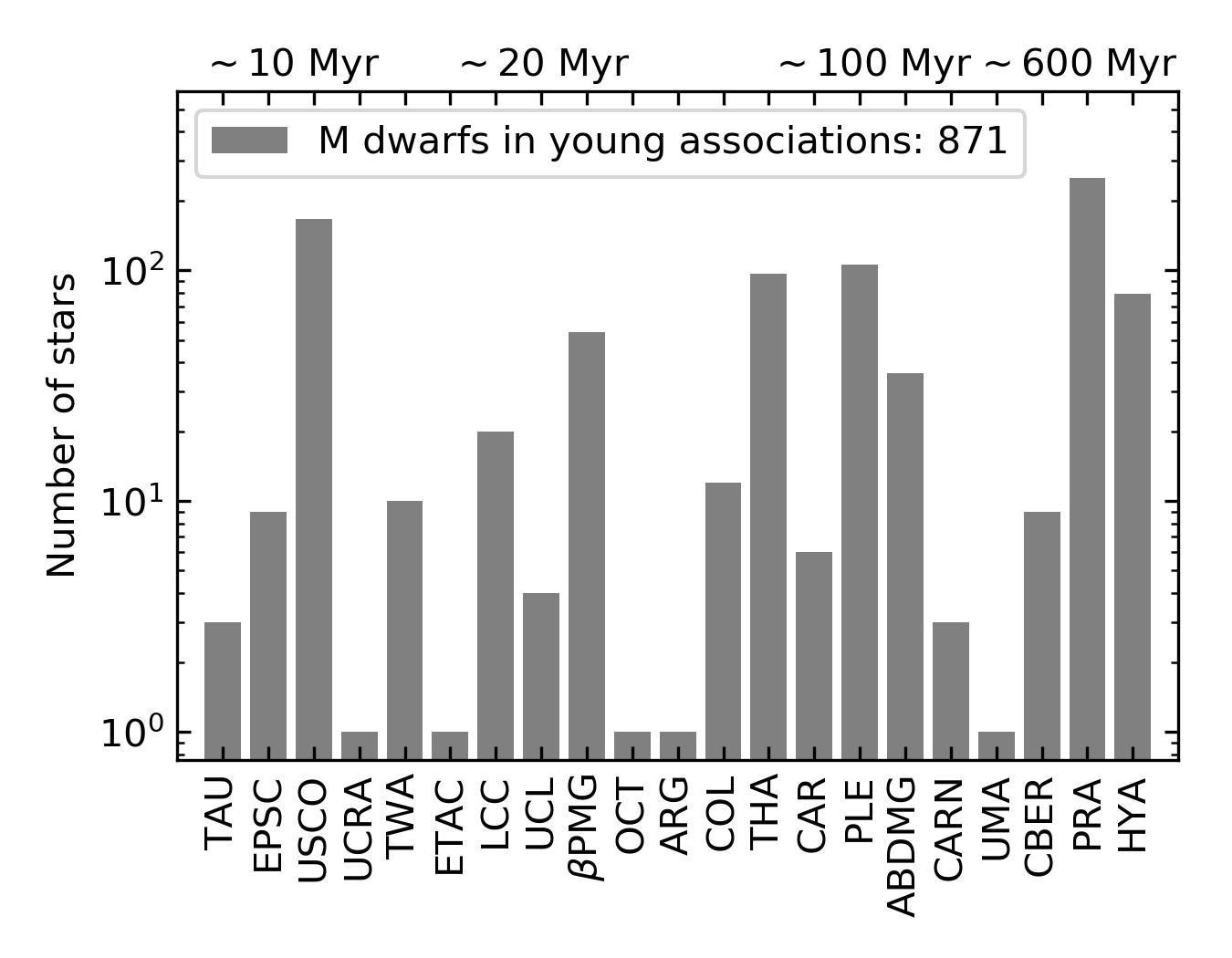}
\caption{Membership distribution for M~dwarfs in our sample, with young associations ordered by age. We include a reference at the top of the figure for the age of each association. Repeated measurements are not included in the total number of stars per association. References for each association are in Table~\ref{table:ya_gagne}.} 
\label{fig:histmg}.
\end{center}
\end{figure}

Based on experience with kinematics and lithium abundances, we expect an average contamination from false members below $10\%$ \citep{Gagne2018}.
In the Appendix \ref{sec:mg_check}, we show the color--magnitude diagrams for each of the young associations with all of the objects used in this study as members. We concluded that we cannot discard any objects with the color--magnitude diagrams.
The distribution of members per young association is shown in Figure~\ref{fig:histmg} and Table~\ref{table:ya_gagne}. 
We found that Praesepe ($\sim 650$\,Myr, $251$ sources) had the largest yield of objects from our sample, followed by Upper Scorpious ($\sim 10$\,Myr, $167$ sources), the Pleiades cluster ($\sim 112$\,Myr, $106$ sources) and Tucana-Horologium ($\sim 45$\,Myr, $97$ sources). 
Therefore, we compiled a significant number of stars covering a range of ages that allows us to calibrate the age-activity relation for ages $<1$\,Gyr.

From the \nmgmem young association members in this work, we found that $708$ have the same original membership assigned in the literature, $17$ had their moving group membership revised and $167$ were not identified as members in our literature search.
From the $184$ ``new" candidate members or members that changed their membership, $2$ were rejected as members using radial velocities from the literature, and $175$ had been previously identified as young association members by studies not collected in this work, i.e. studies where $\halpha$ measurements were not utilized \citep{Roser2011,Galli2018,Gagne2018,Gagne2018a,Goldman2018,Luhman2018,Rebull2018}. 
The remaining \nnewmgmem M~dwarfs, are new candidate members of the young associations identified in this study. In particular, 2MASS~J12323103-7255068 was called a member of LCC by \citet{Goldman2018}, that we revised to be member of EPSC. These stars are summarized in Table~\ref{table:ya_gagne} and listed in Table~\ref{table:newmem}.

\begin{figure*}[ht!]
\begin{center}
\subfloat[$\epsilon$ Chamaeleontis, $3.7\pm 4.6$\,Myr.]{\includegraphics[width=0.43\linewidth,keepaspectratio]{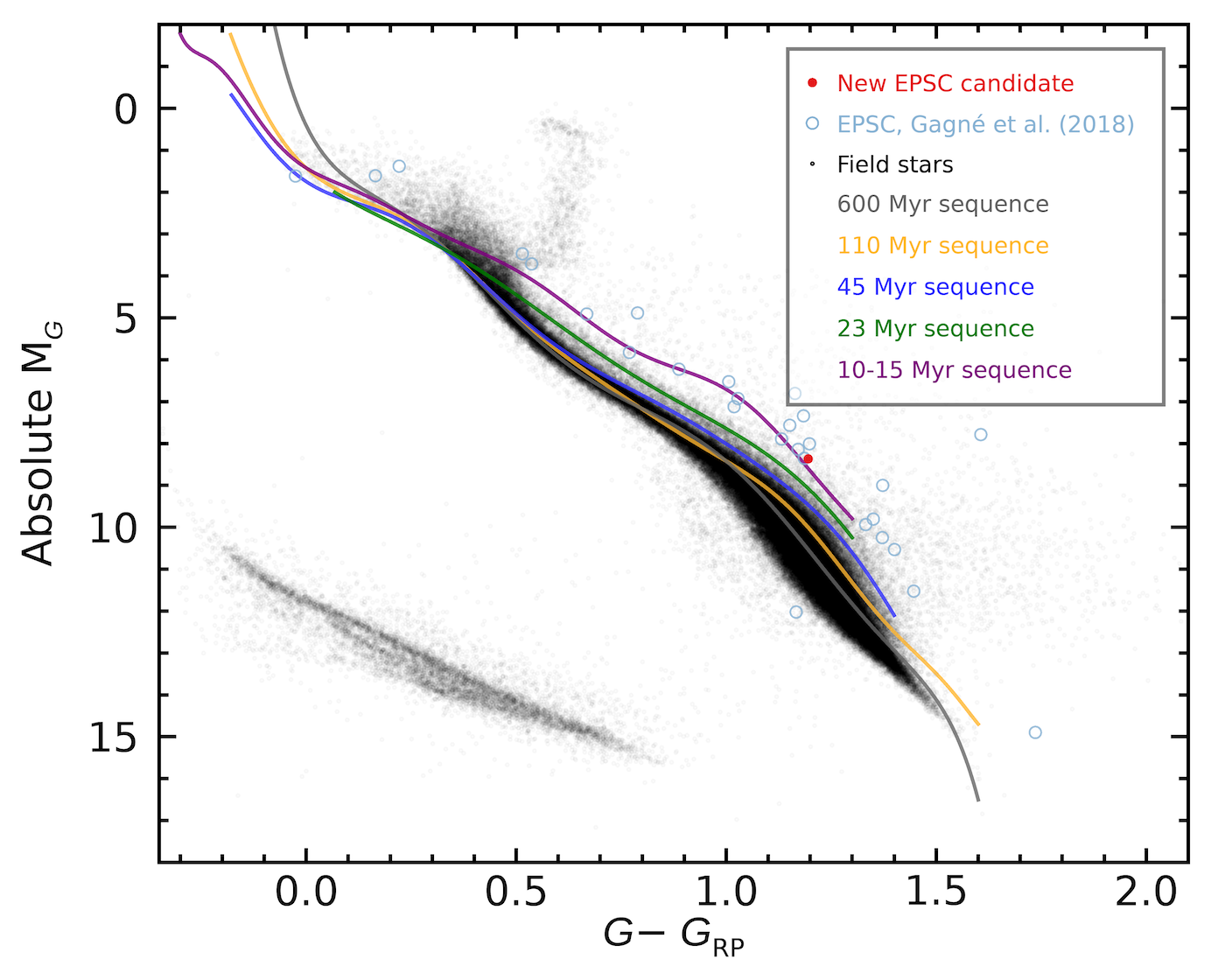}}
\subfloat[Upper Scorpius, $10.0\pm 3.0$\,Myr, Lower Centaurus Crux, $15.0\pm 3.0$\,Myr and Upper Centaurus Lupus, $16.0\pm 2.0$\,Myr.]{\includegraphics[width=0.43\linewidth,keepaspectratio]{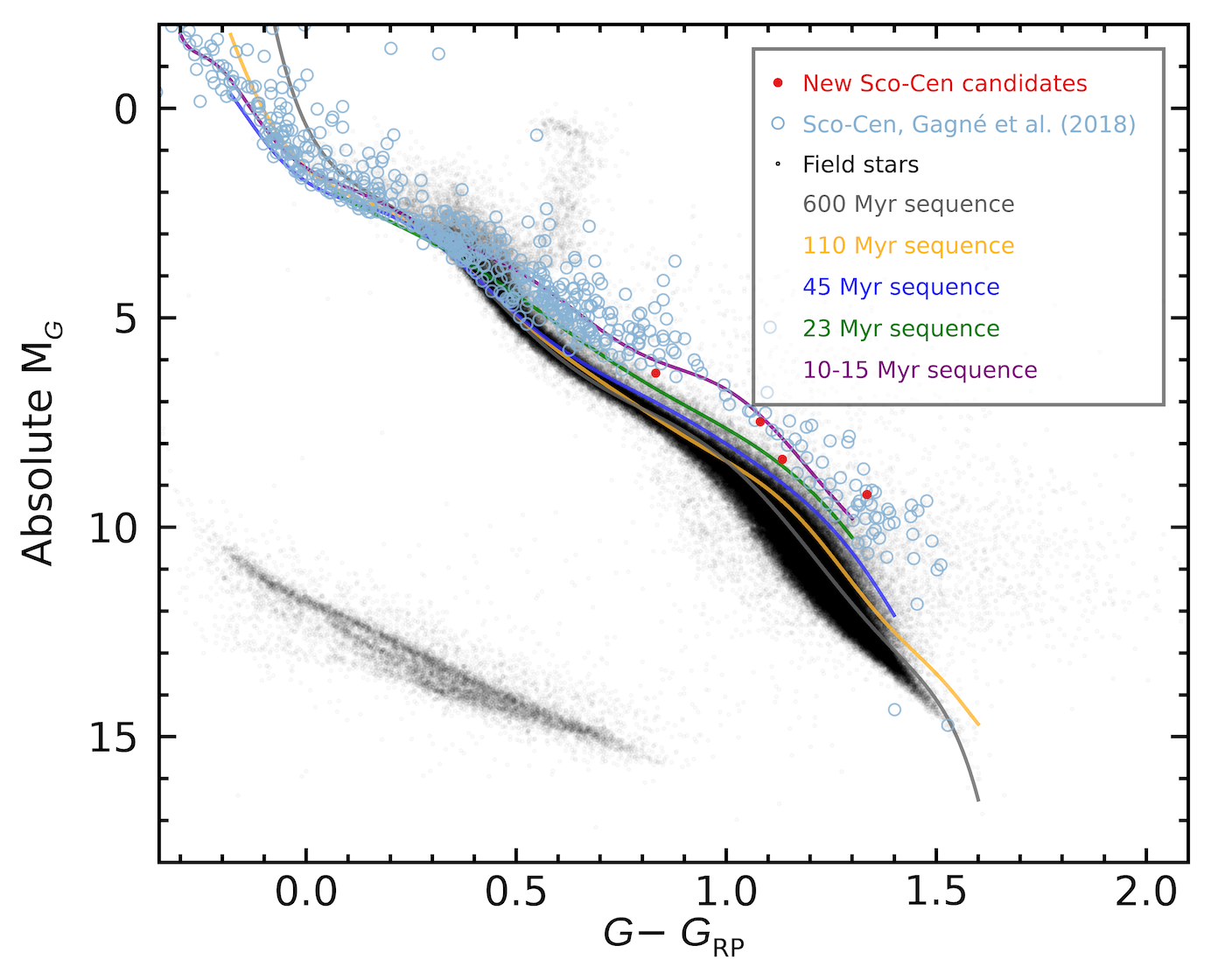}} \\
\subfloat[$\beta$ Pictoris, $24.0\pm 3.0$\,Myr.]{\includegraphics[width=0.43\linewidth,keepaspectratio]{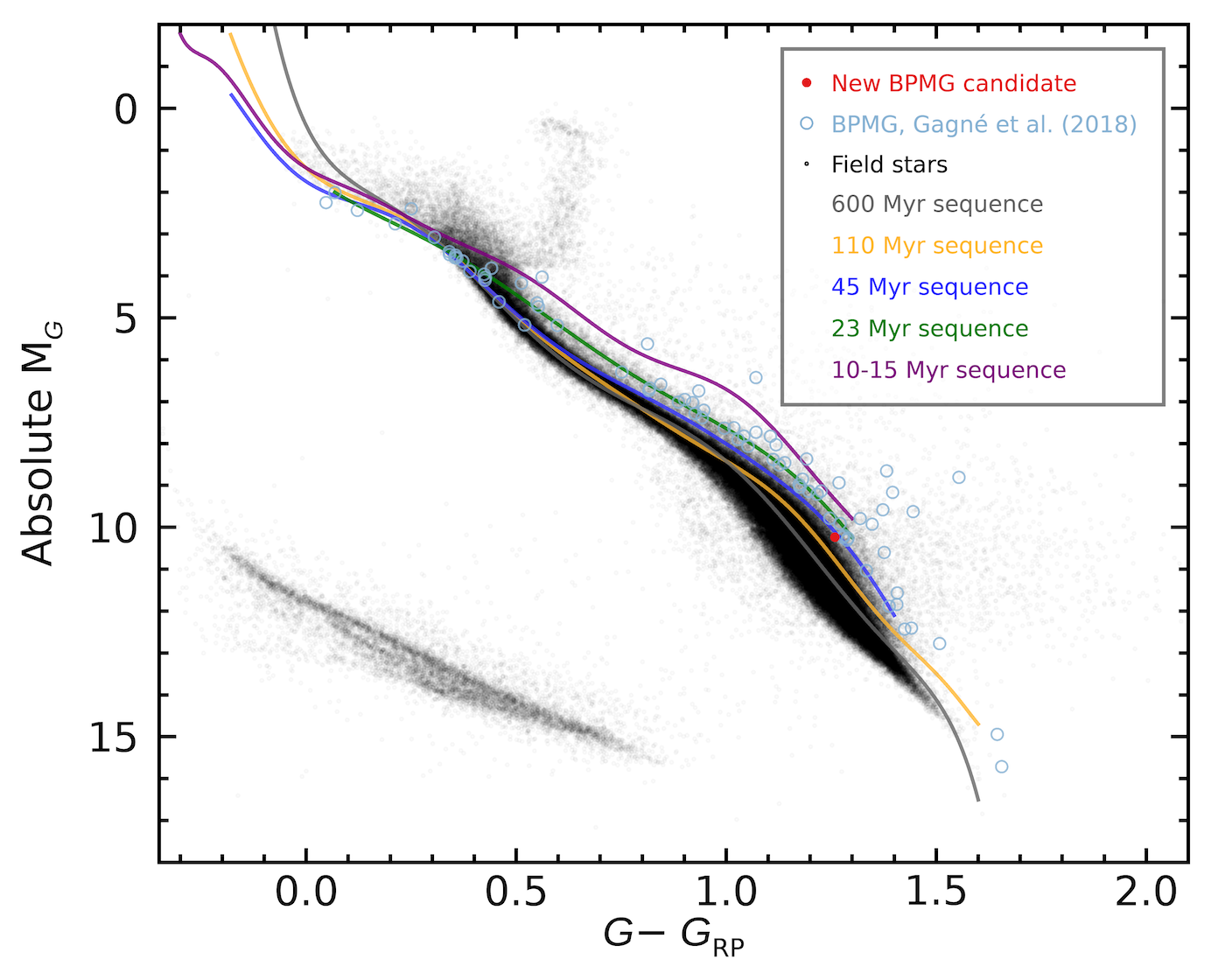}} 
\subfloat[AB Doradus, $149.0\pm 51.0$\,Myr.]{\includegraphics[width=0.43\linewidth,keepaspectratio]{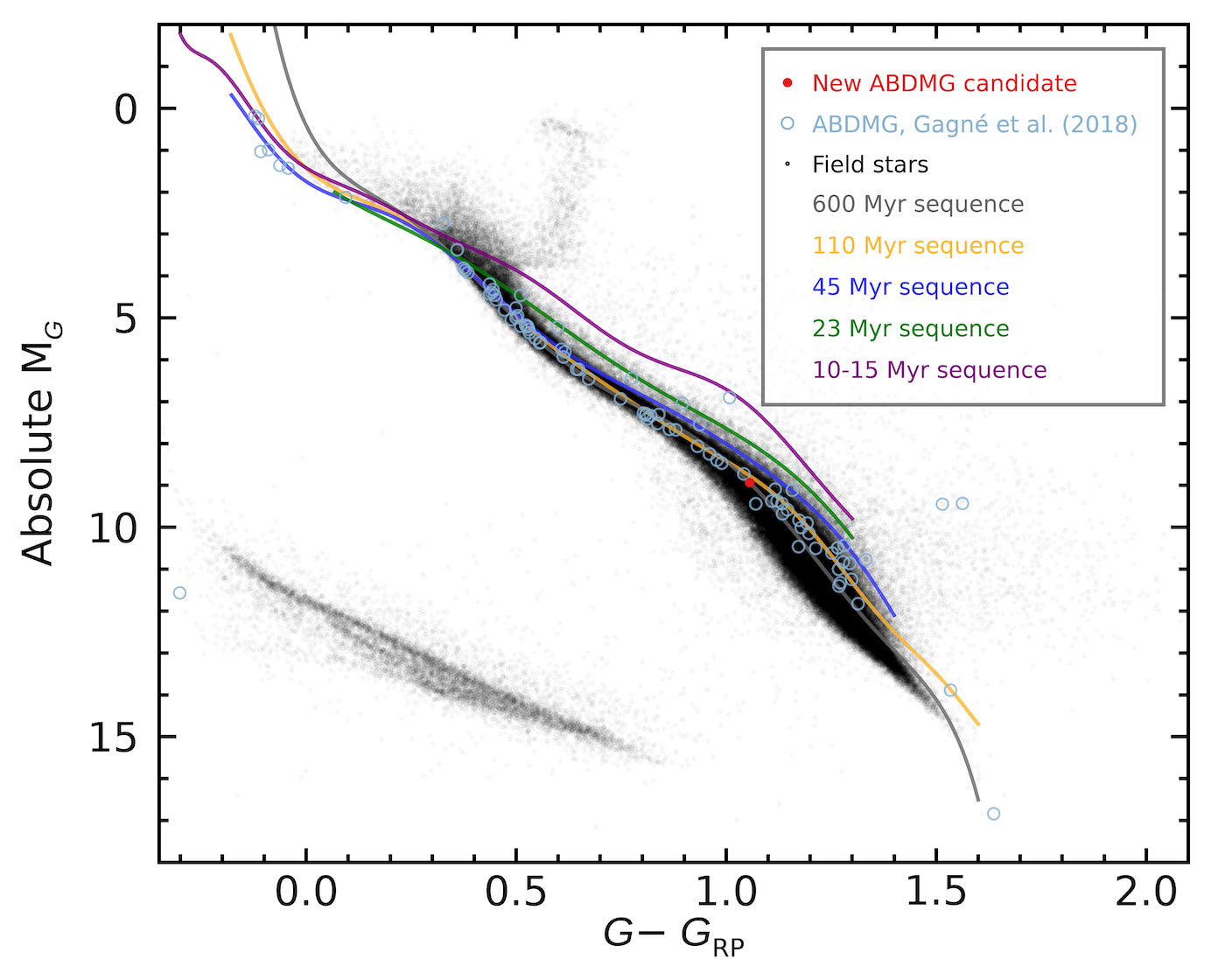}}
\caption{Color--magnitude diagrams to check membership of the newly identified candidate members for each young association (See Table~\ref{table:newmem}). New candidate members are shown as a red points on top of a field sample from \textit{Gaia} DR2 in black. We include empirical sequences based on bona fide members of young association of several ages \citep{Gagne2020} and the members of the groups as light blue empty circles \citep{Gagne2018}. References for each young association are in Table~\ref{table:ya_gagne}. The color--magnitude diagrams do not discard the new candidate members, but more study will be needed before they are confirmed as members. }
\label{fig:modelsgroups}
\end{center}
\end{figure*}

As a further check on the membership probabilities for the new candidate members (or that changed membership group), we compared their position on the \textit{Gaia} color--magnitude diagram to that of empirical sequences based on bona fide members of young association of different ages \citep{Gagne2020}, and members of each association \citep{Gagne2018}.
To compare an individual star's position on a color--magnitude diagram, it is important to evaluate the extinction due to interstellar dust.
We used the extinction maps of \emph{STructuring by Inversion of the Local InterStellar Medium} (STILISM; \citealp{Lallement2014,Capitanio2017,Lallement2018})\footnote{Available at \url{https://stilism.obspm.fr}} and the method of \citet{Gagne2020} to calculate de-reddened {\emph Gaia}~DR2 $G$- and $G_{\rm RP}$-band magnitudes. 
In summary, a template spectrum of the appropriate spectral type is used to calculate the effect of a typical interstellar dust extinction curve on the full {\emph Gaia} bandpasses and stellar spectra, a step that is required because the {\emph Gaia} bandpasses are particularly wide. 
Interstellar extinction will thus typically move M~dwarfs down along isochrones (redder colors and fainter magnitudes) in a {\emph Gaia} absolute $G$ versus $G-G_{\rm RP}$ color--magnitude diagram, but it will move the high-mass stars horizontally to redder colors, and across distinct isochrones.
The position in the color--magnitude diagram of our new candidate members are shown in Figure~\ref{fig:modelsgroups}, and are compared to empirical sequences based on bona fide members of young association of different ages \citep{Gagne2020}, and members of Upper Scorpius, Lower Centaurus Crux, Upper Centaurus Lupus, $\beta$Pictoris, and AB Doradus \citep{Gagne2018}.
We performed a visual inspection of all the high-probability members, and found that all the stars lie around the corresponding age sequences and/or within the scatter of the groups. This comparison corroborates their BANYAN~$\Sigma$ membership classifications when taking into account the scatter of the group. However more study will be needed before they are confirmed as members.
The summary of the catalogs used to compile all these M~dwarf age calibrators with $\haew$ measurements with a $> 90\%$ probability of membership in a young association are given in Table~\ref{table:age_cal}.

\begin{deluxetable}{lccc}[ht!]
\tablewidth{290pt}
\tabletypesize{\scriptsize}
\tablecaption{New candidate members of known young associations found    in this study.\label{table:newmem}}
\tablehead{    \colhead{2MASS Name}     & \colhead{SpT}    & \colhead{Young association}    & \colhead{Source} 
}\startdata 
J06511842-2154268& M$1.5$&ABDMG&\citet{Riaz2006} \\ 
J03273084+2212382& M$4.5$&BPMG&\citet{Jeffers2018} \\ 
J12323103-7255068& M$3.5$&EPSC&\citet{Riaz2006} \\ 
J12554838-5133385& M$2.0$&LCC&\citet{Riaz2006} \\ 
J15264071-6110559& M$3.0$&UCL&\citet{Riaz2006} \\ 
J16544682-3502540& M$4.0$&UCL&\citet{Riaz2006} \\ 
J16040453-2346377& M$4.0$&USCO&\citet{Slesnick2006} \\ 
\enddata 
\end{deluxetable}

\subsection{Identifying and age-dating co-movers with white dwarfs}
\label{subsec:comovers}

With BANYAN~$\Sigma$ we can cover age-calibrators up to $\sim$\,$750$\,Myr. 
For older ages we turned to white dwarfs as chronometers to identify older age calibrators. 
White dwarf cooling models and model atmospheres are robustly developed to estimate precise and reliable cooling ages and masses \citep{Fontaine2001}. These models can be used to calculate total ages with $10-20\%$ precision \citep{Fouesneau2018} and as old as $12$\,Gyr \citep{Cummings2018}. 
Therefore using co-moving white dwarf systems as age calibrators greatly expands the range used for grounding the age-activity relation.

The summary of the catalogs where we found M~dwarf with a white dwarf co-mover, that served as age calibrators are in Table~\ref{table:age_cal}.
To find M~dwarfs co-moving with white dwarfs we cross-matched our literature search sample with the \citet{Fusillo2019} white dwarf catalog using a $10'$ radius.
\citet{Fusillo2019} used a catalog of spectroscopically identified white dwarfs from the Sloan Digital Sky Survey \citep{York2000} to define cuts to select $486,641$ white dwarfs from the \textit{Gaia} DR2 color--magnitude diagram. 
In our study we only used the $260,000$ high-confidence white dwarf candidates ($\rm{P_{WD}} > 0.75$) as suggested by \citet{Fusillo2019}.

To find M~dwarfs co-moving with a white dwarf we used the \nnotinmg~sources which had good \textit{Gaia} data but were rejected as young association members and/or have a high probability ($>90\%$) to be field stars according to BANYAN~$\Sigma$ (Section~\ref{subsec:movinggroups}). 
We did not look for white dwarfs co-moving with M~dwarf members of young associations because the age of the moving group is better constrained than the estimation of the white dwarf age (see Section~\ref{sec:intro}). 
Additionally, companion searches within young associations based solely on kinematics produces many false positives due to their high probability of chance alignment. 
 
We considered an M-dwarf and a white dwarf to be co-moving if: 
\begin{enumerate}
    \item $\pi_{wd}/\sigma_{\pi,wd} > 4$
    \item $\pi_{m}/\sigma_{\pi,m} > 10$
    \item $|\mu_{\alpha,wd} - \mu_{\alpha,m}| < 3\times (\sigma_{\mu_{\alpha,wd}}+\sigma_{\mu_{\alpha,m}})$
    \item $|\mu_{\delta,wd} - \mu_{\delta,m}| < 3\times (\sigma_{\mu_{\delta,wd}}+\sigma_{\mu_{\delta,m}})$
    \item $|\pi_{wd} - \pi_{m}| < 3\times (\sigma_{\pi,wd}+\sigma_{\pi,m})$.
\end{enumerate}
The cut on signal-noise ratio for the white dwarf parallax ($\pi_{wd}/\sigma_{\pi,wd}$), is lower than for M~dwarfs as suggested by \citet{Fusillo2019} and the signal-noise ratio for M~dwarfs ($\pi_{m}/\sigma_{\pi,m}$) follows the suggestions by \citet{Lindegren2018a}. 
For both components of the proper motion ($\mu_{\alpha}$ and $\mu_{\delta}$) and the parallax ($\pi$) we required that the difference between the value for the white dwarf and the M~dwarf in a pair to be smaller than $3\sigma$, where $\sigma$ is the sum of the errors in each parameter. 
In total we found \wdmpairs pairs with this criteria.

To remove false positive companions, we calculated the probability of chance alignment for each M~dwarf by re-assigning the proper motions of all the white dwarfs in \citet{Fusillo2019} and repeating the search for co-movers with the criteria described above, $N$\,$=$\,$1000$ times. 
We then calculated the probability of chance alignment for each M~dwarf as $n_{\rm rand}/N$, where $n_{\rm rand}$ is the number of random co-movers and required it to be smaller than $0.01$ to include a pair in our analysis.
By adding this cut on probability of chance alignment we remove pairs with several close-by white dwarfs which makes a random match more likely. 
Out of the \wdmpairs pairs found, \wdmpairshighchancealignment had a probability of chance alignment higher than $0.01$.

To calibrate the age-activity relation, wide white dwarf-M~dwarf pairs are key to assume both components evolved as single stars without interacting.
We estimated the physical separation between the pairs as $a = 1.22 \theta \times D$, where $a$ is the separation in AU, $\theta$ is the angular separation and $D$ is the distance in parsecs. 
We found that the closest pair is separated by $\sim$\,$500$\,AU, therefore we can assume the two stars in each pair evolved independently \citep{Dhital2015,Skinner2017}.

Once we identified the \wdmpairslowchancealignment M~dwarf-white~dwarf pairs, we used the open source Python package available online \texttt{wdwarfdate} (Kiman et al. in prep.)\footnote{\url{https://wdwarfdate.readthedocs.io/en/latest/}} to estimate their age. 
\texttt{wdwarfdate} estimates ages of white dwarfs in a Bayesian framework from an effective temperature and a surface gravity. 
From the total number of pairs, \wdmpairsage had effective temperature and surface gravity from \cite{Fusillo2019} and were in the ranges where the models we use are valid. 
Using this information and cooling models \citep{Bergeron1995,Fontaine2001,Holberg2006,Kowalski2006,Bergeron2011,Tremblay2011,Blouin2018}\footnote{http://www.astro.umontreal.ca/~bergeron/CoolingModels/} we obtained the cooling ages and the masses of the white dwarfs. 
With a semi-empirical initial-final mass relation \citep{Cummings2018} and the masses of the white dwarfs we obtained the masses of the progenitor stars which were used with the MESA Isochrones \citep{Choi2016,Dotter2016,Paxton2011,Paxton2013,Paxton2015,Paxton2018} to determine the ages of the progenitors. 
By adding the cooling ages and the ages of the progenitors we obtained the total ages. 
In this process we assumed that all the white dwarfs are DA, meaning have a hydrogen-dominated atmosphere, which is a good approximation given that most of the white dwarfs in the galaxy are DA. 
We also assumed solar metallicity and ${\rm v/vcrit}=0$ for the progenitors.
We refer to Kiman et al. (in prep.) for more details on the calculation of the ages of the white dwarfs co-moving with an M~dwarf.
The results are shown in Table~\ref{table:wd_sample}, where we provide the \textit{Gaia} DR2 $\texttt{source\_id}$ for the M~dwarf and the white dwarf in each co-moving pair and the estimated total age. 
Uncertainties are calculated as the $84^{\rm th}$ percentile minus the median as high error and median minus the $16^{\rm th}$ percentile as low error.

\begin{deluxetable}{ccc}[ht!]
\tablewidth{290pt}
\tabletypesize{\scriptsize}
\tablecaption{Ages for the white dwarfs co-moving with an M dwarf. \label{table:wd_sample}}
\tablehead{                        \multicolumn{2}{c}{\textit{Gaia} source id}                         & \colhead{Total age}                         \\ M dwarf & White dwarf & (Gyr) 
}\startdata 
$2543566734628019712$&$2543472279707400320$&$3.05_{-1.15}^{+3.61}$\\ 
$2544030286155342080$&$2544024582438777088$&$1.41_{-0.79}^{+3.5}$\\ 
$2536947571549610368$&$2536960490812885760$&$2.57_{-1.04}^{+3.65}$\\ 
$2536695439789556352$&$2536705752006690304$&$1.94_{-1.03}^{+3.58}$\\ 
$3264871552432918528$&$3264871552432918784$&$1.71_{-1.05}^{+3.58}$\\ 
$676167219784728576$&$676167215489980800$&$2.57_{-1.33}^{+4.31}$\\ 
$703747197659174528$&$703753485491279488$&$0.64_{-0.31}^{+1.84}$\\ 
$636424547365777152$&$636417842920590208$&$2.56_{-0.95}^{+3.36}$\\ 
$799122031706484736$&$799133954536821248$&$3.04_{-1.08}^{+2.97}$\\ 
$793350660811961984$&$793351038769083776$&$2.31_{-0.93}^{+3.33}$\\ 
$743097619303531776$&$743097619303531904$&$2.91_{-1.87}^{+4.42}$\\ 
$4030722598505336192$&$4030722594210006784$&$3.53_{-1.57}^{+3.5}$\\ 
$3898427744542897152$&$3898427744542897408$&$1.67_{-0.48}^{+2.12}$\\ 
$4006695825601458816$&$4006671533266385792$&$5.2_{-1.81}^{+3.18}$\\ 
$3928724924885805568$&$3928724512568932992$&$1.27_{-0.67}^{+2.75}$\\ 
$1465169548332089472$&$1465169548332089600$&$1.94_{-1.17}^{+4.09}$\\ 
$1610798793983536384$&$1610800271452287488$&$2.46_{-1.55}^{+4.15}$\\ 
$4424639574212380800$&$4424639368053787776$&$3.66_{-1.03}^{+2.37}$\\ 
$1321738561431758592$&$1321738565727229184$&$4.87_{-2.06}^{+4.04}$\\ 
$1328907068007155072$&$1328909232670299904$&$4.61_{-1.68}^{+3.08}$\\ 
$4467448891937012992$&$4467448853280423296$&$4.35_{-1.23}^{+2.67}$\\ 
\enddata 
\end{deluxetable}

\section{Measurements and unresolved binaries}
\label{sec:binariesandlhalbol}

\subsection{Calculating $\lhalbol$ for the age-calibrators}
\label{subsec:lhalbol}

The activity strength of M~dwarfs is usually quantified with the ratio of the $\halpha$ luminosity to the bolometric luminosity $\lhalbol$ \citep{Hawley1996,West2008a,Schmidt2015}.
Using this fractional $\halpha$ luminosity removes the dependence on the continuum of the $\haew$ and facilitates comparison of stars of different effective temperatures. 
$\lhalbol$ was calculated using the ``$\chi$'' factor \citep[e.g.][]{Walkowicz2004}, where
\begin{equation*}
    \lhalbol = \chi ({\rm SpT}) \times \haew.
\end{equation*}
The $\chi$ values were empirically calibrated as a function of spectral type using ${\rm M}0-{\rm M}9$ from the Praesepe and Hyades clusters by \citet{Douglas2014}. 
Given that at the ages of these clusters ($\sim$\,$700$\,Myr) M~dwarfs have almost completely converged to the main sequence, we can assume that the $\chi$ values are valid for field dwarfs. 
However, the $\chi$ values may vary for younger stars and/or stars with different metallicities.
In our analysis, we assume that the difference in $\chi$ values for younger stars or different metallicities is not significant.
Note that these might be strong assumptions, but to our knowledge there is no better suited calibration of the $\chi$ values available. Also note that we used the convention that $\haew>0$ means $\halpha$ is in emission and $\haew<0$ means it is in absorption.

To calculate the $\lhalbol$ uncertainty, we generated a normal distribution of $\haew$ with its literature reported error as the standard deviation for each star and calculated the $\lhalbol$ for each value in the normal distribution. 
We adopted the standard deviation of the final distribution as error for $\lhalbol$. We did not include errors of the $\chi$ values in the error calculation. 

\subsection{Unresolved binary identification}
\label{subsec:unresolvedbinaries}

We performed a search for known external factors that could increase the magnetic activity of the star, such as close binarity \citep{Morgan2012,Skinner2017} and accretion disks outside the \citet{White2003} criteria (see Section~\ref{subsubsec:accretors}) between our age-calibrators. 
We used the empirical sequences based on bona fide members for each young association \citep{Gagne2020} to search for photometric binaries by identifying M~dwarfs in the binary sequence, hence at most $0.75$ brighter in absolute magnitude ($M_{\rm G}$) for a given color ($G-G_{\rm RP}$) than the value indicated by the model.
We visually inspected each potential binary in the color--magnitude diagram of its respective young associations, and only selected stars that we could confirm they lived in the binary sequence.
We only applied this method for stars with $G-G_{\rm RP}<1.2$ because the binary sequence is highly scattered for redder stars.
We identified $5$ potential binaries which are indicated in the age-calibrators sample in the column $\texttt{potential\_binary}$ with a $1$. 
These stars were removed from the following analysis.

For M~dwarfs co-moving with a white dwarf in our sample we performed a literature search looking for known binaries or other factors that could increase the magnetic activity. 
All the following cases were removed from the following analysis: 2MASS~J18393839+1623136 was identified as an unresolved X-ray binary \citep{Haakonsen2009} and 2MASS~J13545778+0512391 has an accretion disk around it \citep{Theissen2014}.

\section{Activity Fraction}
\label{sec:activityfraction}

We used the young part of the sample of age calibrators, described in the previous section, to calculate the activity fraction as a function of color for M~dwarfs younger than $1$\,Gyr and compared it with previous activity studies of field dwarfs. In this section we describe how we classified M~dwarfs as active or inactive, and the resulting activity fraction.

\subsection{Fitting the inactive sequence}
\label{subsec:inactivesequence}

\begin{figure}[ht!]
\begin{center}
\includegraphics[width=\linewidth]{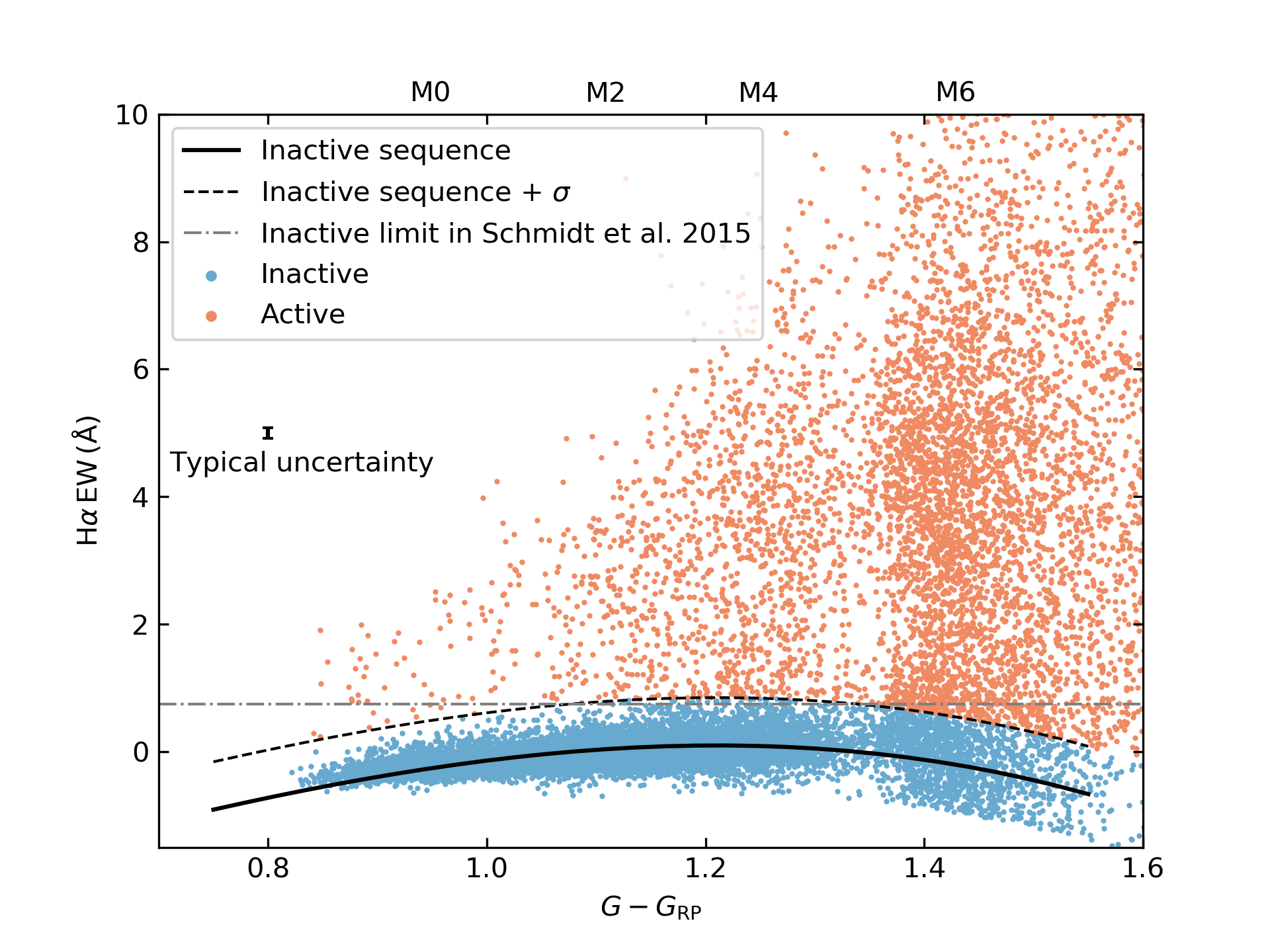}
\caption{Classification of active and inactive M~dwarfs. We fit the inactive sequence shown in a solid black line using the \citet{Kiman2019} sample and defined the boundary between active and inactive stars as the inactive sequence plus $\delta$, where $\delta=0.75$\,${\rm \AA}$, shown as a black dashed line. Typical errorbars are indicated for the active and inactive stars to the left of the plot. As a comparison we include the boundary used by \citet{Schmidt2015} in a gray dashed-dotted line. We show active stars in orange and inactive stars in blue, classified with our criteria.}
\label{fig:inactive_sequence}.
\end{center}
\end{figure}

Previous studies have identified what we define as the inactive sequence in this study, in $\haew$ versus mass or effective temperature for M~dwarfs \citep[e.g.][]{Stauffer1986,Lopez-Santiago2010,Newton2017,Fang2018}.
This inactive sequence is defined as the lower boundary of the $\haew$ versus mass, color or spectral type relation, and it shows an increase in $\haew$ from higher to lower masses (or redder colors).
This increase in $\haew$ is related to a decrease in the $\halpha$ absorption from the photosphere, meaning a decrease in photospheric luminosity \citep{Stauffer1986} and not to an increase in emission. 
We used \citet{Kiman2019} objects to establish the inactive sequence for $\haew$ versus $(G-\grp)$ and with it, the active or inactive category. 
To establish the inactive sequence following \citet{Newton2017}, we did $10$ iterations of a third degree polynomial fit to stars with $\haew < 1{\rm \AA}$. 
In each iteration we rejected the stars that had an $\haew$ higher than the best fit. 
We defined the activity boundary as the best fit to the inactive sequence plus a small increment, $\delta$:
\begin{equation}
\begin{split}
    \haew &=  -2.39\times (G-G_{\rm RP})^3 \\
    &+ 2.84\times (G-G_{\rm RP})^2\\ 
    &+ 3.63\times (G-G_{\rm RP}) - 4.22 + \delta
\end{split}
\end{equation}

\noindent where $\delta = 0.75$\,${\rm \AA}$ and it was added to make our classification in active and inactive compatible with previous definitions \citep{West2011,Schmidt2015}. This limit is valid in the range $0.8<(G-G_{\rm RP})<1.55$ (${\rm M}0 \leq {\rm SpT} \leq {\rm M}7$). 
We considered all the stars which have an $\haew$ above the boundary to be active. All other objects were considered inactive. 
We show the activity boundary, the best fit to the inactive sequence and our classification as active and inactive in Figure~\ref{fig:inactive_sequence}.
As a comparison, we show the limit for active and inactive used by \citet{Schmidt2015}. Among other cuts related to signal to noise, \citet{Schmidt2015} considered inactive stars as those with $\haew<0.75$\,${\rm \AA}$. Their limit agrees with our definition, except for a small deviation for spectral types $<{\rm M}0$ and $>{\rm M}6$.

\subsection{Calculating active fraction as a function of age}

\begin{figure*}[ht!]
\begin{center}
\includegraphics[width=\linewidth]{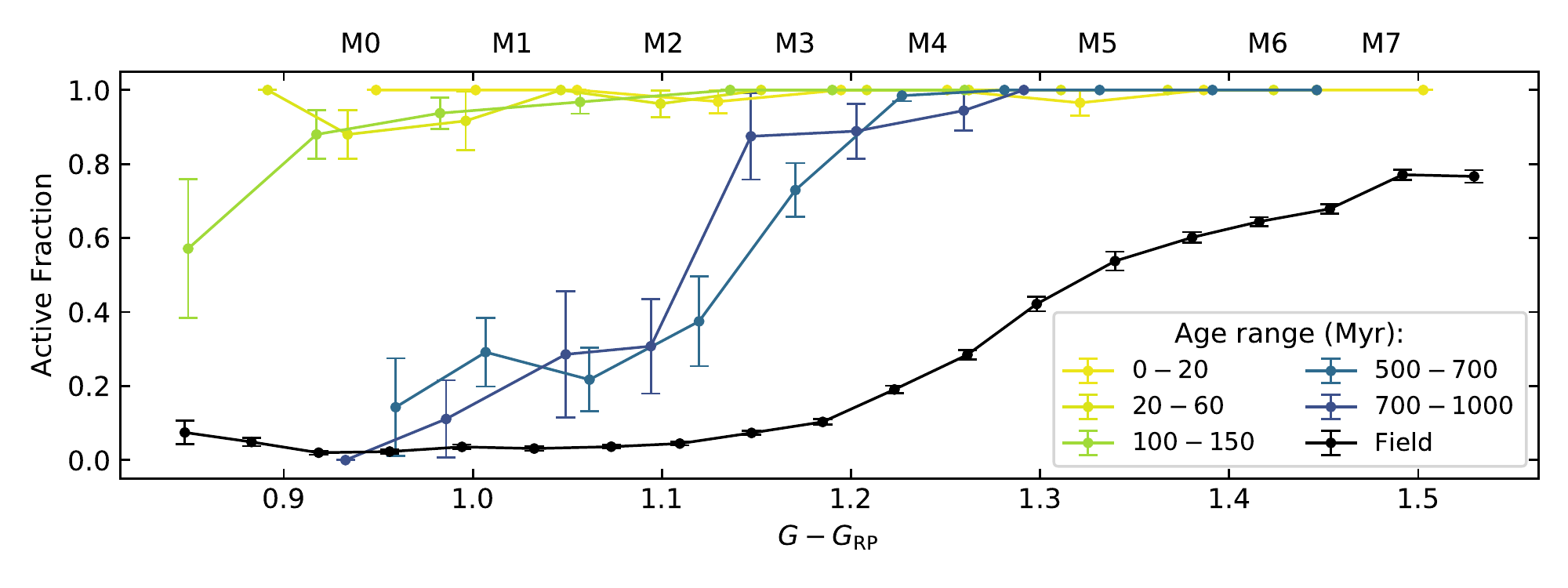}
\caption{Active fraction per bins of the \textit{Gaia} red color for different bins of age. We include the active fraction for field M~dwarfs calculated from the sample \citet{Kiman2019} in black. We noted that the active fraction decreases with age for M dwarfs and that this decrease depends on mass.} 
\label{fig:activityfraction}.
\end{center}
\end{figure*}

Using the definition of active and inactive described above, we calculated the active fraction in bins of the red \textit{Gaia} DR2 color $(G-G_{\rm RP})$ and studied its dependence with age. 
We used the histogram function available in the Python package \texttt{numpy} \citep{oliphant2006guide,van2011numpy} to define the bins in color, because it is optimized for non-normal data.
We repeated this calculation for the following age-bins: $(0-20)$\,Myr, $(20-60)$\,Myr, $(100-150)$\,Myr, $(500-700)$\,Myr and $(700-1000)$\,Myr. 
The results are shown in Figure~\ref{fig:activityfraction}, where we included a reference to the mean spectral type (SpT) for each color at the top of the figure.
We also included the activity fraction for the sample from \citet{Kiman2019} in black, which are assumed to be primarily field stars, and it reproduces the results from previous studies \citep{West2004,West2011,Schmidt2015}.
The errors for each point were calculated based on a binomial distribution as
\begin{equation*}
    \sigma _f = (f\times(1-f))/n
\end{equation*}

\noindent where $f$ and $\sigma _f$ are the active fraction and uncertainty, and $n$ is the number of stars in the bin. 

By analysing the active fraction as a function of age in Figure \ref{fig:activityfraction}, we show that the active fraction of M~dwarfs evolves with age. There is a clear difference in the evolution of the active fraction between early ($<{\rm M}2$) and mid-M~dwarfs (${\rm M}4-{\rm M}7$). 
While the active fraction for early M~dwarfs decreases from $1$ to almost $0$ between $0$\,Myr and $750$\,Myr, for mid-M~dwarfs the active fraction stays close to $1$. 
This difference between early and mid-M~dwarfs agrees with previous studies which showed that later-type M~dwarfs stay active longer than earlier-type stars \citep{West2004,West2011,Schmidt2015}.

The active fraction for early-type M~dwarfs in Figure~\ref{fig:activityfraction} decreases progressively from $0$\,Myr to field age dwarfs, while for mid spectral types (${\rm M}4\leq{\rm SpT}<{\rm M}7$) it seems to decrease more abruptly from $(700-1000)$\,Myr (blue line) to field stars (black line).
Note that the field active fraction is representing a distribution of different ages. 
Therefore the difference between the bins $(700-1000)$\,Myr and field for mid-type M~dwarfs could be indicating that late types stay active longer and we do not have enough age resolution to distinguish the progressive decrease of the active fraction with age. 
This discrepancy also could be indicating that the magnetic activity of mid-type M~dwarfs decreases more abruptly than for early-types. 

Both $(500-700)$\,Myr and $(700-1000)$\,Myr age bins (the two blue lines) are statistically equivalent and present a transition period centered at ${\rm M}2$ where the active fraction increases from $0$ to $1$, for $1.0<(G-\grp)<1.2$. Significantly, this transition is completed close to spectral type $\sim{\rm M}3$, the limit between partially and fully convective M~dwarfs \citep{Chabrier1997}.
The shape of this transition for both bins could be affected by other physical effects which increase the magnetic activity of a star. 
For example, the active fraction could increase because of close companions of the stars which increase their magnetic activity \citep[e.g., ][]{Kraus2011,Morgan2012,Dhital2015,Skinner2017}. For the colors $1.0<(G-\grp)<1.1$ the fraction of active stars is $20\%$, which is close to the multiplicity fraction of M dwarfs of $(23.9\pm 1.4)\%$ at separations $<50$\,AU found by \citet{Winters2019}. We searched the literature for references of binarity and we did not find known binaries for these stars. Also, our sample is not complete, so we cannot compare directly to the results in \citet{Winters2019}.
Follow up observations are necessary to distinguish if these stars are still active at $\sim 700$\,Myr, or they have an unresolved binary keeping them active.

\section{Updated age-activity relation for active M dwarfs}
\label{sec:relation}

\subsection{Characterizing $\halpha$ versus age}
\label{subsec:dataageactivityrel}

\begin{figure*}
\centering 
\subfloat[$\haew$]{%
  \includegraphics[width=0.4\textwidth]{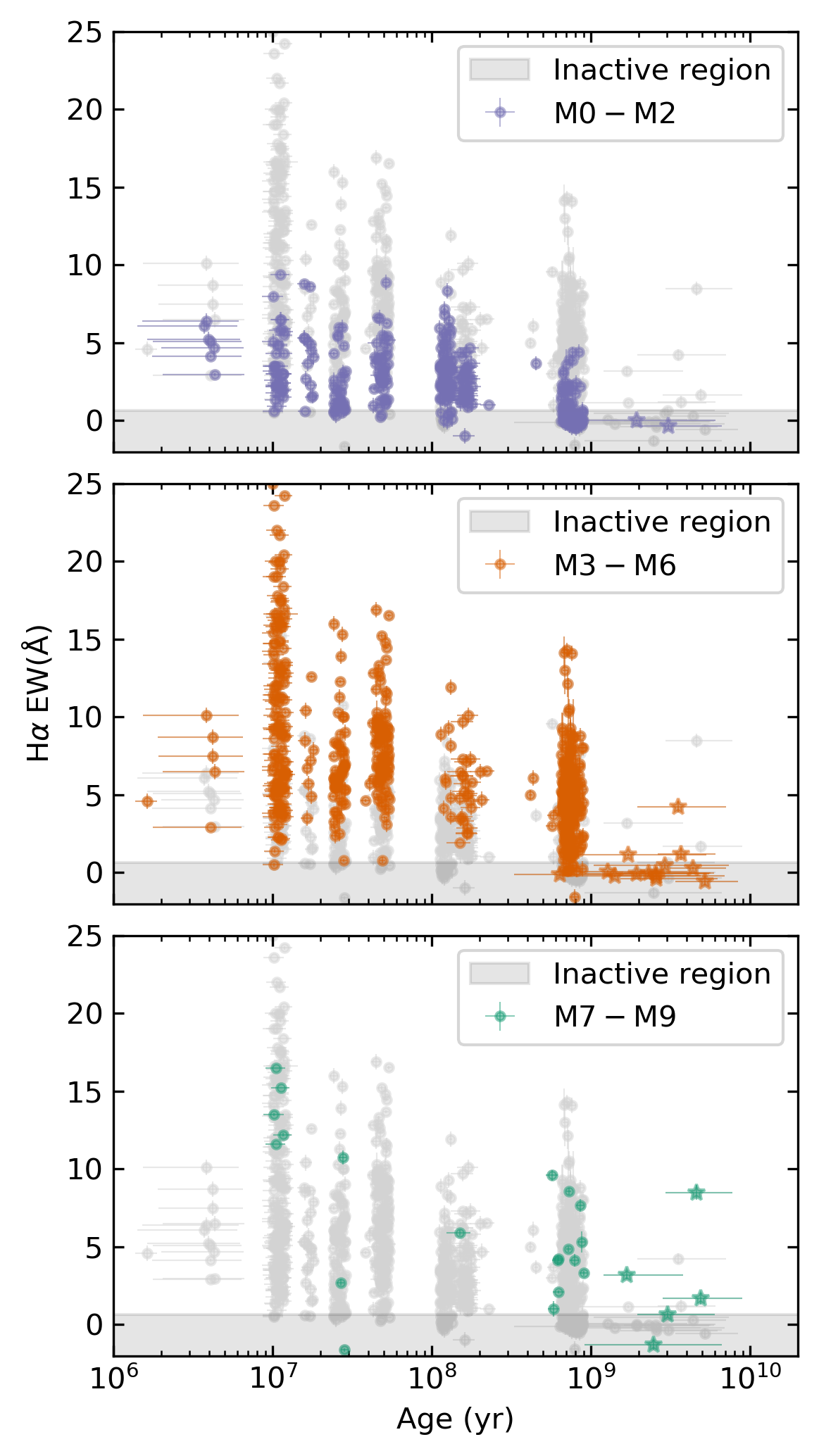}%
  \label{fig:ageactivity_a}%
}\qquad
\subfloat[$\lhalbol$]{%
  \includegraphics[width=0.4\textwidth]{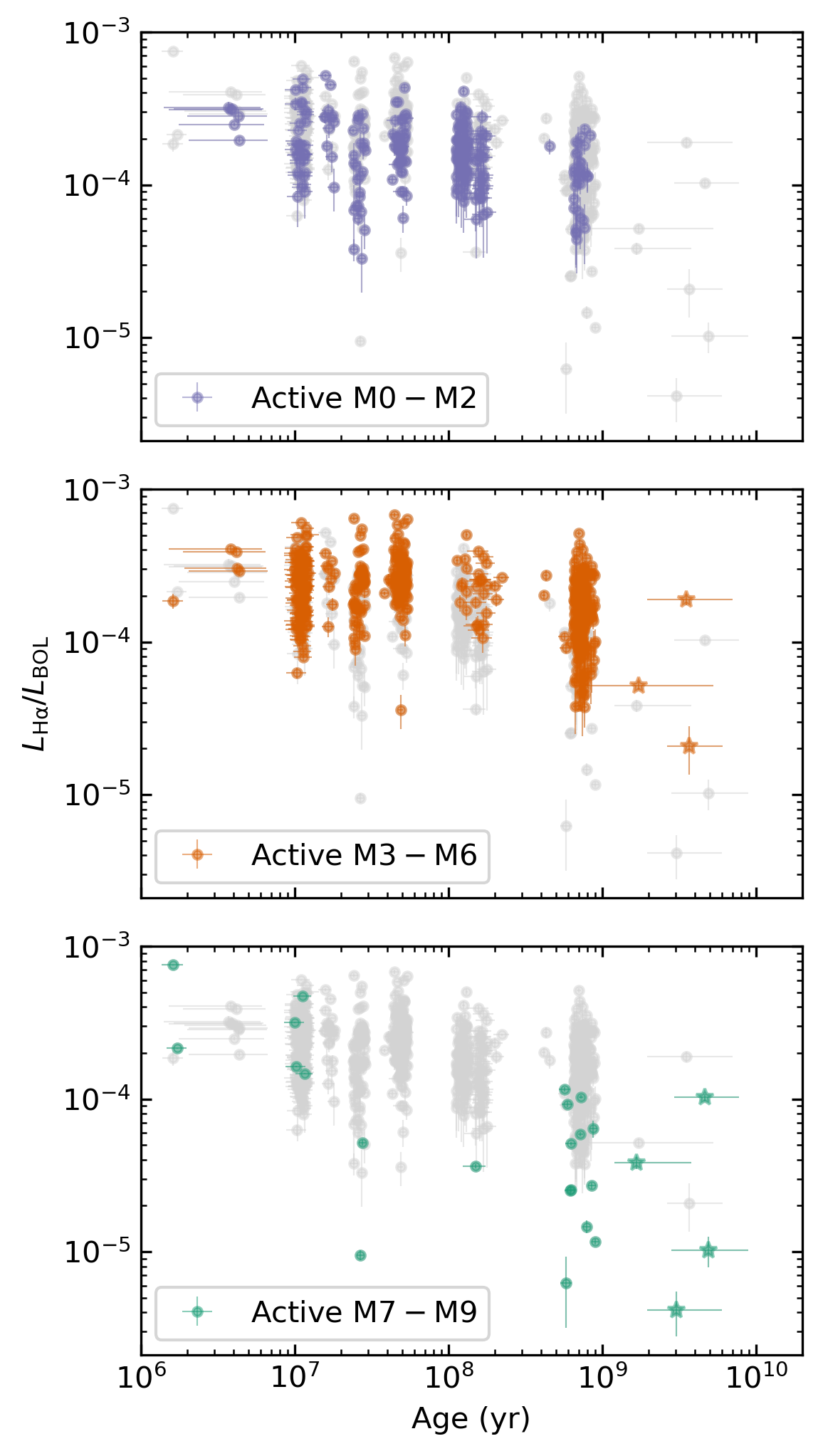}%
  \label{fig:ageactivity_b}%
}
\caption{Age activity relation as indicated by $\halpha$ equivalent width ($\haew$ left panels) and fractional $\halpha$ luminosity ($\lhalbol$ right panels). We divided the sample in three panels according to the spectral type: in purple early partially convective M~dwarfs (${\rm M}0-{\rm M}2$, top panels), in orange mid fully convective M~dwarfs (${\rm M}3-{\rm M}6$, mid panels) and in green ultracool fully convective M~dwarfs (${\rm M}7-{\rm M}9$, bottom panels). M~dwarfs from young associations are shown with a circle, and stars which have a white dwarf companion are shown with a five point star. 
A small random shift has been applied to the age of the stars from young associations to improve the visualization. No shift has been applied to the age of the M~dwarfs with a white dwarf companion. We added a gray area to indicate approximately the stars that are considered inactive. For the precise definition of inactive see Figure \ref{fig:inactive_sequence}. We only show active stars for the $\lhalbol$ relation.}
\label{fig:ageactivity}
\end{figure*}

We used our calibrators for $\haew$ and $\lhalbol$ to study activity strength as a function of age (Figures \ref{fig:ageactivity_a} and \ref{fig:ageactivity_b} respectively). 
We divided the calibrators into three panels according to their spectral type: ${\rm M}0-{\rm M}2$, ${\rm M}3-{\rm M}6$ and ${\rm M}7-{\rm M}9$ which roughly correspond to early-type partially convective, mid-type fully convective, and ultracool fully convective M~dwarfs, respectively. 
In the $\lhalbol$ plots we only show active stars. 
Inactive stars have $\lhalbol$ with too low signal-to-noise to be significant (gray area in Figures~\ref{fig:ageactivity_a}).

There is a large spread of $\haew$ between early, mid and late spectral type M~dwarfs, especially at the younger ages, shown in Figure~\ref{fig:ageactivity_a}.
As indicated by previous studies \citep[e.g.][]{Stauffer1986,West2011,Schmidt2015}, $\haew$ increases from early-to-late M dwarfs of the same age, which explains this effect, and it is removed when calculating $\lhalbol$, as can be seen in Figure~\ref{fig:ageactivity_b}.

We found less than ten mid- and late-type old ($>$\,$1$\,Gyr) active M~dwarfs according to their $\haew$, and none early-types. We could not identify any external factors related to these stars, such as a close binary which would increase their magnetic activity. Therefore, we consider them to be true active stars (see Section~\ref{subsec:unresolvedbinaries}).
The three youngest late-type M~dwarfs in the $\lhalbol$ are not displayed in the $\haew$ plot, because their equivalent width values are higher than $25{\rm \AA}$, and thus outside of the axis limit.

We also note that there are stars in the inactive region at all ages in Figure~\ref{fig:ageactivity_a} ($\haew \lesssim 0.75$\,${\rm \AA}$, gray area), meaning that a small $\haew$ does not necessarily indicate old age for an M~dwarf.
However, we do see an increase in the number of stars with low $\haew$ in the inactive region as age increases.

\begin{figure*}
\centering 
\subfloat[$\haew$]{%
  \includegraphics[width=0.47\textwidth]{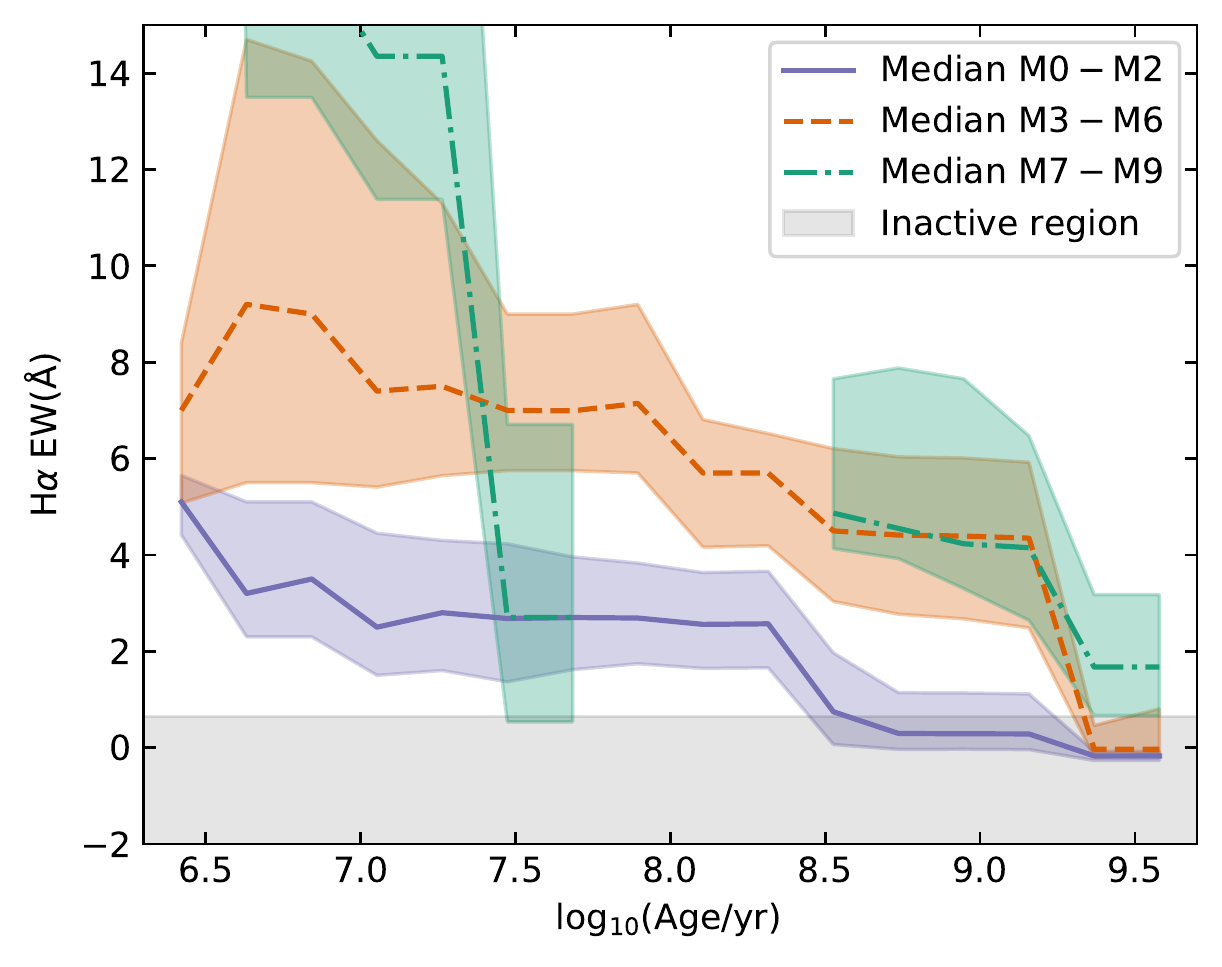}%
  \label{fig:medianageactivity_a}%
}\qquad
\subfloat[$\lhalbol$]{%
  \includegraphics[width=0.47\textwidth]{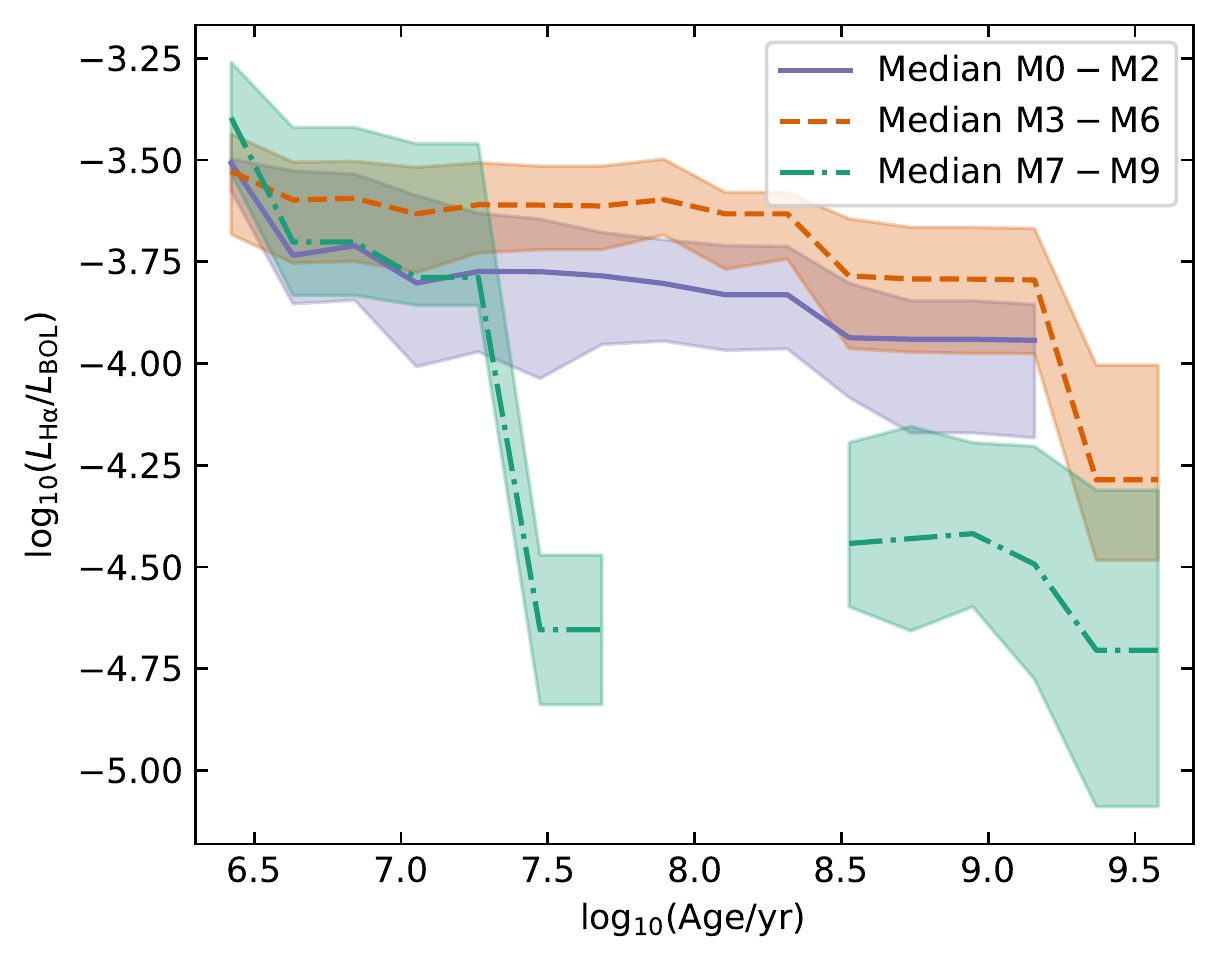}%
  \label{fig:medianageactivity_b}%
}
\caption{Median values and $25$ and $75$ percentile for a sliding window of width $0.8$ in units of $\log _{10}({\rm Age}/{\rm yr})$ for both $\halpha$ equivalent width ($\haew$) and fractional $\halpha$ luminosity ($\lhalbol$). Figure~\ref{fig:ageactivity} shows the data used to make this figure. We note that the magnetic activity for early-type M~dwarfs decreases progressively, while for mid-type M~dwarfs it seems to decrease rapidly after $1$\,Gyr.}
\label{fig:medianageactivity}
\end{figure*}

To study the trends of activity strength with age, we calculated the median values, $25$ and $75$ percentile for a sliding window of width $0.8$ in units of $\log _{10}({\rm Age}/{\rm yr})$ for both $\haew$ and $\lhalbol$. 
Results are shown in Figures~\ref{fig:medianageactivity_a} and \ref{fig:medianageactivity_b} respectively. 
The relation between $\haew$ and age for early M~dwarfs (purple lines in Figure~\ref{fig:medianageactivity_a}), shows a gradually decline from approximately $5$\,${\rm \AA}$ to $0$\,${\rm \AA}$ in $\sim$\,$1$\,Gyr.
For mid spectral types (${\rm M}3-{\rm M}6$, orange lines) we observe a progressive decrease of $\haew$ with age until $\sim$\,$1$\,Gyr. 
After $1$\,Gyr, we observe a large decline in $\halpha$ ($\Delta \haew \sim 4{\rm \AA}$).
There are too few late type dwarfs in our sample to be able to make a robust conclusion. However, we do observe a decline in the value of $\haew$ with age.

$\lhalbol$ presents a similar decay of magnetic activity for early- and mid-types (purple and orange lines) for ages $<1$\,Gyr in Figure~\ref{fig:medianageactivity_b}. 
After this age, early-type M~dwarfs are all inactive in our sample, while mid-type M~dwarfs present a transition to a steeper decrease of the magnetic activity with age. However, we need more old M~dwarfs with $\halpha$ measurements to make a conclusion about the evolution of magnetic activity after $1$\,Gyr.

\begin{figure}[ht]
\begin{center}
\includegraphics[width=\linewidth]{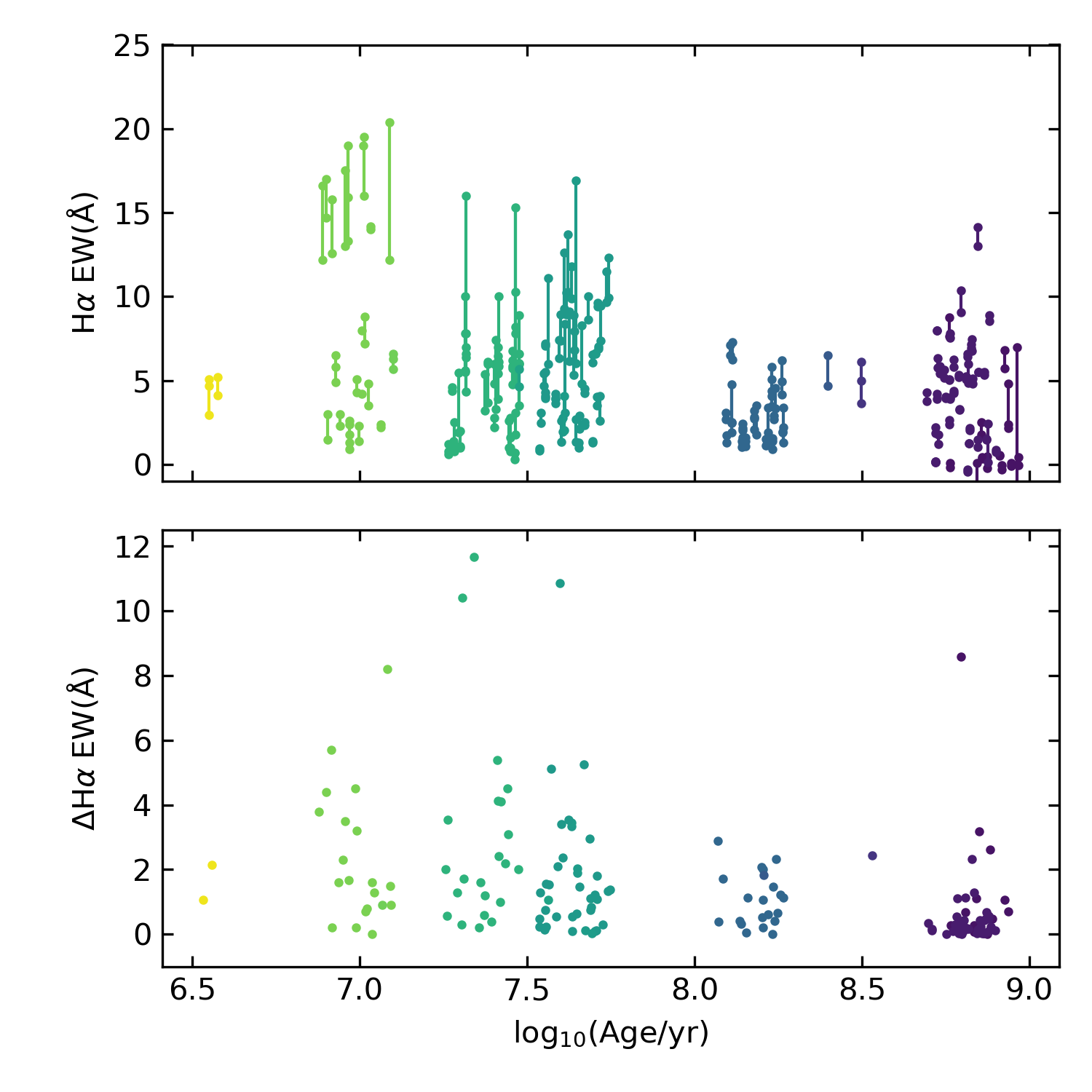}
\caption{$\halpha$ variability from repeated independent measurements of the same stars as a function of age. A small random shift has been applied to the age of each star to improve visualization, so we also color coded the stars according to their age to keep track of it. In the top panel we show all the measurements for each star. Measurements of the same star are connected with a line. In the bottom panel we show the difference between the maximum and the minimum values of $\haew$ for each star. We found that $94\%$ of the repeated stars has a level of $\haew$ variability $\leq 5$\,${\rm \AA}$ at young ages ($<1$\,Gyr).} 
\label{fig:variability}.
\end{center}
\end{figure}

There is significant scatter both in $\haew$ and $\lhalbol$ at each age bin, partially due to the intrinsic variability of $\halpha$ \citep{Lee2010,Bell2012}. 
To study the scatter in more detail, we analysed the M~dwarfs in our sample which have duplicate measurements of $\haew$. 
Some of these duplicated stars come from different studies of the same young association, which did their own $\haew$ measurements. We also noticed that the duplicated age-calibrators are, on average, at a slightly shorter distance than non duplicated age-calibrators. This tendency may be due to these stars being more likely to be selected by studies given the better quality of the data. We see not such tendency with spectra type. Therefore, we are confident that we can use these duplicated measurements to study $\halpha$ variability.
Our sample of age-calibrators contains \nrepearedagecalibrators stars with $2-6$ independent measurements of $\haew$ which are shown in the top panel of Figure~\ref{fig:variability} with the different measurements of $\haew$ corresponding to the same star joined by a line. 
We also show the difference between the maximum and the minimum value of $\haew$ for each star in the bottom panel of Figure~\ref{fig:variability}. We note that $94\%$ of the M~dwarfs in the sample has an intrinsic $\halpha$ variability $\leq 5{\rm \AA}$. 
By doing a literature search of the stars with higher $\Delta \haew$ we found that most of the stars which have $\Delta \haew \geq 10{\rm \AA}$ are known variable stars \citep{Kiraga2012,Schmidt2015,Samus2017}.
This analysis on variability shows that our $3{\rm \AA}$ cut to distinguish compatible measurements was conservative (see Section~\ref{subsec:overlapping}).
We also note that there is correlation between $\Delta \haew$ and $\haew$ measurements in agreement with the results from \citet{Lee2010,Bell2012}.

\subsection{Fitting the age-activity relation}
\label{subsec:modelandfitageactivityrel}

We fit the age-activity relation measured with $\lhalbol$ for all active age-calibrator (\nactivecalibrators stars, Tables~\ref{table:age_cal} and \ref{table:columnsagecal}) without binning by spectral type ranges, given that we lack enough information per bin.
A broken power-law has been used in several previous studies of the X-ray age-activity relation \citep[e.g., ][]{Jackson2012,Booth2017} and for the rotation-activity relation \citep[e.g., ][]{Delfosse1998,Douglas2014,Nunez2016a,Newton2017} for cool dwarfs to model a saturation regime where $\lhalbol$ remains constant, followed by a power-law decay.
Furthermore, \citet{West2008c} found that for M dwarfs, $\lhalbol$ remains constant for young ages and then decays. 
Therefore, we decided to use a broken power-law to fit the age-activity relation. 
See the Appendix~\ref{sec:cross-val} for a comparison between the broken power-law and polynomials of different degrees using the cross-validation method.

We performed a Markov chain Monte Carlo (MCMC) fit to the age-activity relation using \texttt{emcee} \citep{Foreman-Mackey2013} to estimate the parameters of the broken power-law and their uncertainties. 
The broken power-law we fit to the relation was of the form: 
\begin{equation}
    \log _{10}{\left(\frac{L_{{\rm H}\alpha}}{L_{\rm bol}}\right)}_{\rm model} = \begin{cases}
    \alpha _1\log _{10}\frac{t}{t_0} + \beta _1, &t < t_0\\
    \alpha _2\log _{10}\frac{t}{t_0} + \beta _1, &t_0 \leq  t
    \end{cases}
\end{equation}
where $t$ is the age of the star. We used a Gaussian function as the likelihood and uniform priors on each parameter. 
As discussed before, the age-activity relation for $\halpha$ is scattered due in part to the intrinsic variability of this emission line. We included an extra parameter, $\sigma _v$, to model the intrinsic variability of $\halpha$ assuming the variability of $\log _{10}(\lhalbol)$ is the same for all stars, in other words, that the $\halpha$ intrinsic variability is proportional to $\haew$ \citep{Lee2010,Bell2012}. 
The resulting likelihood for our model is: 

\begin{equation}
\begin{split}
    &\log \mathcal{L}= -\frac{1}{2}\times \\
    &\sum _i\frac{\log _{10}\left(\frac{L_{{\rm H}\alpha}}{L_{\rm bol}}\right)_i - \log_{10}\left(\frac{L_{{\rm H}\alpha}}{L_{\rm bol}}\right)_{\rm model}(t_i)}{\sigma _{\log_{10}\left(\frac{L_{{\rm H}\alpha}}{L_{\rm bol}}\right),i}^2 + \sigma _v^2} \\
    &+ \log (\sigma _{\log _{10}\left(\frac{L_{{\rm H}\alpha}}{L_{\rm bol}}\right),i}^2 + \sigma _v^2) \label{eq:likelihood}
\end{split}
\end{equation}

\noindent where we sum over each measurement $i$ of $\lhalbol$ and age.

To ensure that the MCMC had converged, we calculated the autocorrelation time ($\tau _f$) for our likelihood and sampled our posterior by doing $100$\,$\tau _f$ steps to assure $\sim$\,$100$ independent samples\footnote{To calculate the autocorrelation time we followed the tutorial in https://emcee.readthedocs.io/en/stable/tutorials/autocorr/}.

\begin{figure}[ht!]
\begin{center}
\includegraphics[width=\linewidth]{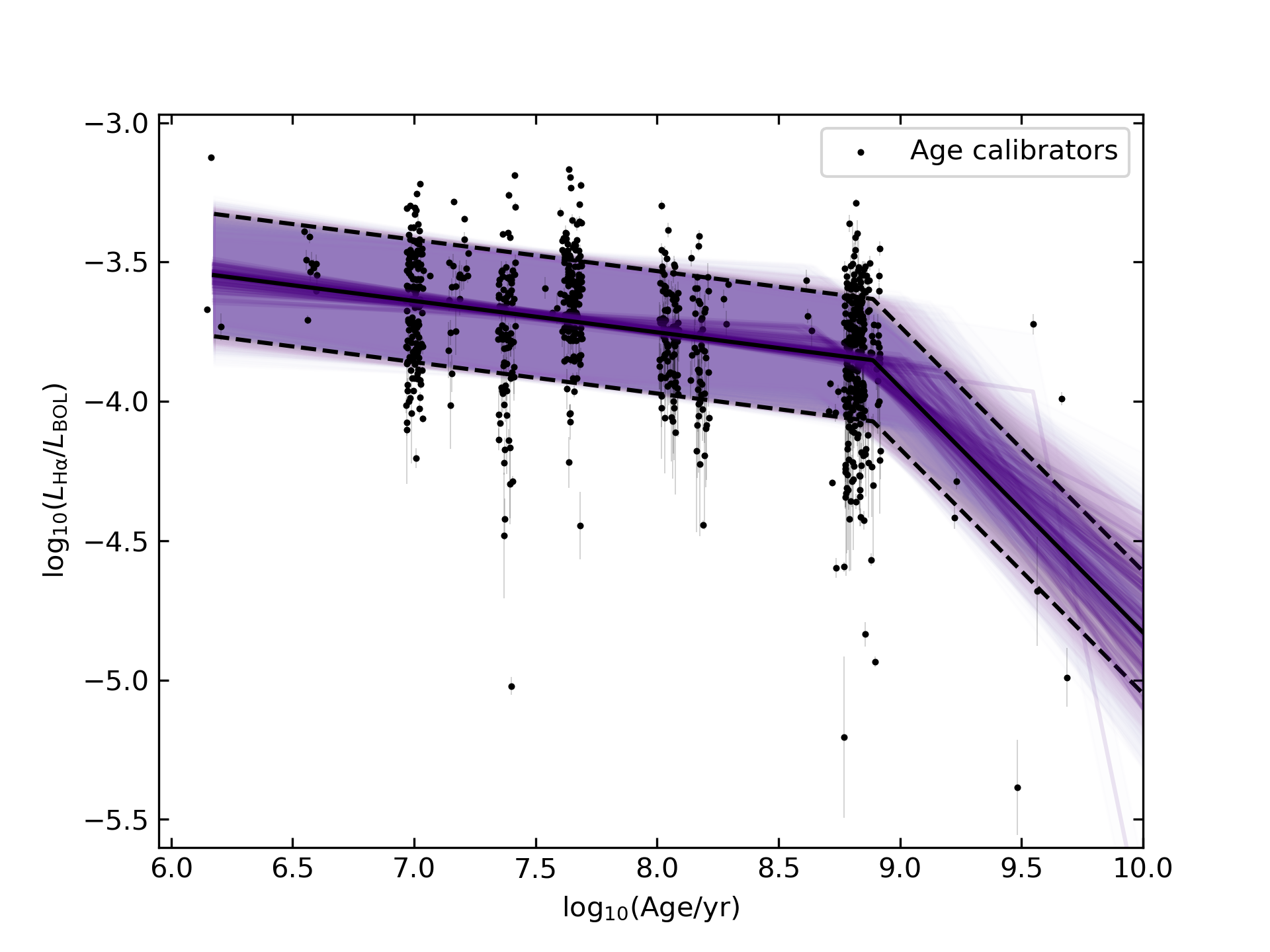}
\caption{Age activity relation for $\lhalbol$ for M~dwarfs. We fit the relation with a broken power-law using a Markov chain Monte Carlo. We show in a black line the maximum likelihood fit and in a dashed black line the fit $\pm \sigma_v$, which models the $\halpha$ variability. We also include $100$ draws from the posterior distributions for the parameters of the fit in purple. The age calibrators are shown as black points and we added a random shift to their ages to facilitate visualization. Our model fits well most of the younger stars ($<1$\,Gyr) and the range of variability includes the denser areas of points. However we do not have enough information to fit the breaking point or the power-law decay of the older stars.}
\label{fig:ageactivity_fit}.
\end{center}
\end{figure}

\begin{figure}[ht!]
\begin{center}
\includegraphics[width=\linewidth]{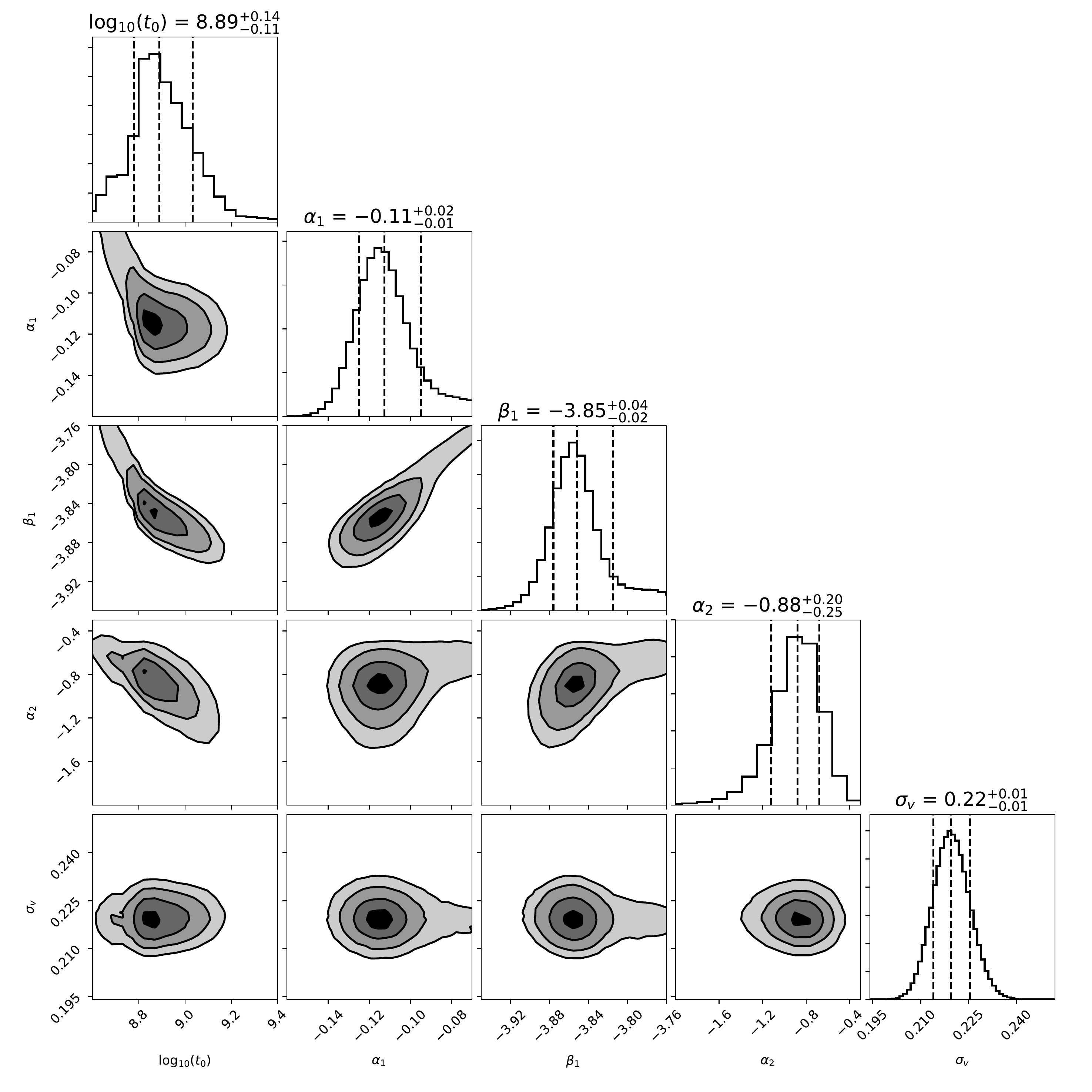}
\caption{Posterior probabilities for each parameter of the broken power-law fitted to the age activity relation in Figure~\ref{fig:ageactivity_fit}.} 
\label{fig:mcmc_res}.
\end{center}
\end{figure}

We show the maximum likelihood of the parameters of the broken power-law and the $\sigma _v$ over the age-calibrators in Figure~\ref{fig:ageactivity_fit}, as well as one hundred random samples from the posterior distributions of the broken power-law parameters.
We show the full posterior distributions and the maximum likelihood values for each parameter in Figure~\ref{fig:mcmc_res}.
As discussed in Section~\ref{subsec:dataageactivityrel}, we found for $\log _{10}(\lhalbol)$ a decrease in activity strength from $1$\,Myr to $t_0\sim 776$\,Myr with a power-law index of $\alpha _1=-0.11^{+0.02}_{-0.01}$, with a variability of $\sigma _v=0.22\pm0.01$.
For ages $>$\,$776$\,Myr, the relation shows a decline in magnetic activity with a power-law index of $\alpha_2=-0.88^{+0.20}_{-0.25}$. 
However, $t_0$ and the power-law index for old stars are based on fewer than $10$ active old stars. Therefore, these parameters are not well constrained with our data, as shown by the purple line fits in Figure~\ref{fig:ageactivity_fit}. 
We do not recommend using this fit for stars with ages $>$\,$776$\,Myr. 
More data for older ages is needed to calibrate the decay.
From Figure~\ref{fig:ageactivity_fit} we can conclude that there is a breaking point at which the magnetic activity starts to decline with a steeper slope according to $\halpha$, and that it is at an age $>$\,$776$\,Myr, as indicated by our $t_0$ parameter. As shown in Figure \ref{fig:ageactivity_b}, the old active stars are of spectral type $\geq{\rm M}3$, therefore this breaking point corresponds only to late-type stars.
We note that this age activity relation was obtained using only active stars. We will develop a model combining $\lhalbol$, active fraction, color and age in future work.

\section{Comparison to previous age-activity relations}
\label{sec:comparisonstudies}

\citet{West2006} studied the $\halpha$ age-activity relation by collecting a sample of M~dwarfs with 3D positions, 3D kinematics and $\halpha$ equivalent widths and luminosities.
They found that $\lhalbol$ decreases with increasing vertical distance from the galactic plane and that the slope of the relation depends on the spectral type.
As vertical distance is an age indicator such that older stars are located farther from the plane  \citep{Wielen1977,Hanninen2002}, their conclusion was that $\lhalbol$ decreases with increasing age. 
They also found that the active fraction of M~dwarfs also decreases with vertical height. 
\citet{West2008a} used a one-dimensional model to simulate the dynamics of stars in the Galaxy and fit the relation between active fraction and vertical height. 
They found that the active lifetime of M~dwarfs increases from $0.8$\,Gyr to $8$\,Gyr with spectral type (${\rm M}0$ to ${\rm M}7$), meaning that later-type M-dwarfs stay active longer than early-type ones. The results from our study of the active fraction as a function of color for different ages (Figure~\ref{fig:activityfraction}) agrees with the results from \citet{West2006} and other works that did a similar analysis \citep[e.g., ][]{Hawley1996,West2011,Schmidt2015}. 
In addition, our age-activity relation (Figures~\ref{fig:ageactivity}, \ref{fig:medianageactivity} and \ref{fig:ageactivity_fit}) qualitatively agrees with the results from \citet{West2008a}.

\begin{figure*}[ht!]
\begin{center}
\includegraphics[width=\linewidth,height=\textheight,keepaspectratio]{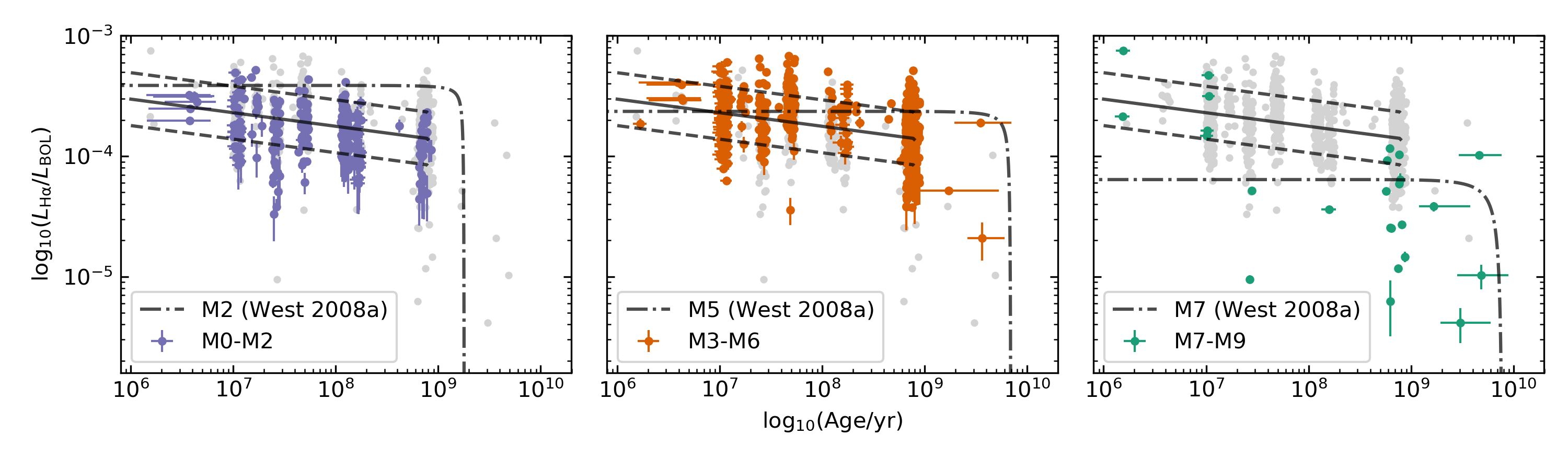}
\caption{Comparison of our result for the age-activity relation for $\lhalbol$ with the results from \citet{West2008c}. We divided our age-calibrators into three spectral type bins: in purple early partially convective M~dwarfs (${\rm M}0-{\rm M}2$), in orange mid fully convective M~dwarfs (${\rm M}3-{\rm M}6$) and in green ultracool fully convective M~dwarfs (${\rm M}7-{\rm M}9$). We show as a dotted-dashed line the age-activity relation obtained by \citet{West2008c} for ${\rm M}2$, ${\rm M}5$ and ${\rm M}7$. We also show our results for the fit for ages $<1$\,Gyr in a black line and our fit to the variability in black dashed lines.}
\label{fig:age_activity_rel_fit}.
\end{center}
\end{figure*}

\citet{West2008c} used a similar simulation as the one described above, to find ages from the vertical height of the stars, and studied the relation between $\lhalbol$ and age for M$2$-M$7$ dwarfs.
These relations are plotted as dotted-dashed lines in Figure~\ref{fig:age_activity_rel_fit} for ${\rm M}2$, ${\rm M}5$ and ${\rm M}7$ over the data for our age calibrators, which are divided according to spectral type range.
We also included our results for the fit from Figure~\ref{fig:ageactivity_fit}, as reference. We do not have enough M~dwarf-white dwarf pairs with $\haew$ measurements to constrain the power-law index for ages $>$\,$1$\,Gyr, so we did not include this part of the fit.
We found that \citet{West2008c} overestimated the value of the activity strength for early-type M~dwarfs and slightly underestimated it for late-type M~dwarfs. 
Their relation also indicates that $\lhalbol$ remains constant for young ages up to the decline for all M~dwarfs. 
Our fit does a better job at describing the data, and unlike \citet{West2008c}, we found that $\lhalbol$ for early- and mid-type M~dwarfs decreases progressively for ages $<$\,$1$\,Gyr with a slope $\alpha_1=-0.11^{+0.02}_{-0.01}$. 
As shown by the relation between $\haew$ and age for each spectral type bin in Figure~\ref{fig:medianageactivity_a}, the magnetic activity of old ($>1$\,Gyr) mid- and late-type M~dwarfs seems to decline more rapidly than young stars, in agreement with \citet{West2008c}. However, we found that early-type M~dwarfs become progressively inactive unlike their result, which indicates a steep decline.  


The age-activity relation has been studied for other magnetic activity indicators. \citet{Stelzer2013} using a $90\%$ complete sample of M~dwarfs within $10$\,pc, found that the fluxes of $\halpha$, X-ray and UV are correlated. This correlation was expected given that the three parameters are magnetic activity indicators.
Therefore we also compared our results to X-ray and UV age-activity relations. \citet{Jackson2012} studied the age-activity relation using X-ray luminosity for young F, G and K dwarfs from known young associations ($<1$\,Gyr) and found a saturated region for ages $<100$\,Myr  where the X-ray luminosity remains constant and then declines with a power-law index between $\alpha=-1.09\pm0.28$ and $-1.40\pm0.11$.
\citet{Booth2017} also studied the dependence of X-ray luminosity for a sample of F, G, K and M dwarfs with ages obtained from astroseismology or from white dwarf co-movers, such that the stars had an age $>1$\,Gyr.
They found that X-ray luminosity decreases with age with a power-law index of $\alpha=-2.8\pm0.72$.
Instead of a saturated region where $\lhalbol$ remains constant, we found that magnetic activity for stars with ages $<1$\,Gyr decreases with a small power-law index of $\alpha _1=-0.11^{+0.02}_{-0.01}$ and for stars $>1$\,Gyr of $\alpha_2=-0.88^{+0.20}_{-0.25}$ (Figures~\ref{fig:ageactivity_fit} and \ref{fig:mcmc_res}).
Although $\alpha _2$ is not well constrained by our model, our results seem to agree with \citet{Jackson2012} and \citet{Booth2017} in that the decrease of magnetic activity is faster for old stars that for young stars ($<1$\,Gyr).

\citet{Schneider2018} studied a sample of M~dwarfs from young associations ($<1$\,Gyr) and compared it to field dwarfs to characterize the UV age-activity relation. 
They found that mid- to late- M~dwarfs ($0.08-0.35$\,${\rm M\odot}$) remain relatively active throughout their lifetimes according to UV flux density, with only a small decrease of magnetic activity from young to field ages. They also found that early-Ms ($0.35-0.6$\,${\rm M\odot}$) have a much more significant decrease of magnetic activity over the same age range.
This result agrees with our result for the evolution of the magnetic activity using $\halpha$. For the active fraction as a function of $(G-\grp)$ color for different age bins in Figure~\ref{fig:activityfraction}, we found that the active fraction of early-type M~dwarfs decreases from $1$ to almost zero for field dwarfs while the active fraction for mid-type M~dwarfs remains close to one for ages $<1$\,Gyr and declines at an older age. 

\section{Conclusions}
\label{sec:conclusions}

In this study, we analyze the age-dependence of the magnetic activity of M~dwarfs from three complementary perspectives: (1) the dependence of the active fraction of spectral subtypes with age, (2) the dependence of $\halpha$ equivalent width ($\haew$) with age, and (3) the dependence of fractional $\halpha$ luminosity ($\lhalbol$) with age.
We compiled a sample of \nagecal compatible single ${\rm M}0-{\rm M}9$ dwarfs ($1,121$ in total, including repeated measurements and not compatible) to serve as age calibrators, based on \nlitsearch M~dwarfs with $\haew$ measurements collected from the literature, focusing on M~dwarf members of young associations or co-moving with a white dwarf.
We cross-matched our sample with \textit{Gaia} DR2 to obtain proper motions and parallaxes and found that \percentingaia~of our sample was in \textit{Gaia} DR2. 
From this sample we identified \nmgmem M~dwarf members of known young associations ($<1$\,Gyr) using \textit{Gaia} DR2 kinematics and parallaxes, and BANYAN~$\Sigma$ \citep{Gagne2018}, as well as \wdmpairsage M~dwarfs co-moving with a white dwarf from \citet{Fusillo2019}. 
The age for M~dwarf members of young associations was obtained from the estimated age of their association. For M~dwarfs co-moving with a white dwarf, the age was calculated with the Python package $\texttt{wdwarfdate}$ (Kiman et al. in prep) available online\footnote{\url{https://wdwarfdate.readthedocs.io/en/latest/}}. 
We have made the code used in this work available on Zenodo\footnote{\url{https://doi.org/10.5281/zenodo.4660208}} and GitHub\footnote{\url{https://github.com/rkiman/M-dwarfs-Age-Activity-Relation}}.

We present the results of our analysis as follows:
\begin{enumerate}
    \item From the \nmgmem identified members of young associations in this study, \nnewmgmem are new candidate members (Table~\ref{table:newmem}).
    \item We studied the variability of $\halpha$ at young ages for \nrepearedagecalibrators M~dwarfs with $2-6$ independent $\haew$ measurements. We found that $94\%$ of the sample has a $\Delta \haew\leq 5 {\rm \AA}$ for ages $<$\,$1$\,Gyr, where $\Delta \haew$ is the difference between the maximum and the minimum value of $\haew$ (see Figure~\ref{fig:variability}).
    \item We confirmed that both $\haew$ and $\lhalbol$ decrease with age for spectral types ${\rm M}0-{\rm M}9$. We lacked a large enough sample to determine a precise trend in late M~dwarfs ($>{\rm M}7$). \citet{Kiman2019} found a dependence of $\lhalbol$ on vertical action dispersion, which is also a proxy for age \citep{Wielen1977,Hanninen2002}. Figure $20$ of \citet{Kiman2019}, shows that $\log _{10}(\lhalbol)$ decreases from $-3.5$ to $-5$ for both mid ($ {\rm M}5\leq {\rm SpT}\leq {\rm M}8$) and late spectral types (${\rm SpT} \geq {\rm M}8$), which agrees with the dependence of $\lhalbol$ with age we found for M~dwarfs co-moving with white dwarfs (see Figure~\ref{fig:ageactivity_b}). This result indicates that mid-type M dwarfs ($\sim$\,${\rm M}3 - {\rm M}6$) have a similar $\halpha$ age-activity relation to ultracool dwarfs ($>{\rm M}6$).
    \item We classified our age-calibrators as active or inactive according to their $\haew$. Using this classification we calculated the active fraction per color bin. We confirmed that the active fraction increases with color, which is a proxy for decreasing mass, from ${\rm M}0-{\rm M}7$ in agreement with \citet{West2004} and \citet{Schmidt2015}. Moreover, by calculating the active fraction per color per age bin, we found that the active fraction varies with age according to spectral type: the active fraction of early M~dwarfs ($<{\rm M}3$) decreases gradually from $1$ to close to $0$ between $0-750$\,Myr, while later types stay active longer, such that their active fraction stays close to unity for ages $<1$\,Gyr (see Figure \ref{fig:activityfraction}). 
    \item We found that the active fraction for early-type M~dwarfs in the age-bin $700-1000$\,Myr is close to the field value ($\gg 1$\,Gyr), while mid-types have an active fraction close to unity in the age-bin $700-1000$\,Myr and $0.6$ for the field. The difference in active fraction between early and mid-types after $1$\,Gyr could be indicating that late types stay active longer and we do not have enough age resolution to distinguish the progressive decrease of the active fraction with age. This discrepancy also could be indicating that the magnetic activity of mid-type M~dwarfs decreases rapidly after $1$\,Gyr (See Figure \ref{fig:activityfraction}). 
    \item By comparing $\haew$ and $\lhalbol$ as a function of age, we found that the magnetic activity strength of early- and mid-type M~dwarfs ($<{\rm M}7$) gradually decreases during the first Gyr of their lives. After $\sim$\,$1$\,Gyr, early- and mid-types seem to behave differently. \changedtext{We found only two early M~dwarfs ($<{\rm M}3$), which are both inactive and that continue the trend of a gradual decrease of magnetic activity. For mid-type M~dwarfs we found $14$ old stars out of which $11$ are inactive and present a large decline in $\halpha$ ($\Delta \haew \sim 4$\,${\rm \AA}$), which seems to indicate that the magnetic activity of mid-type M~dwarfs decreases rapidly after $\sim$\,$1$\,Gyr.} However, higher numbers of old stars ($>1$\,Gyr) are needed to make a robust conclusion.
    \item We found that the power-law index for the relation between $\lhalbol$ and age -- using all spectral types to fit the data -- is $\alpha_1=-0.11^{+0.02}_{-0.01}$ for ages $ \lesssim 776$\,Myr. For older ages we only have $\lhalbol$ measurements for spectral types $\geq {\rm M}3$ (early-types found are inactive). We found a power-law index of $\alpha_2=-0.88^{+0.20}_{-0.25}$ for ages $ \gtrsim 776$\,Myr, yet it is poorly constrained since it was fit on a sample of $<10$ old active stars; hence we do not recommend using this relation for stars with ages of $ \gtrsim 776$\,Myr (see Figures~\ref{fig:ageactivity_fit} and \ref{fig:mcmc_res}). We conclude that spectral types $\geq {\rm M}3$ show a sharp decrease in $\halpha$ at ages $\gtrsim 776$\,Myr. More $\haew$ measurements of M~dwarfs co-moving with white dwarfs are needed to fit the age-activity relation for older ages. 
\end{enumerate}

In this study we did not take into account metallicity to calibrate the age-activity relation.
Previous studies have shown that the metallicity is a source of scatter for the relation between magnetic activity indicators and age: as the metallicity increases, the convection becomes more efficient which enhances chromospheric activity \citep[e.g., ][]{Lyra2005,Lorenzo-Oliveira2016}.
Metallicity should not affect the age-relation obtained from young associations given than the ones used in this study have similar values \citep{Malo2014a}. While metallicity might affect the relation for older stars, we are not setting a strong constraint in the age-relation for older stars, so it does not modify our conclusions.

This work represents a big advancement on the understanding of M dwarfs magnetic activity and evolution. In addition, this work is key for studying star-planet interactions, identifying habitable exoplanets and identifying true planet signals with the radial velocity method, which can be affected if the host star is magnetically active \citep[ex.][]{Robertson2013}.
In future work, we plan to model the relation between $\haew$ and age using the calibration from this study. 
We will use this model in a Bayesian algorithm to estimate ages of individual M~dwarfs from their $\haew$ combined with other age indicators to improve the precision and accuracy of M~dwarf ages.

\section{Acknowledgements}

The authors would like to thank Evgenya Shkolnik and Amelia Bayo for helpful references of $\halpha$ equivalent width measurements and Jeff Andrews and Alejandro N\'u\~{n}ez, Siyi Xu for great discussions. The authors would also like to thank John Bochanski for the helpful comments on the paper.

Support for this project was provided by a PSC-CUNY Award, jointly funded by The Professional Staff Congress and The City University of New York.

This material is based upon work supported by the National Science Foundation under Grant No. 1614527.

This work has been supported by NASA K2 Guest Observer program under award 80NSSC19K0106.

This work was supported by the SDSS Faculty and Student Team (FAST) initiative.

Support for this work was provided by the William E Macaulay Honors College of The City University of New York.

\software{TOPCAT \citep{Taylor2005}; \texttt{emcee} \citep{Foreman-Mackey2013}; \texttt{scipy} \citep{2020SciPy-NMeth}; \texttt{numpy} \citep{oliphant2006guide,van2011numpy}; \texttt{matplotlib} \citep{Hunter2007};
\texttt{Astropy} \citep{astropy:2013, astropy:2018}; \texttt{wdwarfdate} (Kiman et al. in prep.)}

\appendix

\section{Color--magnitude diagrams of young associations members used in this study}
\label{sec:mg_check}

In this section we compare the position in the \textit{Gaia} color--magnitude diagram of the stars used as members of young associations in this study, with the position of candidate members of each young association \citep{Gagne2018}, and empirical sequences based on bona fide members of young association of different ages \citep{Gagne2020}. The color--magnitude diagrams for each young association are in Figure \ref{fig:cmd_all_all}. We conclude that the stars used in this study as members are not discarded with the color--magnitude diagrams: although some stars do not follow the corresponding age sequence, they present a similar scatter as the candidate members (light blue empty circles). More study of these systems will be needed before they are confirmed as members.

\begin{figure*}[ht!]
\begin{center}
\subfloat[Taurus, $1.5\pm 0.5$\,Myr.]{\includegraphics[width=0.43\linewidth,keepaspectratio]{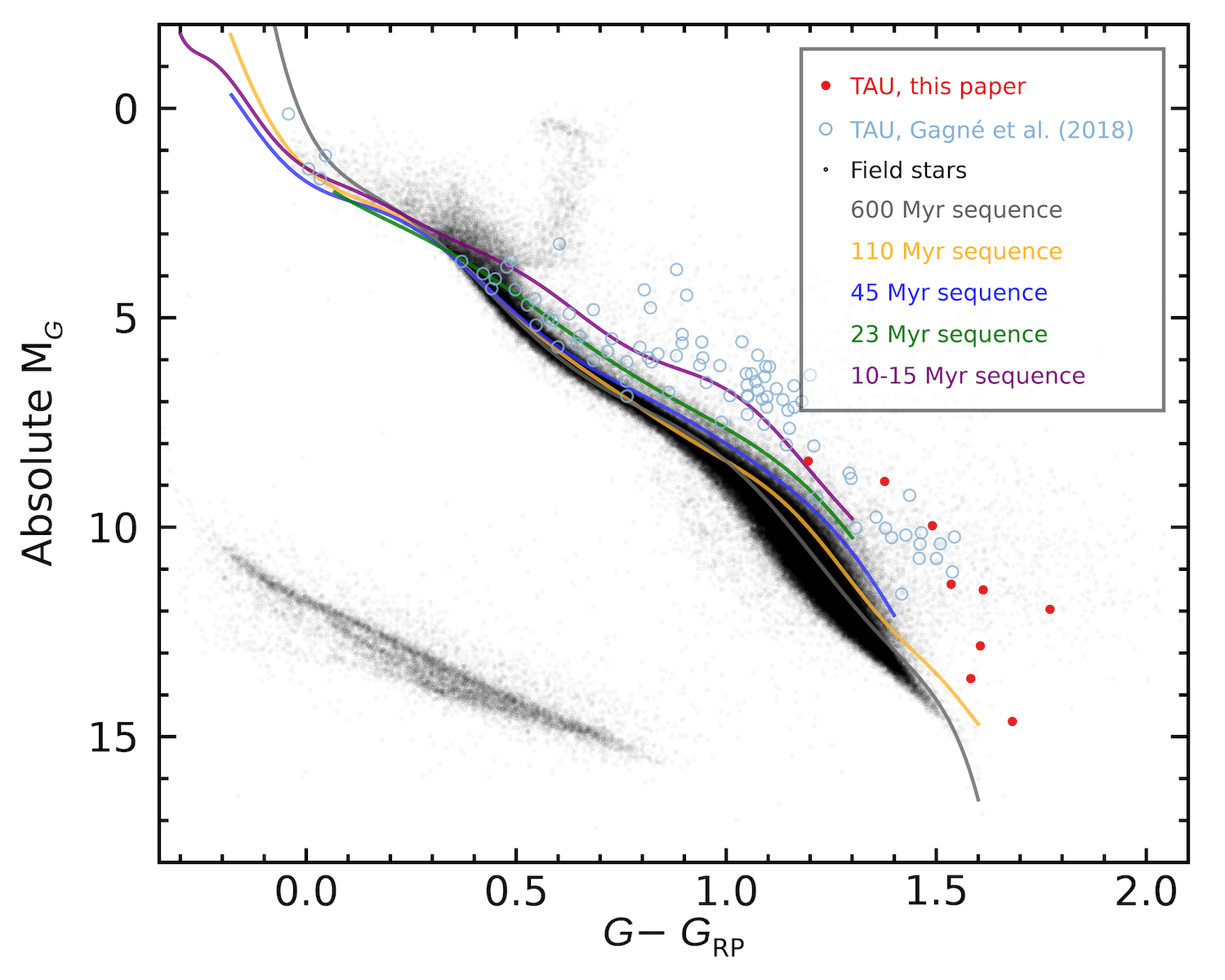}} 
\subfloat[$\epsilon$ Chamaeleontis, $3.7\pm 4.6$\,Myr.]{\includegraphics[width=0.43\linewidth,keepaspectratio]{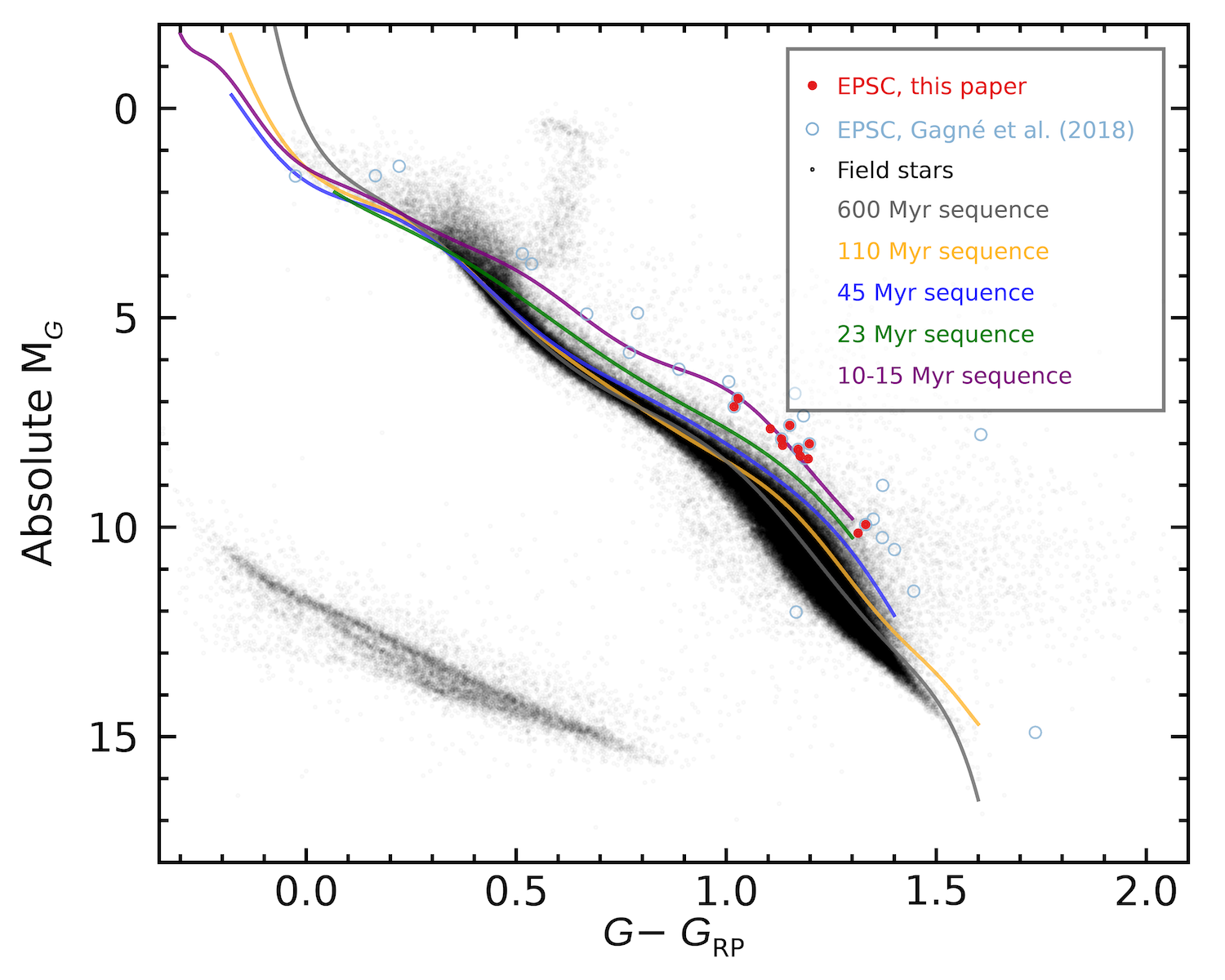}}\\
\subfloat[TW Hya, $10.0\pm 3.0$\,Myr.]{\includegraphics[width=0.43\linewidth,keepaspectratio]{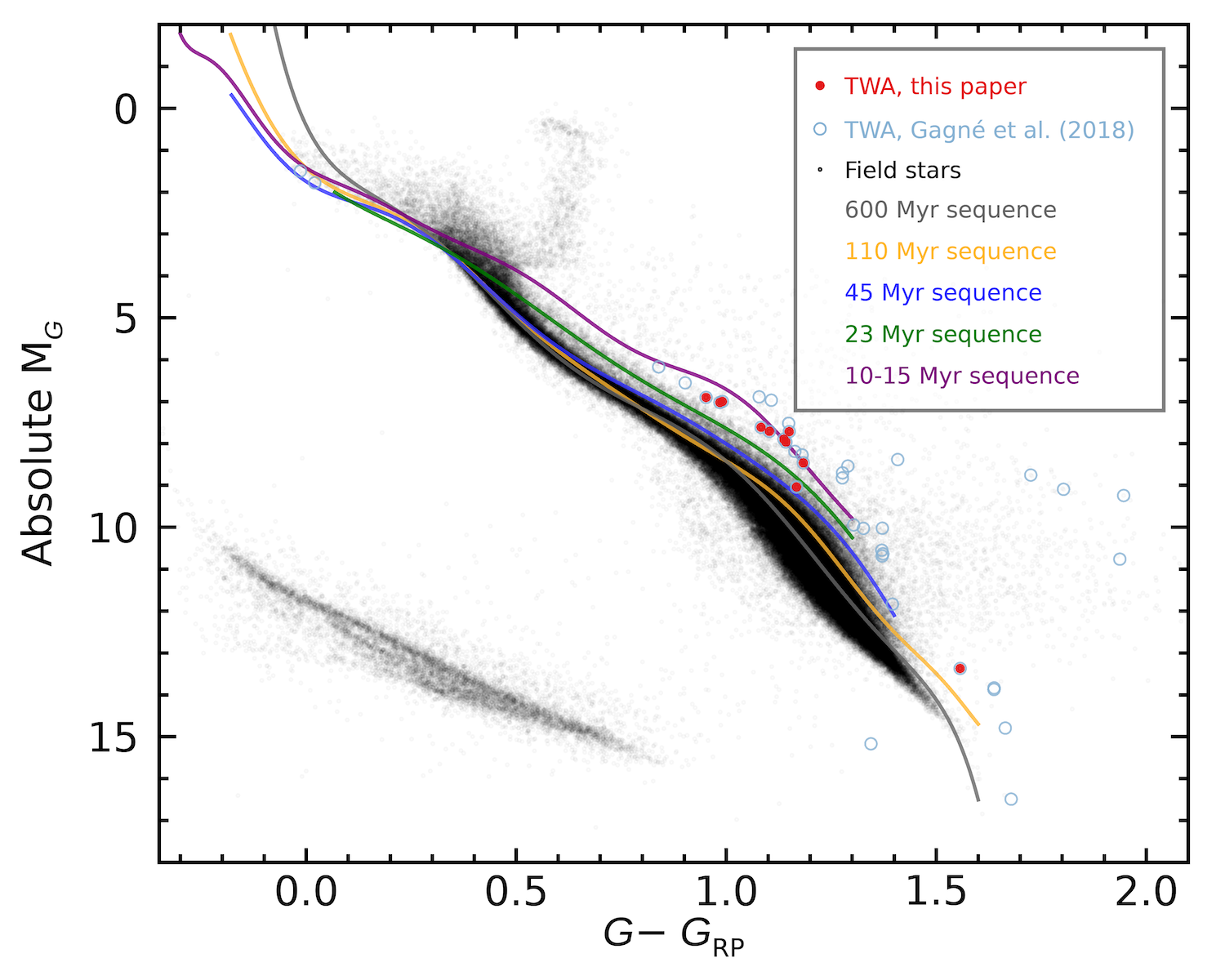}}
\subfloat[Upper Scorpius, $10.0\pm 3.0$\,Myr.]{\includegraphics[width=0.43\linewidth,keepaspectratio]{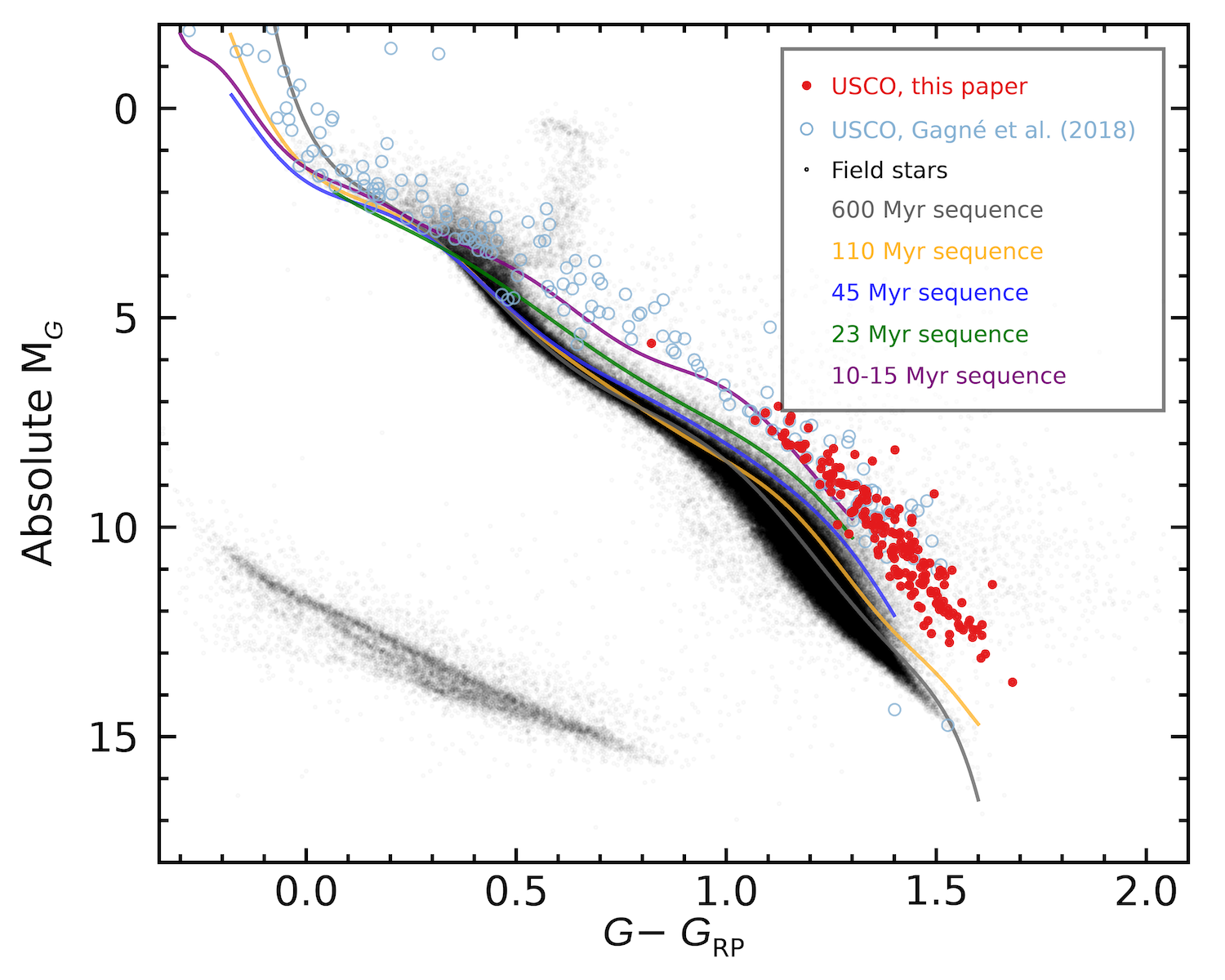}}\\
\subfloat[Upper CrA, $\sim 10$\,Myr.]{\includegraphics[width=0.43\linewidth,keepaspectratio]{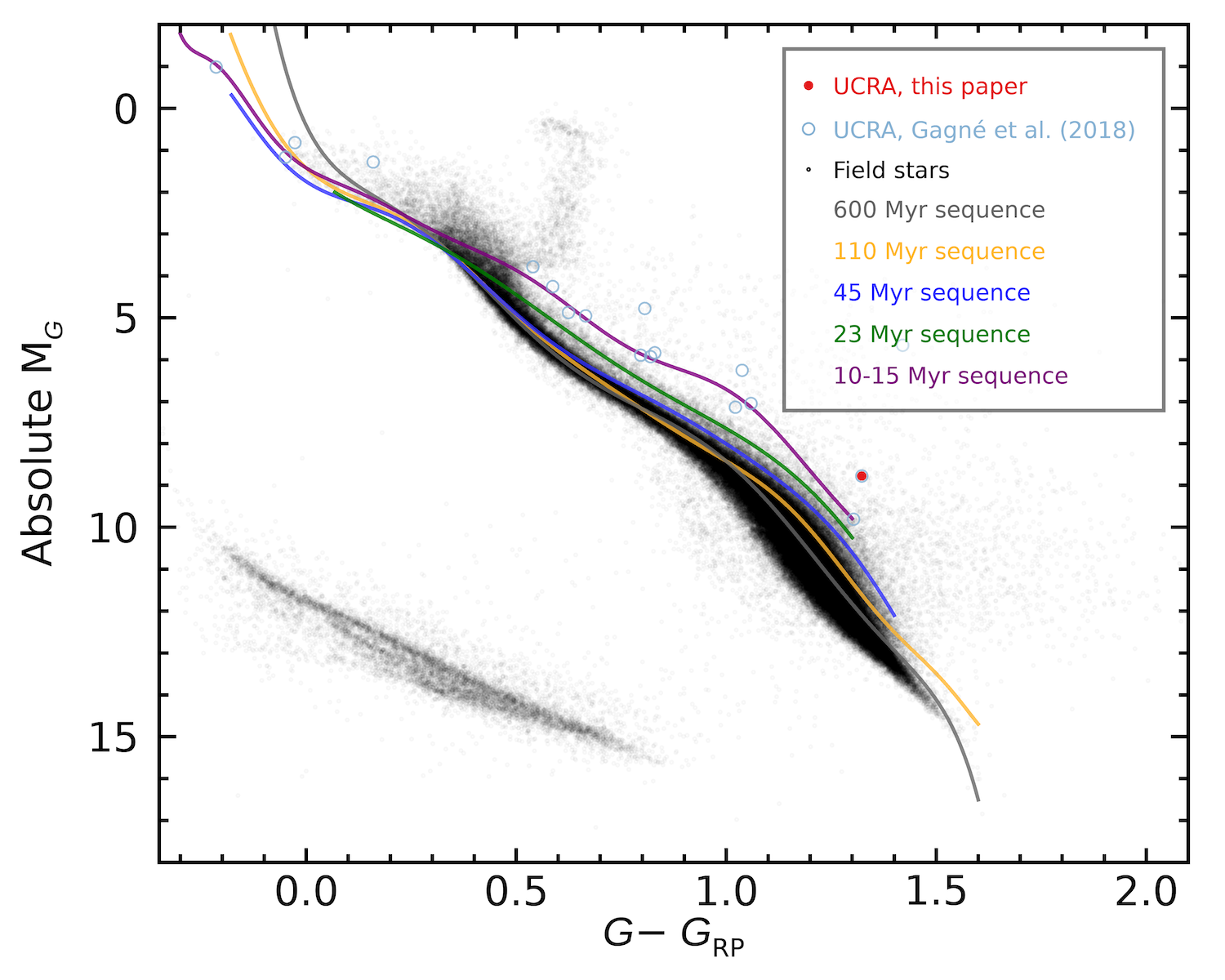}}
\subfloat[$\eta$ Chamaeleontis, $11.0\pm 3.0$\,Myr.]{\includegraphics[width=0.43\linewidth,keepaspectratio]{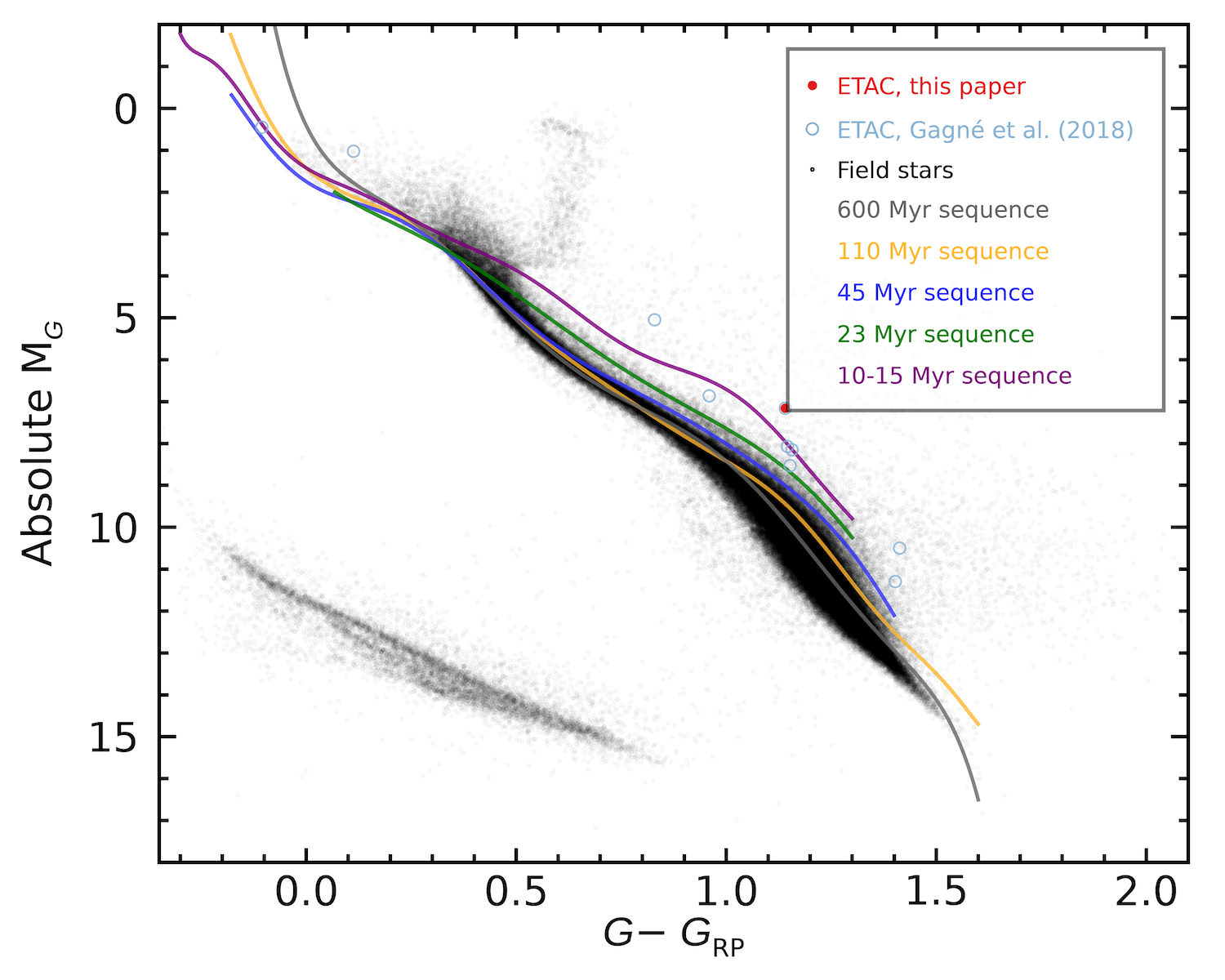}} 
\caption{CMDs comparing stars used as members of young association in this study (red circles) with candidate members of each association (light blue empty circles) \citep{Gagne2018}. We also show a sample of field stars from \textit{Gaia} DR2 in black, and the empirical sequences based on bona fide members of young association for the ages of $10-15$, $23$, $45$, $110$ and $600$\,Myr \citep{Gagne2020}. The position in the CMD does not discard as members any of stars used in this study.}
\label{fig:cmd_all_all}
\end{center}
\end{figure*}

\begin{figure*}[ht!] \ContinuedFloat
\begin{center}
\subfloat[Lower Centaurus Crux, $15.0\pm 3.0$\,Myr.]{\includegraphics[width=0.43\linewidth,keepaspectratio]{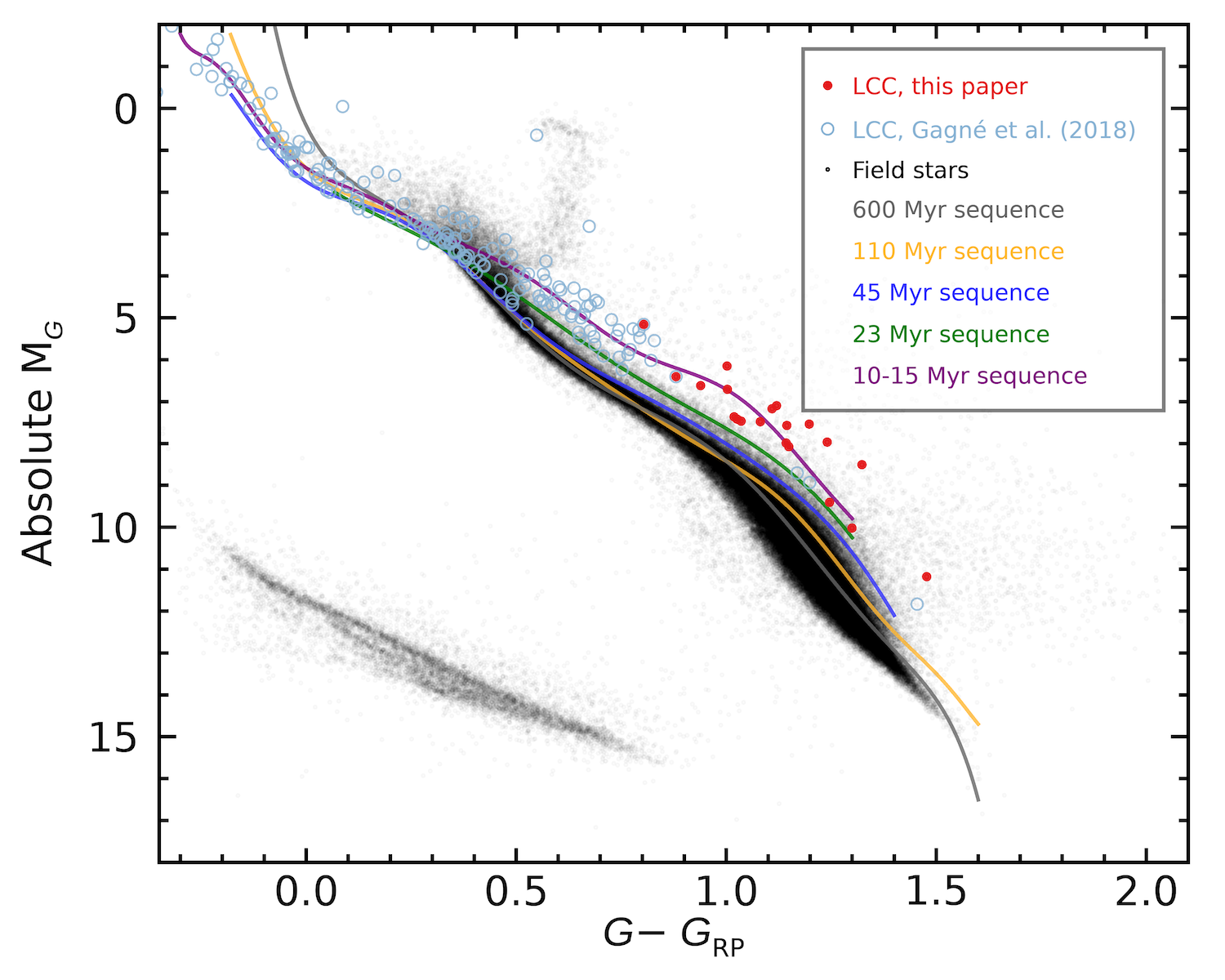}}
\subfloat[Upper Centaurus Lupus, $16.0\pm 2.0$\,Myr.]{\includegraphics[width=0.43\linewidth,keepaspectratio]{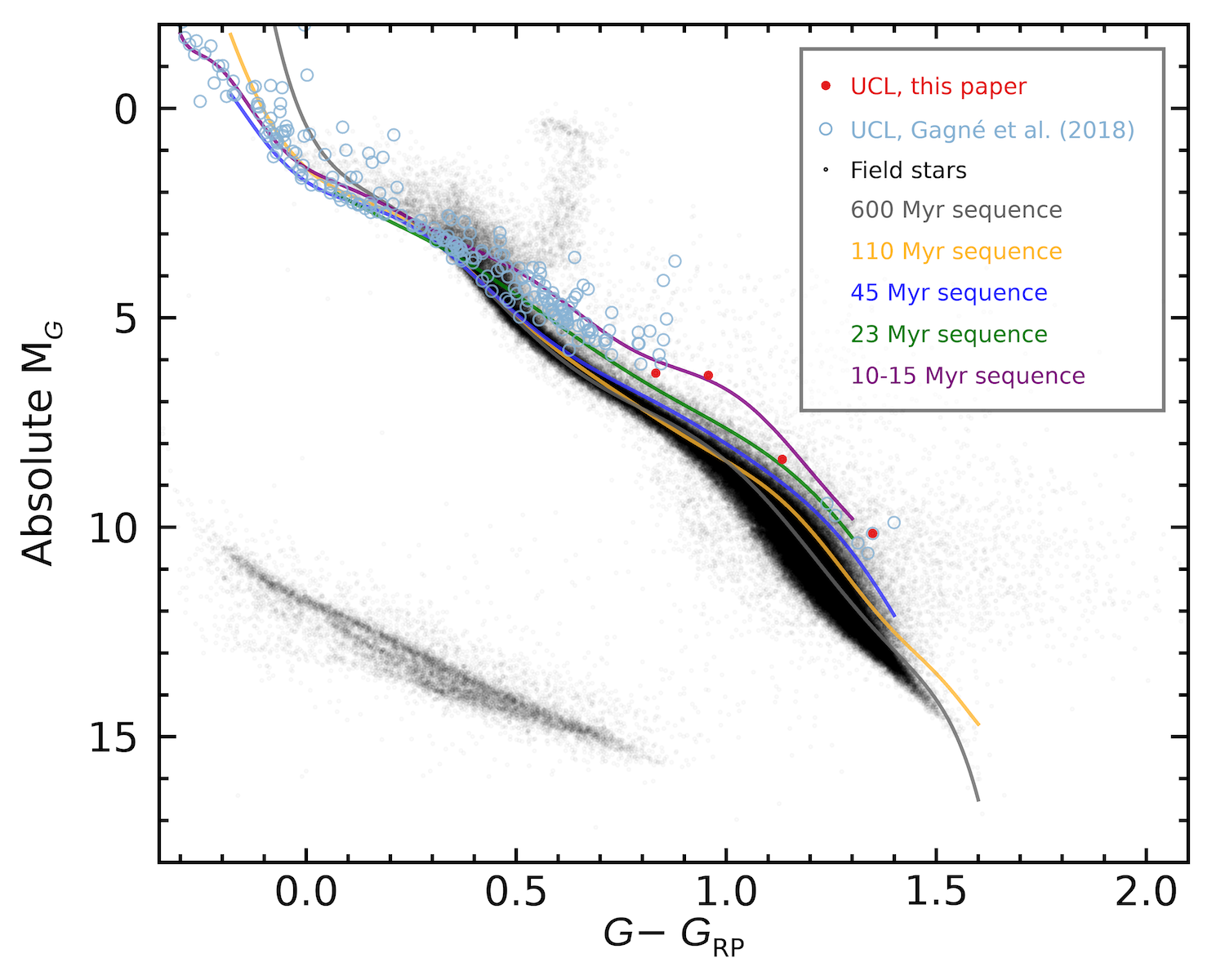}}\\
\subfloat[$\beta$ Pictoris, $24.0\pm 3.0$\,Myr.]{\includegraphics[width=0.43\linewidth,keepaspectratio]{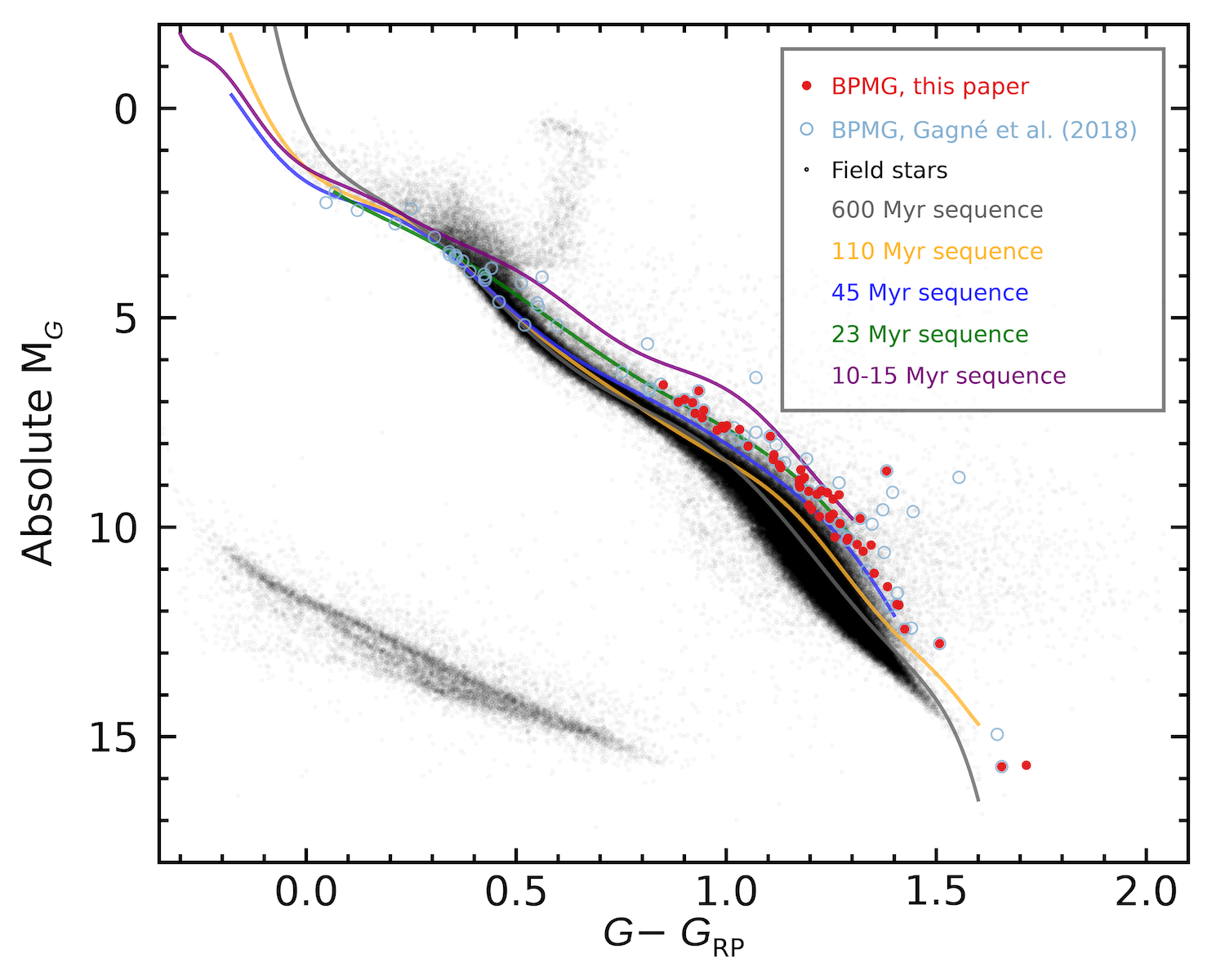}}
\subfloat[Octans, $35.0\pm 5.0$\,Myr.]{\includegraphics[width=0.43\linewidth,keepaspectratio]{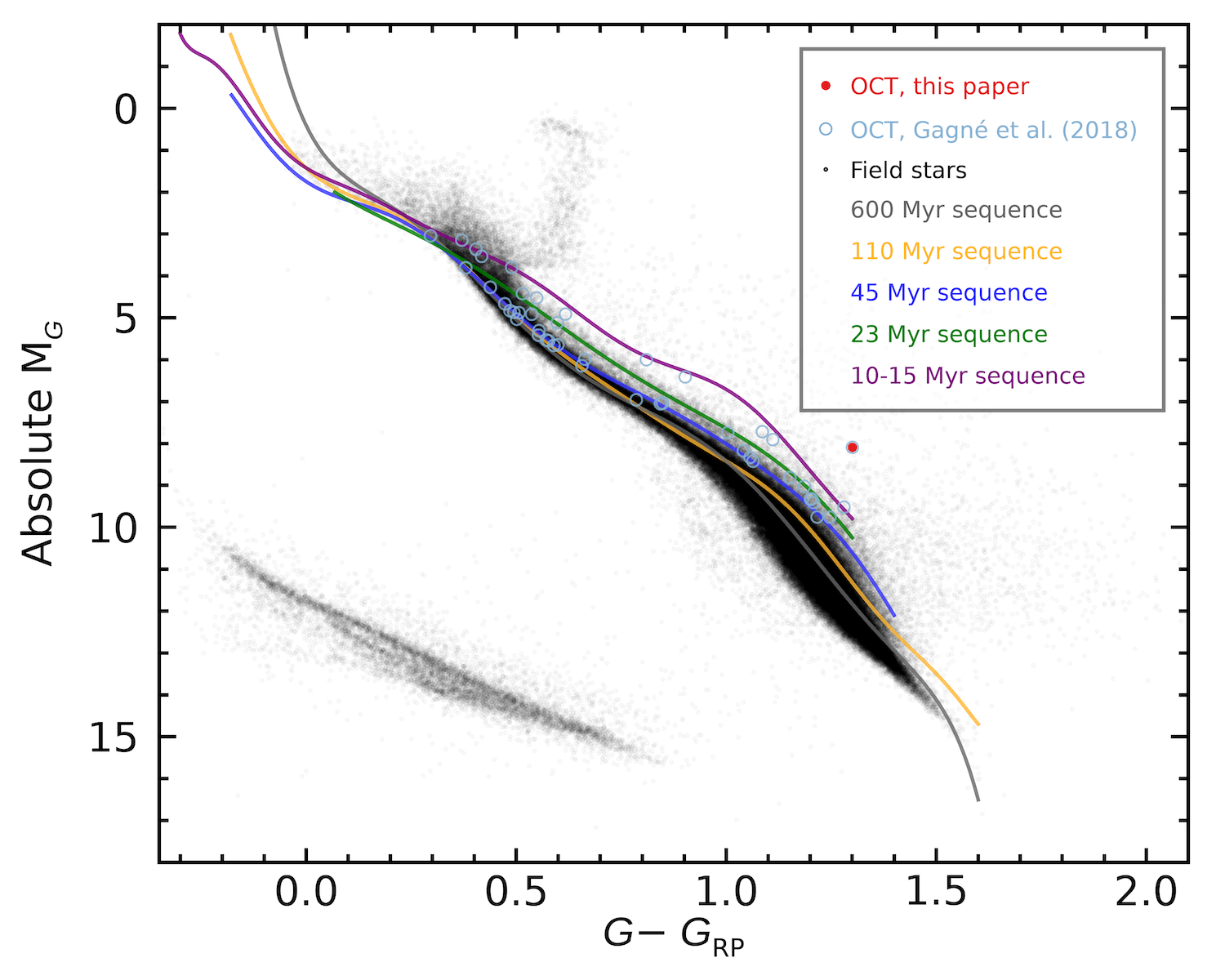}} \\
\subfloat[Argus, $40-50$\,Myr.]{\includegraphics[width=0.43\linewidth,keepaspectratio]{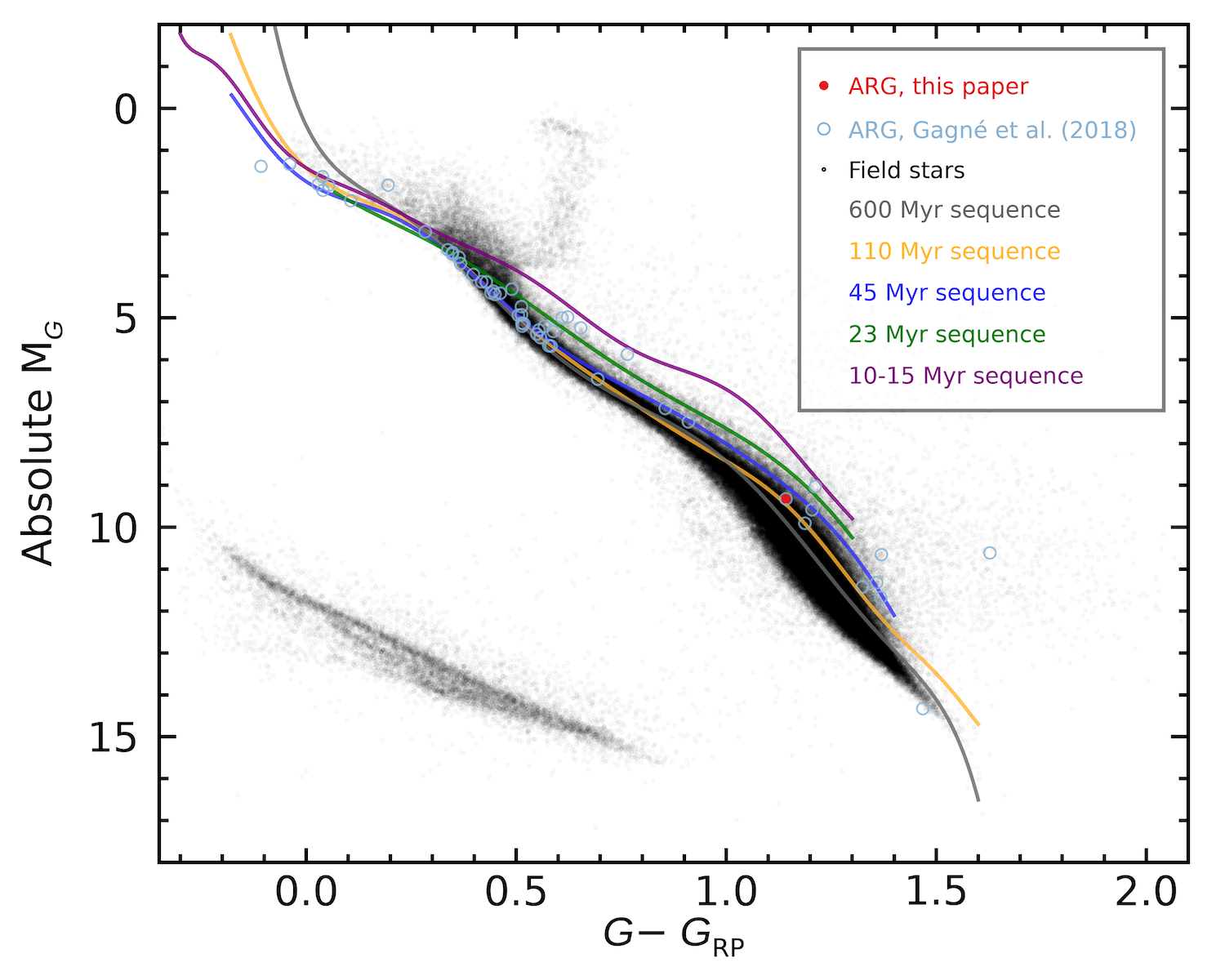}}
\subfloat[Columba, $42.0\pm 6.0$\,Myr.]{\includegraphics[width=0.43\linewidth,keepaspectratio]{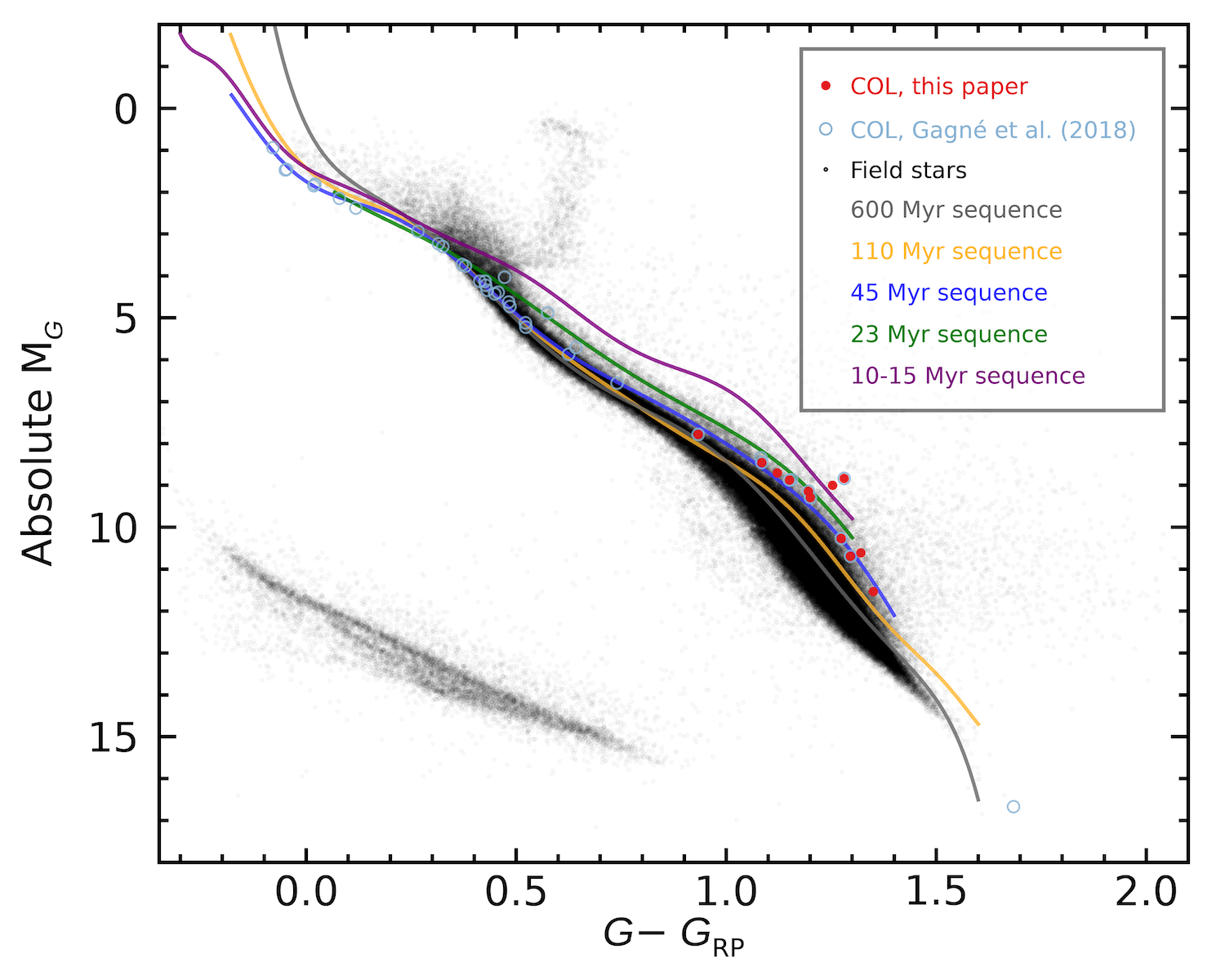}} 
\caption{CMDs comparing stars used as members of young association in this study (red circles) with candidate members of each association (light blue empty circles) \citep{Gagne2018}. We also show a sample of field stars from \textit{Gaia} DR2 in black, and the empirical sequences based on bona fide members of young association for the ages of $10-15$, $23$, $45$, $110$ and $600$\,Myr \citep{Gagne2020}. The position in the CMD does not discard as members any of stars used in this study.}
\end{center}
\end{figure*}

\begin{figure*}[ht!] \ContinuedFloat
\begin{center}
\subfloat[Carina, $45.0\pm 11.0$\,Myr.]{\includegraphics[width=0.43\linewidth,keepaspectratio]{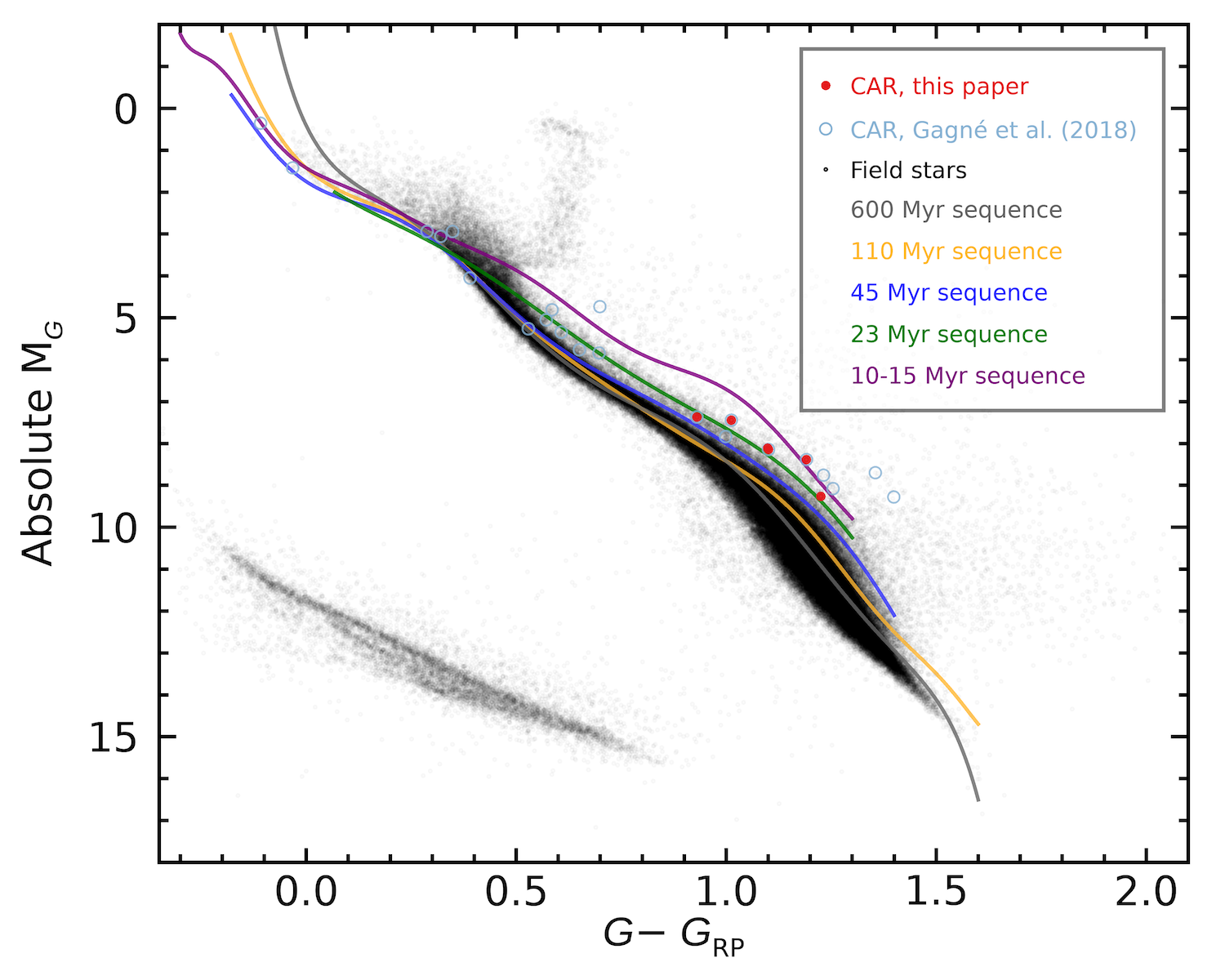}}
\subfloat[Tucana-Horologium association, $45.0\pm 4.0$\,Myr.]{\includegraphics[width=0.43\linewidth,keepaspectratio]{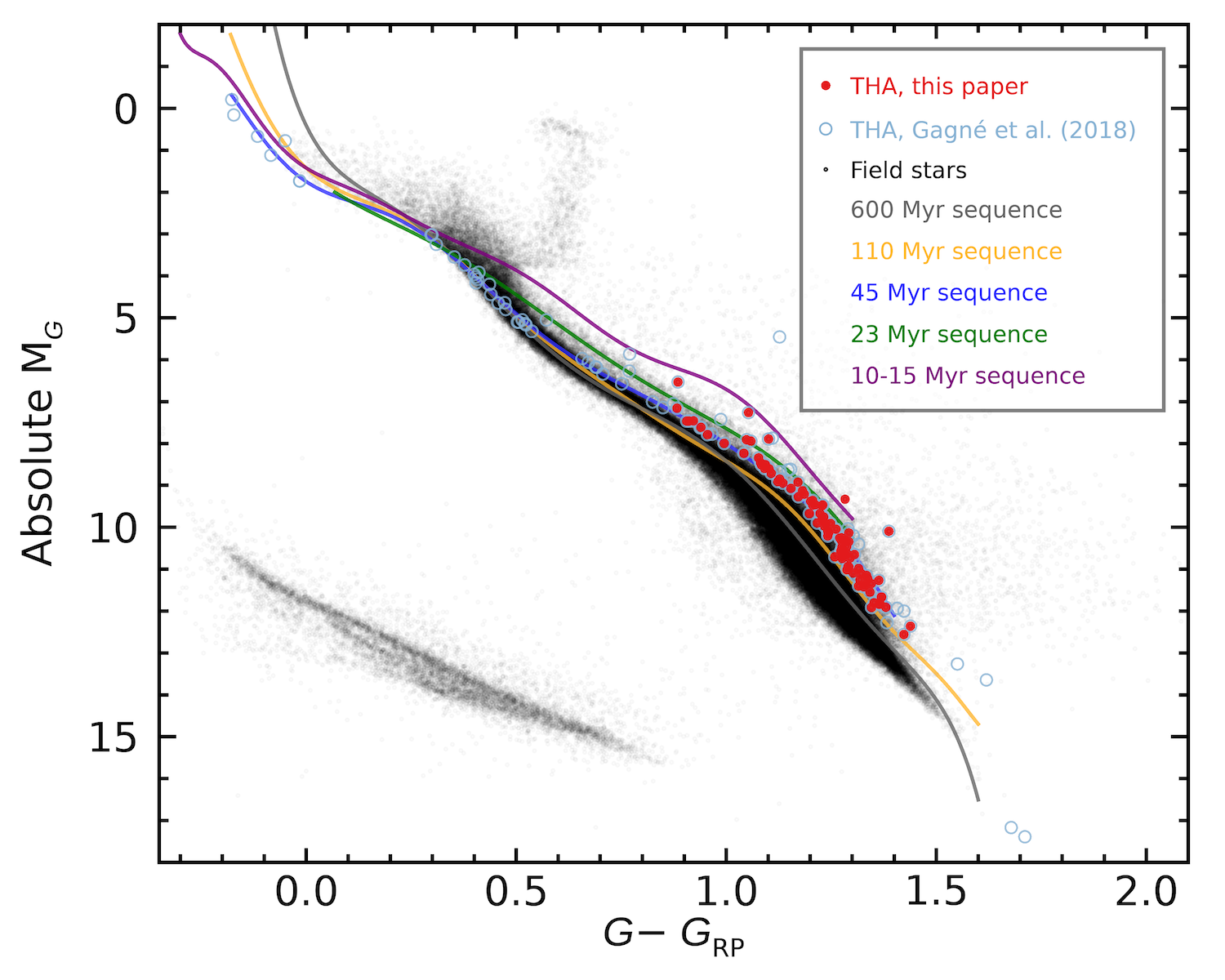}} \\
\subfloat[Pleiades cluster, $112\pm 5$\,Myr.]{\includegraphics[width=0.43\linewidth,keepaspectratio]{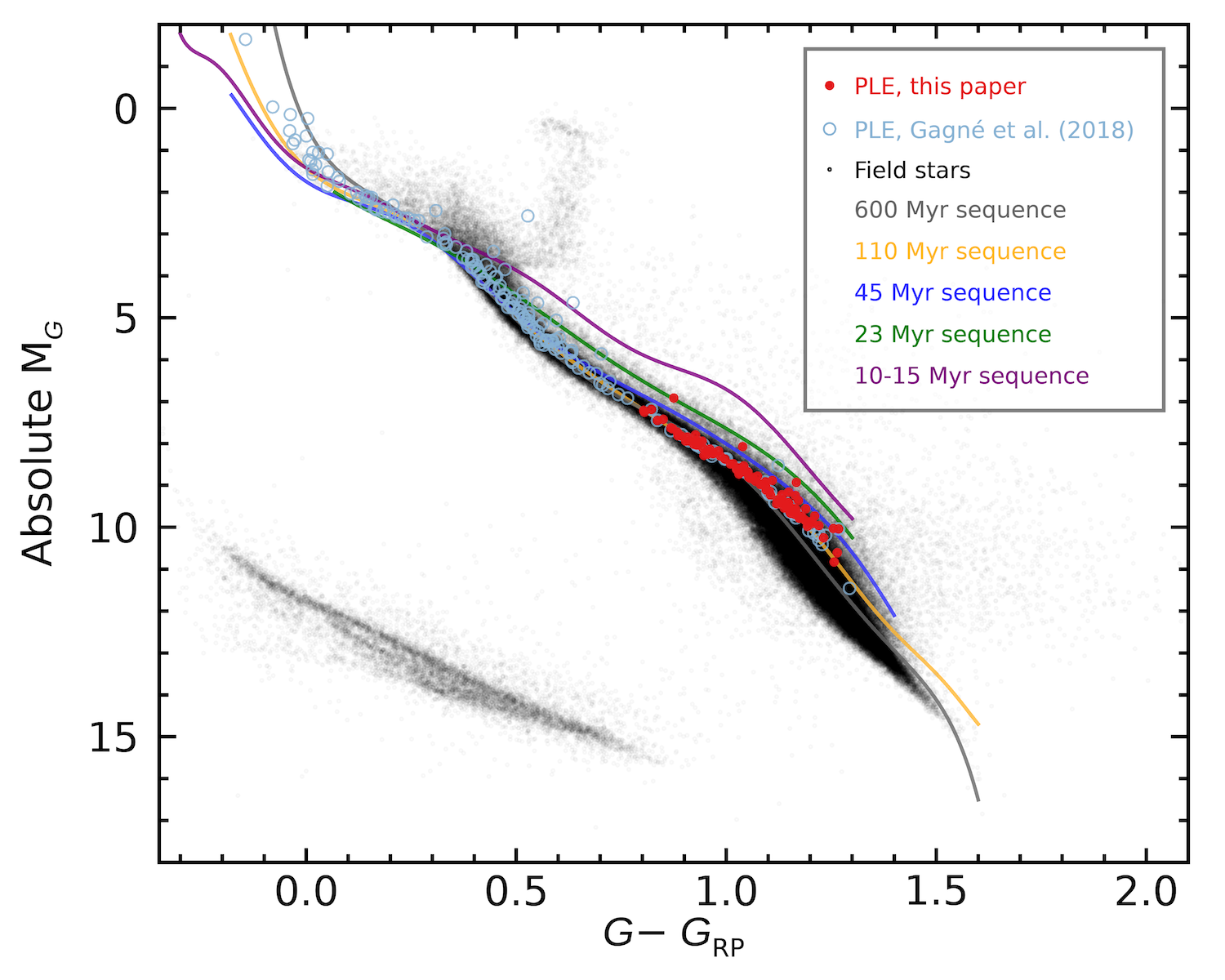}}
\subfloat[AB Doradus, $149.0\pm 51.0$\,Myr.]{\includegraphics[width=0.43\linewidth,keepaspectratio]{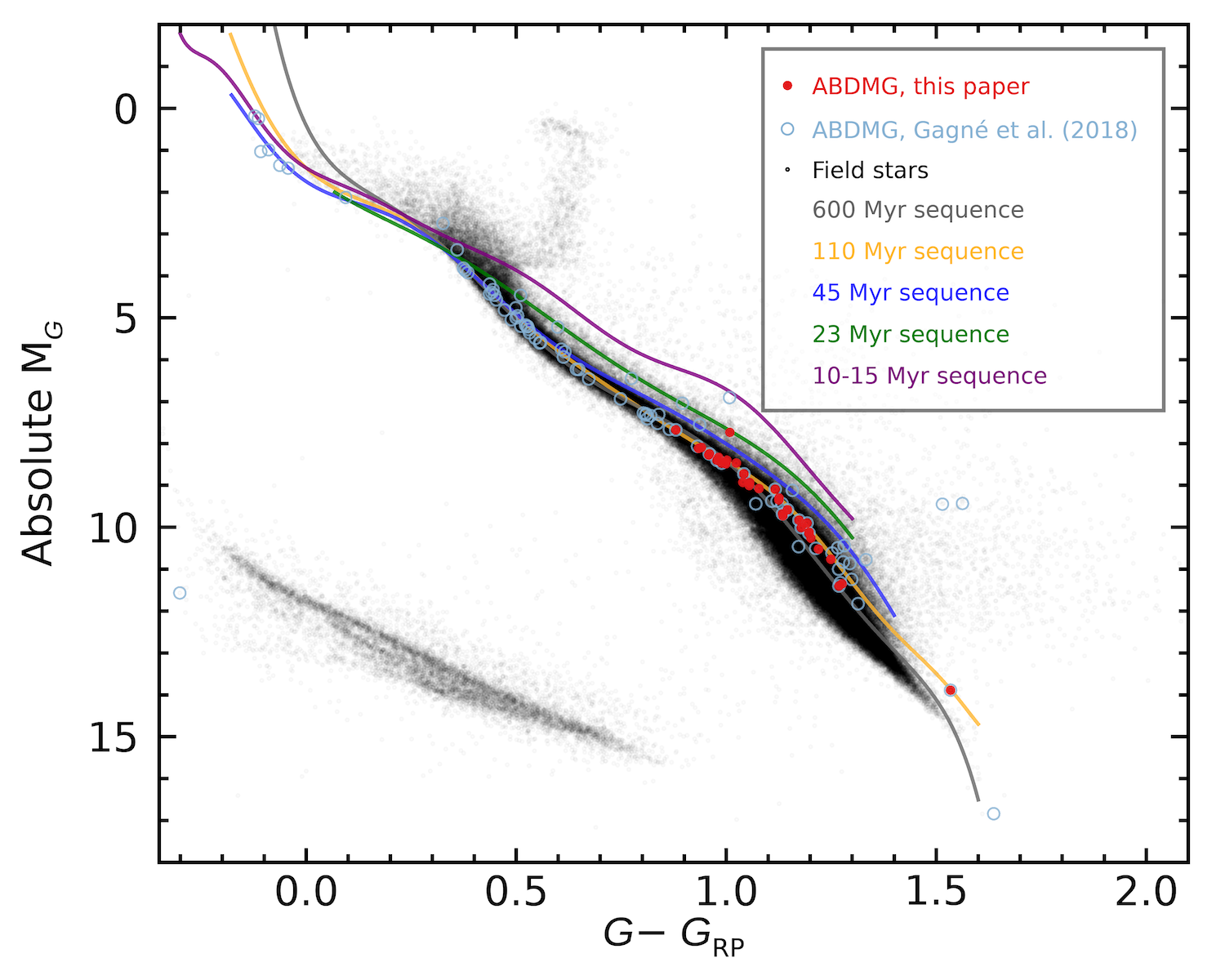}} \\
\subfloat[Carina-Near, $200.0\pm 50.0$\,Myr.]{\includegraphics[width=0.43\linewidth,keepaspectratio]{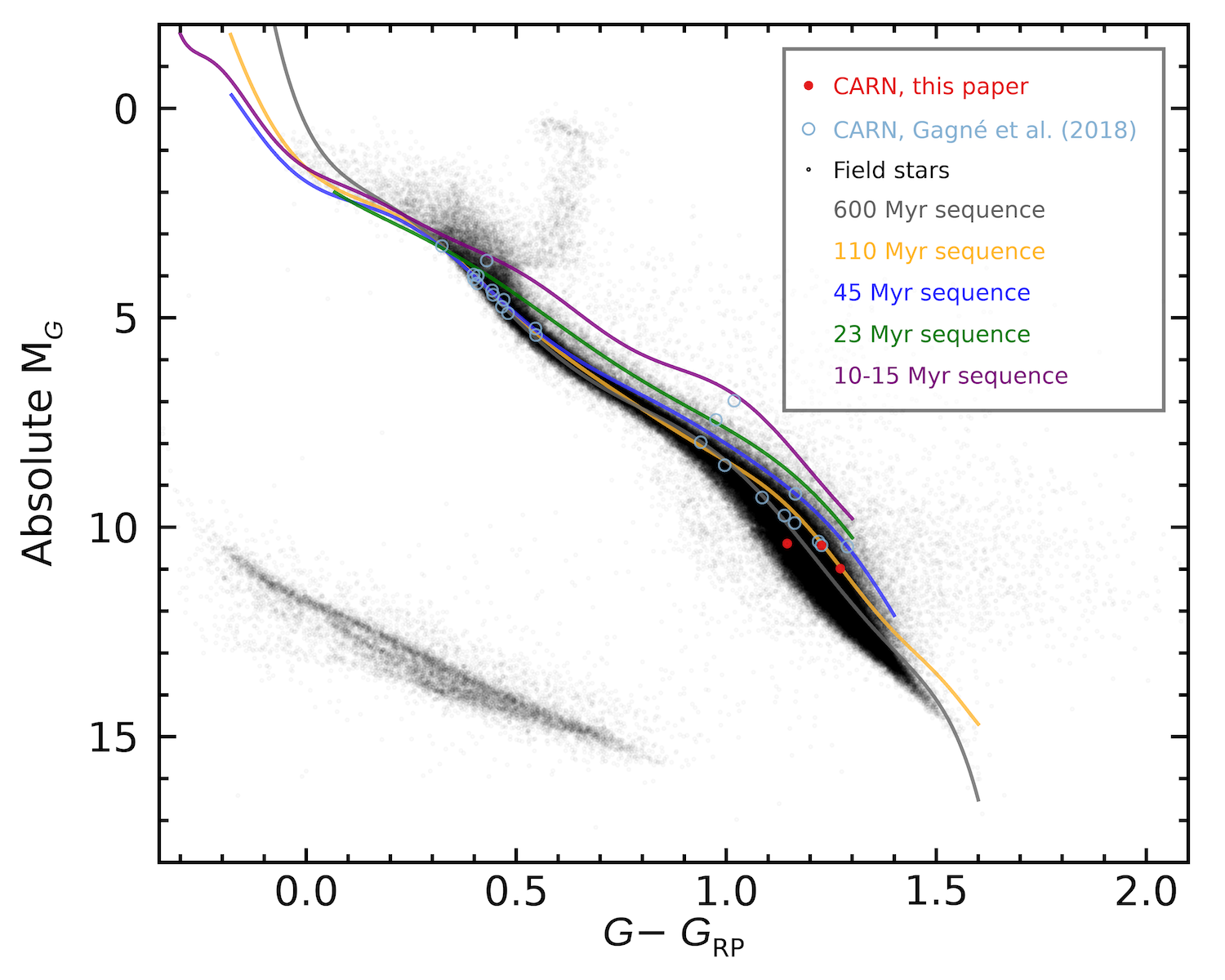}}
\subfloat[core of the Ursa Major cluster, $414.0\pm 23.0$\,Myr.]{\includegraphics[width=0.43\linewidth,keepaspectratio]{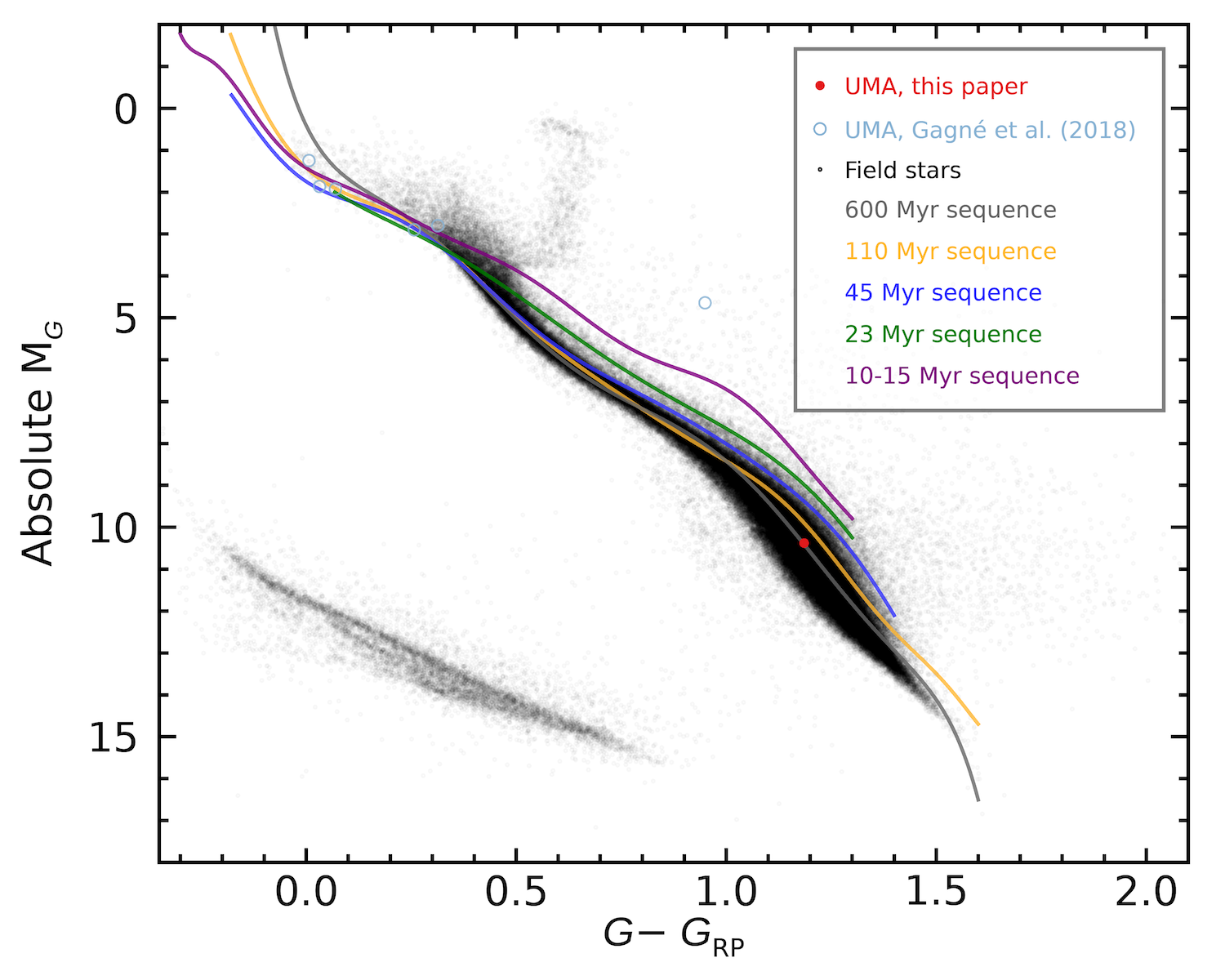}} 
\caption{CMDs comparing stars used as members of young association in this study (red circles) with candidate members of each association (light blue empty circles) \citep{Gagne2018}. We also show a sample of field stars from \textit{Gaia} DR2 in black, and the empirical sequences based on bona fide members of young association for the ages of $10-15$, $23$, $45$, $110$ and $600$\,Myr \citep{Gagne2020}. The position in the CMD does not discard as members any of stars used in this study.}
\end{center}
\end{figure*}

\begin{figure*}[ht!] \ContinuedFloat
\begin{center}
\subfloat[Coma Berenices, $562.0\pm 98.0$\,Myr.]{\includegraphics[width=0.43\linewidth,keepaspectratio]{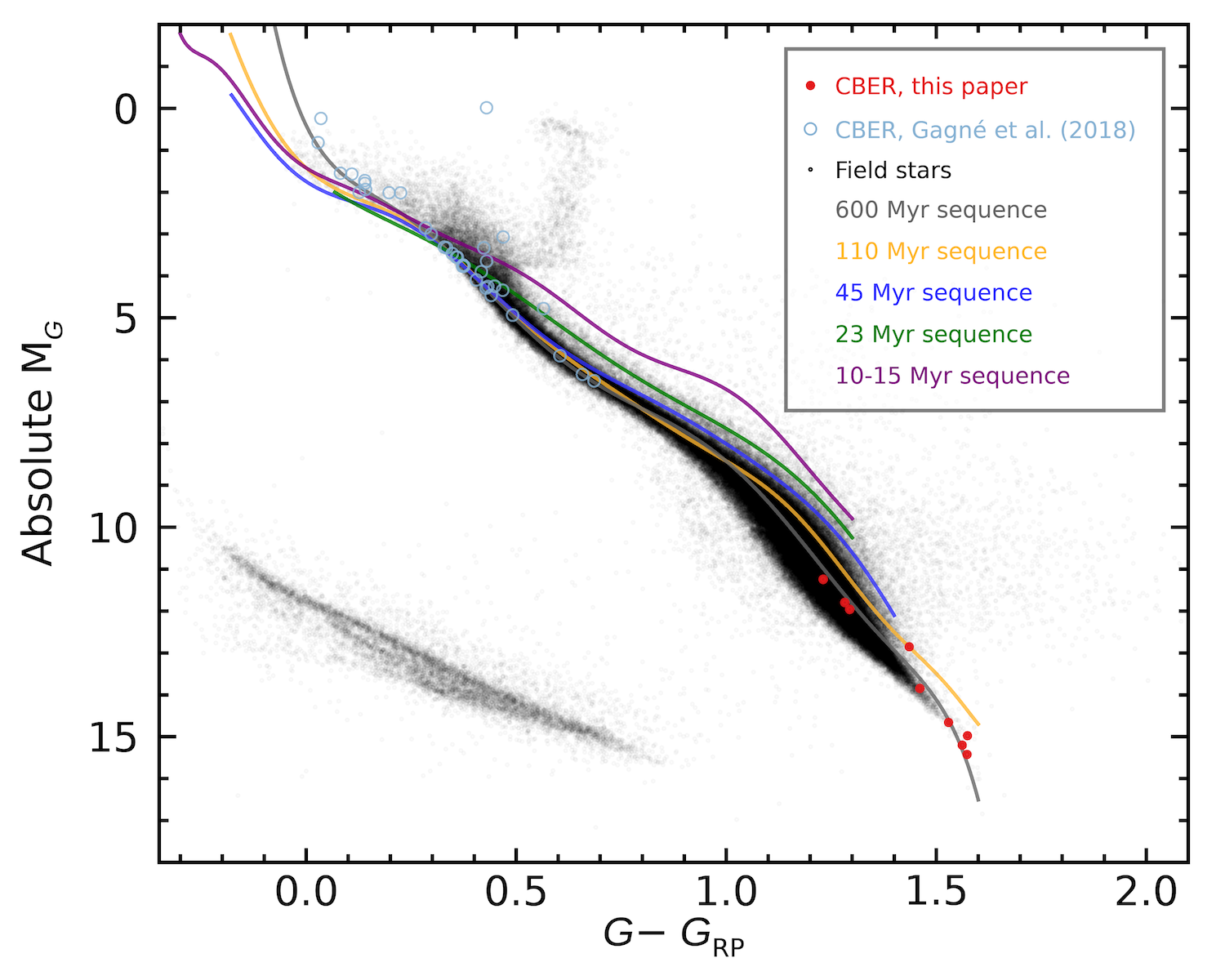}}
\subfloat[Praesepe cluster, $650.0\pm 50.0$\,Myr.]{\includegraphics[width=0.43\linewidth,keepaspectratio]{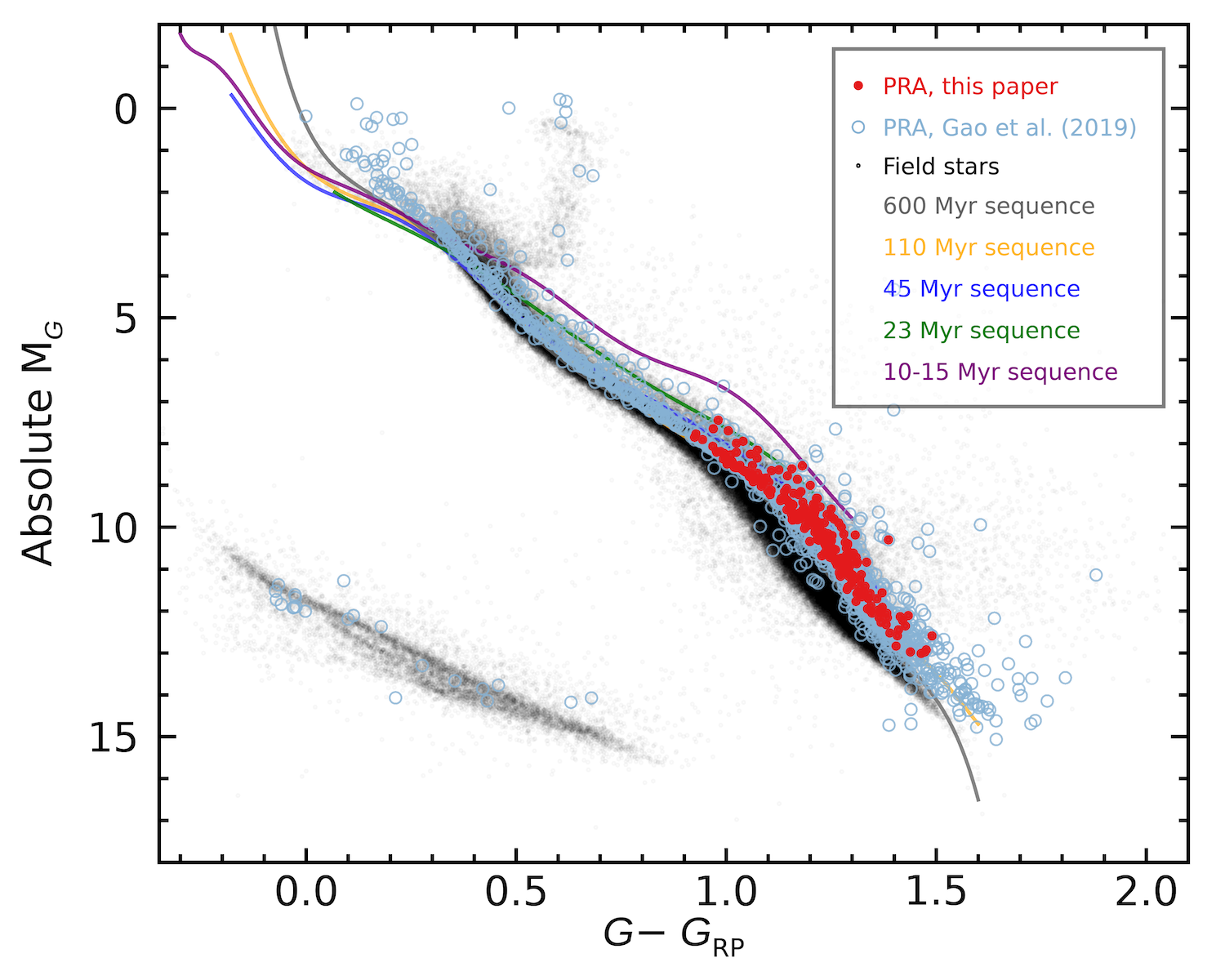}} \\
\subfloat[Hyades cluster, $750.0\pm 100.0$\,Myr.]{\includegraphics[width=0.43\linewidth,keepaspectratio]{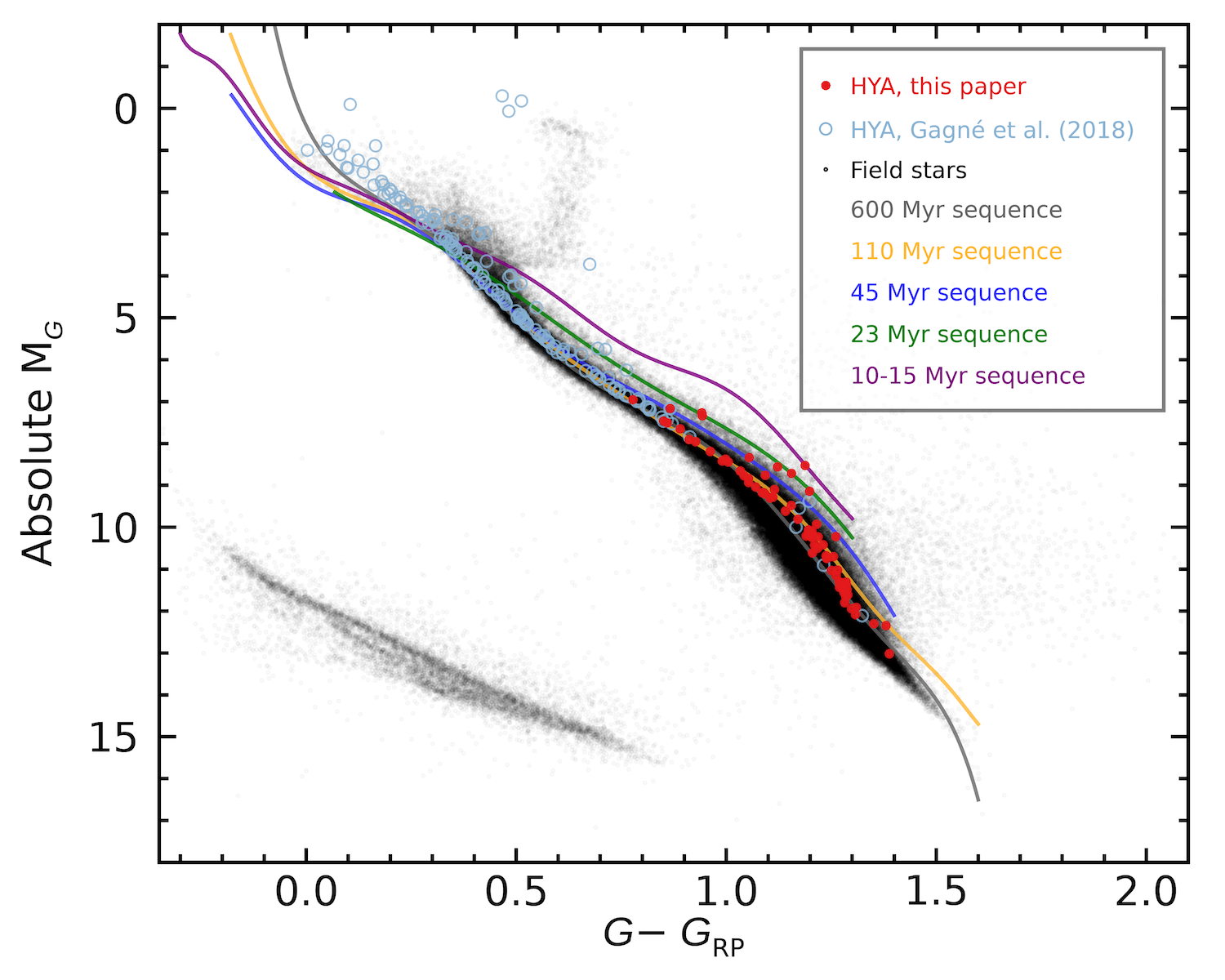}}
\caption{CMDs comparing stars used as members of young association in this study (red circles) with candidate members of each association (light blue empty circles) \citep{Gagne2018}. We also show a sample of field stars from \textit{Gaia} DR2 in black, and the empirical sequences based on bona fide members of young association for the ages of $10-15$, $23$, $45$, $110$ and $600$\,Myr \citep{Gagne2020}. The position in the CMD does not discard as members any of stars used in this study.}
\end{center}
\end{figure*}

\clearpage

\section{Comparison models to fit the age-activity relation}
\label{sec:cross-val}

Previous studies have used the broken power-law to fit age-activity relations (See Section \ref{subsec:modelandfitageactivityrel}).
As we are interested in the predictive power of our model, we tested the broken power-law against polynomials of degrees $1$ to $6$ using a cross-validation method.
In this method, we leave one of the calibration stars out and fit the rest with one of the models we want to compare.
Then we use the fitted model to predict the value of the left out element and we calculate how close the predicted value is to the true value. 
The total score for each model is defined as:
\begin{equation}
    {\rm total\,score} = \sum_k \frac{(\haew _k-f(t_k))^2}{\sigma _{\haew _k}^2}
\end{equation}
where $\haew _k$ and $t_k$ are the equivalent width and the age of the excluded $k$-element respectively and $f$ represents the model being tested so $f(t_k)$ predicts the value of $\haew$.
The model with the lowest score does the best job at predicting new data. 

Applying the described cross-validation method with the \texttt{scipy} Python package \citep{2020SciPy-NMeth}, we found that a first degree polynomial and a broken power-law had the lowest cross-validation scores, indicating that those models can more accurately predict ages based on $\halpha$.
Therefore we confirm the choice of a broken power-law to fit the age-activity relation.

\bibliographystyle{aasjournal}
\bibliography{references.bib,references_extra.bib}

\end{document}